\documentclass[11pt,english]{article}
\usepackage[OT1]{fontenc}
\usepackage[latin9]{inputenc}
\usepackage{geometry}
\usepackage{amsmath}
\usepackage{amssymb}

\makeatletter
\usepackage{amsmath}
\usepackage{amsthm}
\usepackage{amssymb}

\theoremstyle{plain}    
\newtheorem{theorem}{Theorem}[section] 

\theoremstyle{plain}    
\newtheorem{theorem-seq}{Theorem}

\newenvironment{myproof}[1][Proof.]{\par
\pushQED{\qed}%
\normalfont \topsep6\p@\@plus6\p@\relax
\trivlist
\item[\hskip\labelsep
\bfseries
#1]\ignorespaces
}{%
\popQED\endtrivlist\@endpefalse
}

\newcommand{\eqdef}{\stackrel{\rm{def}}{=}}
\DeclareMathOperator{\poly}{poly}

\newcommand{\E}{\mathbb{E}}
\newcommand{\V}{\mathbb{V}}

\newcommand{\N}{\mathbb{N}}

\newcommand{\F}{\mathbb{F}}

\newcommand{\B}{\left\{ 0,1 \right \}}

\newcommand{\NP}{\mathbf{NP}}

 \theoremstyle{plain}
 \theoremstyle{definition}
 \newtheorem{remark}[theorem]{Remark}
 \theoremstyle{definition}    
 \newtheorem{notation}[theorem]{Notation} 
 \theoremstyle{definition}
 \newtheorem{definition}[theorem]{Definition}
 \theoremstyle{plain}    
 \newtheorem{lemma}[theorem]{Lemma}  
 \theoremstyle{plain}    
 \newtheorem{proposition}[theorem]{Proposition} 
 \theoremstyle{plain}    
 \newtheorem*{proposition*}{Proposition} 
 \theoremstyle{plain}    
 \newtheorem{claim}[theorem]{Claim}
 \theoremstyle{plain}    
 \newtheorem{fact}[theorem]{Fact} 
 \theoremstyle{plain}    
 \newtheorem*{theorem*}{Theorem} 
 \theoremstyle{plain}    
 \newtheorem{corollary}[theorem]{Corollary}  
 \theoremstyle{plain}    
 \newtheorem*{lemma*}{Lemma} 
 \theoremstyle{plain}    
 \newtheorem*{claim*}{Claim}
 \theoremstyle{definition}    
 \newtheorem*{acknowledgement*}{Acknowledgement} 
 \theoremstyle{definition}
 \newtheorem*{definition*}{Definition}
 \theoremstyle{plain}    
 \newtheorem*{fact*}{Fact} 

\makeatother

\usepackage{babel}
\usepackage{hyperref}

\newcommand{\remove}[1]{}

\makeatother

\usepackage{babel}
\begin{document}

\title{Derandomized Parallel Repetition via Structured PCPs}

\author{Irit Dinur%
\thanks{Weizmann Institute of Science, ISRAEL. Email: \texttt{irit.dinur@weizmann.ac.il}.
Research supported in part by the Israel Science Foundation and by
the Binational Science Foundation and by an ERC grant.%
} \and Or Meir%
\thanks{Weizmann Institute of Science, ISRAEL. Research supported in part
by the Israel Science Foundation (grant No. 1041/08) and by the Adams
Fellowship Program of the Israel Academy of Sciences and Humanities.
Email: \texttt{or.meir@weizmann.ac.il}.%
}}
\maketitle
\begin{abstract}
A PCP is a proof system for NP in which the proof can be checked by
a probabilistic verifier. The verifier is only allowed to read a very
small portion of the proof, and in return is allowed to err with some
bounded probability. The probability that the verifier accepts a proof
of a false claim is called the soundness error, and is an important
parameter of a PCP system that one seeks to minimize. Constructing
PCPs with sub-constant soundness error and, at the same time, a minimal
number of queries into the proof (namely two) is especially important
due to applications for inapproximability.

In this work we construct such PCP verifiers, i.e., PCPs that make
only two queries and have sub-constant soundness error. Our construction
can be viewed as a combinatorial alternative to the ``manifold vs.
point'' construction, which is the basis for all the constructions
in the literature for this parameter range. The ``manifold vs. point''
PCP is based on a low degree test, while our construction is based
on a direct product test. We also extend our construction to yield
a decodable PCP (dPCP) with the same parameters. By plugging in this
dPCP into the scheme of Dinur and Harsha (FOCS 2009) one gets an alternative
construction of the result of Moshkovitz and Raz (FOCS 2008), namely:
a construction of two-query PCPs with small soundness error and small
alphabet size.

Our construction of a PCP is based on extending the derandomized direct
product test of Impagliazzo, Kabanets and Wigderson (STOC 09) to a
derandomized parallel repetition theorem. More accurately, our PCP
construction is obtained in two steps. We first prove a derandomized
parallel repetition theorem for specially structured PCPs. Then, we
show that any PCP can be transformed into one that has the required
structure, by embedding it on a de-Bruijn graph.\newpage{}
\end{abstract}
\tableofcontents{}\newpage{}

\section{Introduction}

\global\long\def\smooth{\gamma}
\global\long\def\iab{\Gamma}
\global\long\def\oab{\Sigma}
\global\long\def\csat{\textsc{CircuitSat}}
\global\long\global\long\def\set#1{\left\{  #1\right\}  }
 \global\long\global\long\def\eqdef{\stackrel{{\rm def}}{=}}
 \global\long\global\long\def\B{\left\{  0,1\right\}  }
 \global\long\global\long\def\NP{\mathbf{NP}}
 \global\long\global\long\def\aand{ \hbox{\rm {\,\, and\,\,}}}
\global\long\def\and{\aand}
 \global\long\global\long\def\N{\mathbb{N}}
 \global\long\global\long\def\span{\mbox{span}}
 \global\long\global\long\def\unsat{{\rm UNSAT}}
 \global\long\global\long\def\sat{{\rm SAT}}
 \global\long\global\long\def\l{{\rm left}}
 \global\long\global\long\def\r{{\rm right}}
 \global\long\global\long\def\E{\mathbb{E}}
 \global\long\global\long\def\V{\mathbb{V}}
 \global\long\global\long\def\eps{\varepsilon}
 \global\long\global\long\def\aps{\alpha}
 \global\long\global\long\def\ap{\alpha}
 \global\long\global\long\def\as{\alpha}
 \global\long\global\long\def\hp{h}
 \global\long\global\long\def\hps{h}
 \global\long\global\long\def\hs{h}
 \global\long\global\long\def\cps{c}
 \global\long\global\long\def\cs{c}
 \global\long\global\long\def\card#1{\left|#1\right|}
 \global\long\global\long\def\K{\mathcal{K}}
 \global\long\global\long\def\F{\mathbb{F}}
 \global\long\global\long\def\poly{{\rm poly}}
 \global\long\global\long\def\DB{\mathcal{\mathcal{DB}}}
 \global\long\global\long\def\DBs{\DB_{\Lambda,m}}
 \global\long\global\long\def\DBf{\DB_{\F,m}}
 \global\long\global\long\def\inote#1{({\bf Irit: }{#1})}
 \global\long\global\long\def\apx#1{\stackrel{#1}{\approx}}
 \global\long\global\long\def\napx#1{\stackrel{#1}{\not\approx}}
 \global\long\global\long\def\card#1{\left|#1\right|}
 \global\long\global\long\def\D{\mathcal{D}}
 \def\e 0{\eps_0} \global\long\global\long\def\sett#1#2{\left\{  #1\left|\;\vphantom{#1#2}\right.#2\right\}  }
The PCP theorem \cite{AS98,ALMSS98} says that every language in NP
can be verified by a polynomial-time verifier that queries proof of
polynomial length in a constant number of locations. The verifier
is guaranteed to always accept a correct proof, and to accept a proof
of a false claim with bounded probability (called the \emph{soundness
error}). Following the proof of the PCP theorem, research has been
directed towards strengthening the PCP theorem in terms of the important
parameters, such as the proof length, the number of queries, the soundness
error, and the randomness complexity of the verifier.

In parallel, there is a line of work attempting to expand the variety
of techniques at our disposal for constructing PCPs. Here the aim
is to gain a deeper and more intuitive understanding of why PCP theorems
hold. One of the threads in this direction is replacing algebraic
constructions by combinatorial ones. This is motivated by the intuition
that algebra is not an essential component of PCPs, indeed the definition
of PCPs involves no algebra at all. Of course, one may also hope that
the discovery of new techniques may lead to new results.

For the ``basic'' PCP theorem \cite{AS98,ALMSS98} there have been
alternative combinatorial proofs \cite{DR06,D07}. It is still a challenge
to match stronger PCP theorems with combinatorial constructions. Such
is the work of the second author \cite{M09_efficient_PCPs} on PCPs
with efficient verifiers. In this paper we seek to do so for PCPs
in the small soundness error regime.

In this work we give a new construction of a PCP with sub-constant
soundness error and two queries. This setting is particularly important
for inapproximability, as will be discussed shortly below. In addition,
our PCP maintains the polynomial proof length and logarithmic randomness
complexity of the original PCP theorem of~\cite{AS98,ALMSS98}. Formally,
we prove
\begin{theorem}
[\label{thm:main}Two-query PCP with small soundness]There exists
a constant $\kappa>0$ such that for every function $\varepsilon:\N\to\left(0,1\right)$
satisfying $1/n^{\kappa}\le\varepsilon(n)\le1/\poly\log n$ the following
holds: Every language $L\in\NP$ has a two-query PCP system with perfect
completeness, soundness error $1/\poly\log n$, alphabet size $2^{1/\poly\left(\varepsilon\right)}$,
proof length $\poly\left(n\right)$, and randomness complexity $O(\log n)$.
Furthermore, the verifier in this PCP system makes only `projection'
queries.
\end{theorem}
This theorem matches the parameters of the folklore ``manifold vs.
point'' construction which has been the only construction in the
literature for this parameter range. The technical heart of that construction
is a sub-constant error low degree test~\cite{RS97,AS03}, see full
details in \cite{MR08}.

Our proof of Theorem~\ref{thm:main} is based on the elegant derandomized
direct product test of \cite{IKW09}. In a nutshell, our construction
is based on applying this test to obtain a ``derandomized parallel
repetition theorem''. While it is not clear how to do this for an
arbitrary PCP, it turns out to be possible for PCPs with certain structure.
We show how to convert any PCP to a PCP with the required structure,
and then prove a ``derandomized parallel repetition theorem'' for
such PCPs, thereby getting Theorem~\ref{thm:main}. The derandomized
parallel repetition theorem relies on a reduction from the derandomized
direct product test of \cite{IKW09}.

\paragraph{The Moshkovitz-Raz Construction.}

Recently, Moshkovitz and Raz~\cite{MR08} constructed even stronger
PCPs. Specifically, they managed to remove the limitation $\varepsilon(n)\le1/\poly\log n$
from Theorem~\ref{thm:main}, thus allowing any function $\varepsilon(n)\ge1/n^{\kappa}$.
This allows constructing PCPs with sub-constant error and \emph{any}
alphabet size smaller than~$2^{\poly\log n}$, at the expense of
a suitable increase in the soundness error. Being able to reduce the
alphabet size has strong consequences for inapproximability, see \cite{MR08}
for details. The technique of \cite{MR08} (as explained in the later
simplification of \cite{DH09}) is essentially based on the composition
of certain PCP constructions. In fact, their main building block is
the ``manifold vs. point'' construction mentioned above.

Our construction can be extended to yield a so-called decodable PCP~\cite{DH09},
which is an object slightly stronger than a PCP. This can be plugged
into the scheme of \cite{DH09} to give a nearly%
\footnote{It is debatable whether our use of ``linear structure'' disqualifies
the result from being considered purely combinatorial.%
} combinatorial proof of the following 
result of \cite{MR08}. Namely,
\begin{theorem}
[\label{thm:MR}\cite{MR08}]There exists a constant $\kappa>0$
such that for every function $\varepsilon(n)\ge1/n^{\kappa}$ the
following holds: Every language $L\in\NP$ has a two-query PCP system
with perfect completeness, soundness error $\varepsilon$, alphabet
size at most $2^{1/\poly\left(\varepsilon\right)}$, proof length
$\poly\left(n\right)$, and randomness complexity $O(\log n)$. Furthermore,
the verifier in this PCP system makes only `projection' queries.
\end{theorem}
We note that the result of \cite{MR08} is in fact even stronger than
claimed above since their verifier has almost-linear proof length
(specifically $n^{1+o(1)}$), and has randomness complexity of only
$(1+o(1))\log n$ random bits, see also Remark~\ref{rem:MR-short}.

\paragraph*{Organization of the introduction. }

In the following four sections we outline the background and main
ideas of this work. We start by describing the parallel repetition
technique in general and its relation with direct product tests. We
proceed to describe our technique of derandomized parallel repetition.
We then describe our notion of ``PCPs with linear structure'', to
which the derandomized parallel repetition is applied.

After the foregoing outline, we discuss relevant works and possible
future directions, and describe the organization of this work.

\subsection*{Parallel repetition and Direct Products}

A natural approach to reducing the soundness error of a PCP verifier
is by running it several times independently, and accepting only if
all runs accept. This is called \emph{sequential repetition}. Obviously,
if the verifier is invoked $k$ times the soundness error drops exponentially
in $k$. However, the total number of queries made into the proof
grows $k$-fold, and in particular, it is greater than $2$. Since
our focus is on constructing PCPs that make only two queries, we can
not afford sequential repetition.

In order to decrease the soundness error while maintaining the query
complexity, one may use \emph{parallel repetition}. For the rest of
this discussion, we consider only PCPs that use only two queries.
Let us briefly recall what parallel repetition means in this context.
As in the case of sequential repetition, one starts out with a PCP
with constant soundness error, and then amplifies the rejection probability
by repetition of the verifier. However, in order to save on queries,
the prover is expected to give the $k$-wise direct product encoding
of the original proof. Formally, if $\pi:[n]\to\Sigma$ describes
the original proof then its direct product encoding, denoted by $\pi^{\otimes k}$,
is the function $\pi^{\otimes k}:[n]^{k}\to\Sigma^{k}$ defined by
\[
\pi^{\otimes k}(x_{1},\ldots,x_{k})=(\pi(x_{1}),\ldots,\pi(x_{k})).
\]
The new verifier will simulate the original verifier on $k$ independent
runs, but will read only \emph{two} symbols from the new proof, which
together contain answers to $k$ independent runs of the original
verifier.

Of course, there is no a priori guarantee that the given proof is
a direct product encoding $\pi^{\otimes k}$ of any underlying proof~$\pi$,
as intended in the construction. This is the main difficulty in proving
the celebrated parallel repetition due to Raz \cite{R98} that shows
that the the soundness error does go down exponentially with $k$.

One may try to circumvent the difficulty in analyzing the parallel
repetition theorem by augmenting it with a direct product test. That
is, making the verifier \emph{test} that the given proof $\Pi$ is
a direct product encoding of some string $\pi$, and only then running
the original parallel repetition verifier. This can sometimes be done
without even incurring extra queries. Motivated by this approach Goldreich
and Safra~\cite{GS00} suggested and studied the following question:
\\

\textbf{DP testing:} Given a function $F:[n]^{k}\to\Sigma^{k}$ test
that it is close to $f^{\otimes k}$ for some $f:[n]\to\Sigma$.\\

\noindent Let us now describe a two query direct product test. From
now on let us make the simplifying assumption that the function $F:[n]^{k}\to\Sigma^{k}$
to be tested is given as a function of $k$-sized subsets rather than
tuples, meaning that $F(x_{1},\ldots,x_{k})$ is the same for any
permutation of $x_{1},\ldots,x_{k}$. The test chooses two random
$k$-subsets $B_{1},B_{2}\in{[n] \choose k}$ that intersect on a
subset $A=B_{1}\cap B_{2}$ of a certain prescribed size and accept
if and only if $F\left(B_{1}\right)_{|A}=F\left(B_{2}\right)_{|A}$.
This test was analyzed further in several works, see \cite{GS00,DR06,DG08,IKW09}.
\begin{remark}
An expert reader may note that the above direct product test is not
a projection test, while we need a projection test for Theorem~\ref{thm:main}.
Indeed, in our actual proof we use a variant of the above direct product
test which is a projection test (see Section~\ref{sub:prelim-IKW}
for details).
\end{remark}

\subsection*{Derandomized Direct Product Testing}

Recall that our goal is to construct PCPs with sub-constant soundness
error. Note, however, that since the parallel repetition increases
the proof length exponentially in $k$ (and the randomness of the
verifier grows $k$-fold), one can only afford to make a constant
number of repetitions if one wishes to maintain polynomial proof length
and logarithmic randomness complexity. On the other hand, obtaining
sub-constant soundness error requires a super-constant number of repetitions.

This leads to the derandomization question, addressed already 15 years
ago~\cite{FK95}. Can one recycle randomness of the verifier in the
parallel repetition scheme without losing too much in soundness error?

Motivated by this question, Impagliazzo, Kabanets, and Wigderson \cite{IKW09}
introduced a method for analyzing the direct product test which allowed
them to derandomize it. Namely, they exhibited a relatively small
collection of subsets $\K\subset{[n] \choose k}$, and considered
the restriction of the direct product encoding $f^{\otimes k}$ to
this collection. They then showed that this form of derandomized direct
product can be tested using the above test. The collection $\K$ is
as follows: identify $[n]$ with a vector space $\F^{m}$, let $k=\card{\F}^{d}$
for constant $d$, and let $\K$ be the set of all $d$-dimensional
linear subspaces.

A natural next step is to use the derandomized direct product of \cite{IKW09}
to obtain a derandomized parallel repetition theorem. Recall that
the parallel repetition verifier works by simulating $k$ independent
invocations of the original verifier on $\pi$, and querying the (supposed)
direct product $\Pi$ on the resulting $k$-tuples of queries. However,
in the derandomized setting, the $k$-tuples of queries generated
by the verifier may fall outside $\K$. This is the main difficulty
that we address in this work.

This is where the structure of the PCP comes to our aid. We show that
for PCPs with a certain linear structure, the $k$-tuples of queries
can be made in a way that is compatible with the derandomized direct
product test of \cite{IKW09}. More specifically, the $k$-tuples
of queries always belong to the collection~$\mathcal{K}$, and are
distributed like queries of the derandomized direct product test.
This allows us to prove a derandomized parallel repetition theorem
for the particular case of PCPs with linear structure. Our main theorem
is proved by constructing PCPs with linear structure (discussed next),
and applying the derandomized parallel repetition theorem.

\subsection*{PCPs with Linear Structure}

We turn to discuss PCPs with linear structure. The \emph{underlying
graph structure} of a two-query PCP is a graph defined as follows.
The vertices are the proof coordinates, and the edges correspond to
all possible query pairs of the verifier. (See also Section~\ref{subsec:PCP}).
We say that a graph has \emph{linear structure} if the vertices can
be identified with a vector space $\F^{m}$ and the edges, which clearly
can be viewed as a subset of $\F^{2m}$, form a linear subspace of
$\F^{2m}$ (see also Definition~\ref{def:linear-structure}). A two-query
PCP has linear structure if its underlying graph has linear structure.

As mentioned above, an additional contribution of this work is the
construction of PCPs with linear structure. That is, we prove the
following result.
\begin{theorem}
[\label{thm:structure}PCPs with linear structure]Every language
$L\in\NP$ has a two-query PCP system with a linear structure which
has perfect completeness, soundness error $1-1/\poly\log n$, constant
alphabet size, proof length $\poly\left(n\right)$, and randomness
complexity $O(\log n)$.
\end{theorem}
We believe that Theorem~\ref{thm:structure} is interesting in its
own right: For known PCPs, the underlying graph structure is quite
difficult to describe, mostly due to the fact that PCP constructions
are invariably based on composition. In principle, however, the fact
that a PCP is a ``complex'' object need not prevent the underlying
graph from being simple. In analogy, certain Ramanujan expanders~\cite{LPS88}
are Cayley graphs that are very easy to describe, even if the proof
of their expansion is not quite so easy. It is therefore interesting
to study whether there exist PCPs with simple underlying graphs.

Philosophically, the more structured the PCP, the stronger is the
implied statement about the class NP, and the easier it is to exploit
for applications. Indeed, the structure of a PCP system has been used
in several previous works. For example, Khot constructs~\cite{K06}
a PCP with quasi-random structure in order to establish the hardness
of minimum bisection. Dinur~\cite{D07} imposes an expansion structure
on a PCP to obtain amplification.

We prove Theorem~\ref{thm:structure} by embedding a given PCP into
the de Bruijn graph and relying on the algebraic structure of this
graph. We remark that the de Bruijn graph has been used in constructions
of PCPs before, e.g. \cite{PS94,BFLS91}, in similar contexts. We
believe that structured PCPs are an object worthy of further study.
One may view their applicability towards proving Theorem~\ref{thm:main}
as supporting evidence. An interesting question which we leave open
is whether Theorem~\ref{thm:structure} can be strengthened so as
to get \emph{constant} soundness error. By simply plugging such a
PCP into our derandomized parallel repetition theorem one would get
a direct proof of the aforementioned result of \cite{MR08}, \emph{without}
using two-query composition.
\begin{remark}
Our notion of PCPs with linear structure should not be confused with
the notion of ``linear PCPPs'' that appeared in the literature before
(see \cite{BHLM09}, and the related ``linear inner verifier'' of
\cite{GS00}). A linear PCPP is, roughly, a PCP system for checking
the membership of a vector in a given linear subspace, in which the
proof is required to be a linear function of the aforementioned vector.
This requirement is unrelated to our definition, which does not restrict
the claim to be verified or the proof, and on the other hand restricts
the query structure of the PCP verifier.
\end{remark}

\subsection*{Decodable PCPs}

We extend our results to also yield a new construction of \emph{decodable
PCPs} (dPCPs). A dPCP gives a way to encode NP witnesses so that a
verifier (called a decoder in this context) is able to both locally
test their validity as well as to locally decode bits from the encoded
NP witness. Decodable PCPs%
\footnote{Decodable PCPs generalize the notion of ``locally decode/reject codes''
of \cite{MR08} and the even earlier notion of ``LDF readers'' of
\cite{DFKRS99}.%
} were introduced in \cite{DH09} towards simplifying and modularizing
the work of \cite{MR08} on two-query PCPs with small soundness. In
\cite{DH09} the result of \cite{MR08} was reproved assuming the
existence of two building blocks, a PCP and a dPCP, which were used
as a black box. Until this work there has been only one known construction
of a dPCP, based on the manifold vs. point construction. In this work
we give a new construction of a dPCP which is obtained by applying
derandomized parallel repetition in an analogous way to Theorem~\ref{thm:main}.
We prove
\begin{theorem}
[\label{thm:main-dPCP}dPCP, informal version]There exists a two-query
PCP decoder with perfect completeness, soundness error $1/\poly\log n$,
list size $\poly\log n$, proof alphabet $2^{\poly\log n}$, proof
length $\poly\left(n\right)$, and randomness complexity $O(\log n)$.
\end{theorem}
The notion of dPCPs is described in detail in Section~\ref{sec:dPCP},
and in particular in Section~\ref{sub:dPCP}. Theorem~\ref{thm:main-dPCP}
is stated and proved in Section~\ref{sub:dPCP-construction} based
on two main lemmas, which are proved in Sections~\ref{sec:dPCP-embedding-linear-structure}
and~\ref{sec:dPCP-derandomized-parallel-repetition}.

In order to prove this theorem we generalize each of the steps of
the proof of Theorem~\ref{thm:main}. First, we construct a dPCP
with linear structure but with relatively high soundness error in
an analogous way to our proof of Theorem~\ref{thm:structure} (PCPs
with linear structure). Next, we apply derandomized parallel repetition
to get the desired~dPCP. The two steps are described in Sections~\ref{sec:dPCP-embedding-linear-structure}
and~\ref{sec:dPCP-derandomized-parallel-repetition} respectively.

An additional contribution of this work is an extension of the definitions
of \cite{DH09}, of dPCPs that work with low soundness error, to one
that works with high soundness error. This is necessary because plugging
in a higher value for the soundness error parameter into the existing
definition of~\cite{DH09} turns out to be useless. Instead, we give
a variant which we call uniquely decodable PCPs (udPCPs). We show
that udPCPs are in fact equivalent to PCPs of Proximity (PCPPs). This
allows us to rely on known constructions of PCPPs~\cite{BGHSV06,DR06}
as our starting point. For more details see Section~\ref{sub:dPCP}.

Together, Theorem~\ref{thm:main} and Theorem~\ref{thm:main-dPCP}
imply Theorem~\ref{thm:MR} (the \cite{MR08} result). This is sketched
in Section~\ref{sub:MR}.
\begin{remark}
In fact, Theorem~\ref{thm:main-dPCP} can be proved for any soundness
error $\varepsilon(n)$ satisfying $1/n^{\kappa}\le\varepsilon(n)\le1/\poly\log n$
(for some constant $\kappa>0$. As in Theorem~\ref{thm:main}, the
alphabet size in such case is $2^{1/\poly\left(\varepsilon\right)}$,
and furthermore the list size becomes $1/\poly\left(\varepsilon\right)$.
However, in this paper we only prove Theorem~\ref{thm:main-dPCP}
for $\varepsilon(n)=1/\poly\log n$, since this is all we need to
in order to prove Theorem~\ref{thm:MR} (the \cite{MR08} result).
\end{remark}

\subsection*{Related Work and Future directions}

Our final construction of a two-query PCP has exponential relation
between the alphabet size and the error probability (that is, $\card{\Sigma}=2^{1/\poly\left(\varepsilon\right)}$).
In general, one can hope for a polynomial relation, and this is the
so-called ``sliding scale'' conjecture of \cite{BGLR93}. Our approach
is inherently limited to an exponential relation both because of a
lower bound on direct product testing from \cite{DG08}, and, more
generally, because of the following lower bound of Feige and Kilian~\cite{FK95}
on parallel repetition of games. Feige and Kilian prove that for every
PCP system and $k=O(\log n)$ invocations of the original verifier,
if one insists on the parallel repetition using only $O(\log n)$
random bits, then the soundness error must be at least $1/\poly\log n$
(and not $1/\poly(n)$ as one might hope). For the choice of $k=O(\log n)$,
our work matches the \cite{FK95} lower bound by exhibiting a derandomized
parallel repetition theorem, albeit only for PCPs with linear structure,
that achieves a matching upper bound of $1/\poly\log n$ on the soundness
error.

Nevertheless, for three queries we are in a completely different ball-game,
and no lower bound is known. It would be interesting to find a derandomized
direct product test with three queries with lower soundness error,
and to try and adapt it to a PCP. We note that there are ``algebraic''
constructions~\cite{RS97,DFKRS99} that make only three queries and
have much better relationship between the error and the alphabet size.

It has already been mentioned that while our result matches the soundness
error and alphabet size of the \cite{MR08} result, it does not attain
nearly linear proof length. Improving our result in this respect is
another interesting direction.

\subsection*{Structure of the paper}

The paper has two main parts, the first part is concerned with proving
the main result for PCPs, and the second part generalizes this result
to dPCPs. 
\begin{itemize}
\item \textbf{Part 1.} The structure of the proof is ``top to bottom''.
Our main theorem for PCPs is based on two main steps: (i) embedding
a PCP into a PCP with linear structure, and (ii) a derandomized parallel
repetition theorem for such PCPs. We begin, in Section~\ref{sec:main},
by stating the two main lemmas corresponding to the two steps above,
and then proving the main theorem, assuming correctness of the lemmas.
We then proceed to prove each main lemma. In Section~\ref{sec:embedding-linear-structure}
we show how to embed a PCP into one with linear structure (by routing
it on a de Bruijn like graph). In Section~\ref{sec:derandomized-parallel-repetition}
we prove the ``derandomized parallel repetition'' theorem for PCPs
with linear structure. This is done by reduction to the derandomized
direct product test of ~\cite{IKW09}. More accurately, our analysis
relies on a specialized variant of this test which we call an $S$-test,
which is analyzed in Section~\ref{sec:S-test-analysis}. 
\item \textbf{Part 2.} The second part of the paper adapts our PCP construction
to a dPCP. In Section~ \ref{sec:dPCP} we discuss and define dPCPs,
and prove Theorem~\ref{thm:main-dPCP}. We also show how to use this
theorem to derive the \cite{MR08} result (Theorem~\ref{thm:MR})
as a corollary. The two main steps in the proof of Theorem~\ref{thm:main-dPCP}
are described in Sections~\ref{sec:dPCP-embedding-linear-structure}
and \ref{sec:dPCP-derandomized-parallel-repetition} and are analogous
to the two main steps of proving Theorem~\ref{thm:main}. 
\item Finally, we analyze the specialized direct product test (called the
S-test) in Section~\ref{sec:S-test-analysis}, based on the work
of~\cite{IKW09}.
\end{itemize}

\section{\label{sec:prelim}Preliminaries}

Let $g:U\to\Sigma$ be an arbitrary function, and let $A\subset U$
be a subset. We denote by $g_{|A}$ the restriction of $g$ (as a
function) to $A$. We also use the following convention.
\begin{notation}
\label{not:function-distance}Given two functions $f,g:U\to\Sigma$,
we denote $f\apx{\alpha}g$ ($f\napx{\alpha}g$) to mean that they
differ on at most (more than) $\alpha$~fraction of the elements
of $U$.
\end{notation}
We refer to a $d$-dimensional linear subspace of an underlying vector
space simply as a $d$-\emph{subspace}. For two linear subspaces $A_{1}$
and $A_{2}$, the standard notation $A_{1}+A_{2}$ denotes the smallest
linear subspace containing both of them. We say that $A_{1},A_{2}$
are \emph{independent} if and only if $A_{1}\cap A_{2}=\set 0$. If
$A_{1}$ and $A_{2}$ are disjoint, the standard notation $A_{1}\oplus A_{2}$
is used to denotes $A_{1}+A_{2}$.

Let $G=\left(V,E\right)$ be a directed graph. For each edge $e\in E$
we denote by $\l\left(e\right)$ and $\r\left(e\right)$ the left
and right endpoints of $e$ respectively. That is, if we view the
edge $e\in E$ as a pair in $V\times V$, then $\l\left(e\right)$
and $\r\left(e\right)$ are the first and second elements of the pair
$e$ respectively. Given a set of edges $E_{0}\subseteq E$, we denote
by $\l\left(E_{0}\right)$ and $\r(E_{0})$ the set of left endpoints
and right endpoints of the edges in $E_{0}$ respectively.

\subsection{\label{sub:prelim-IKW}Direct product testing \cite{IKW09}}

Let us briefly describe the setting in which we use the derandomized
direct product test of \cite{IKW09}. In \cite{IKW09} the main derandomized
direct product test is a so-called ``V-test''. We consider a variation
of this test that appears in \cite[Section 6.3]{IKW09} to which we
refer as the ``P-test'' (P for projection).

Given a string $\pi\in\Sigma^{\ell}$, we define its (derandomized)
P-direct product $\Pi$ as follows: We identify $\left[\ell\right]$
with $\F^{m}$, where $\F$ is a finite field and $m\in\N$, and think
of $\pi$ as an assignment that maps the points in $\F^{m}$ to $\Sigma$.
We also fix $d_{0}<d_{1}\in\N$. Now, we define $\Pi$ to be the assignment
that assigns each $d_{0}$- and $d_{1}$-subspace $W$ of $\F^{m}$
to the function $\pi_{|W}:W\to\Sigma$ (recall that $\pi_{|W}$ is
the restriction of $\pi$ to $W$).

We now consider the task of testing whether a given assignment $\Pi$
is the P-direct product of some string $\pi:\F^{m}\to\Sigma$. In
those settings, we are given an assignment to subspaces, i.e. a function
$\Pi$ that on input a $d_{0}$-subspace $A\subset\F^{m}$ (respectively
$d_{1}$-subspace $B\subset\F^{m}$), answers with a function $a:A\to\Sigma$
(respectively, $b:\F^{m}\to\Sigma$). We wish to test whether $\Pi$
is a P-direct product of some $\pi:\F^{m}\to\Sigma$, and to this
end we invoke the P-test, described in Figure~\ref{fig:P-test}.
\begin{figure}
\centering \fbox{%
\begin{minipage}[c]{5in}%
 \centering 
\begin{enumerate}
\item Choose a uniformly distributed $d_{1}$-subspace $B\subseteq\F^{m}$. 
\item Choose a uniformly distributed $d_{0}$-subspace $A\subseteq B$. 
\item Accept if and only if $\Pi\left(B\right)_{|A}=\Pi(A)$. \end{enumerate}
\end{minipage}} \caption{\label{fig:P-test}The P-test}
\end{figure}

It is easy to see that if $\Pi$ is a P-direct product then the P-test
always accepts. Furthermore, it can be shown that if $\Pi$ is ``far''
from being a P-direct product, then the P-test rejects with high probability.
Formally, we have the following result.
\begin{theorem}
[\label{thm:IKW}Soundness of the P-test\cite{IKW09}] There exists
a universal constant $\hp\in\N$ such that the following holds: Let
$\varepsilon\ge\hp\cdot d_{0}\cdot\left|\F\right|^{-d_{0}/\hp}$,
$\ap\eqdef\hp\cdot d_{0}\cdot\left|\F\right|^{-d_{0}/\hp}$. Assume
that $d_{1}\ge\hp\cdot d_{0}$, $m\ge\hp\cdot d_{1}$. Suppose that
an assignment $\Pi$ passes the P-test with probability at least $\varepsilon$.
Then, there exists an assignment $\pi$ such that 
\begin{equation}
\Pr\left[\Pi\left(B\right)_{|A}=\Pi\left(A\right)\aand\Pi\left(B\right)\apx{\ap}\pi_{|B}\aand\Pi\left(A\right)\apx{\ap}\pi_{|A}\right]=\Omega(\varepsilon^{4}),\label{eq:IKW}
\end{equation}
where the probability is over $A,B$ chosen as in the P-test.
\end{theorem}
Theorem~\ref{thm:IKW} can be proved by adapting the analysis of~\cite{IKW09}
(in particular, Sections~3.4 and~4) to the setting of the $P$-test,
while relying on a lemma of~\cite{IKW09}. For completeness, the
proof is given in Appendix~\ref{sec:Proof-of-IKW}.

\paragraph*{Working with randomized assignments.}

As observed by \cite{IKW09}, Theorem~\ref{thm:IKW} works in even
stronger settings. Suppose that $\Pi$ is a randomized function, i.e.,
a function of both its input and some additional randomness. Then,
Theorem~\ref{thm:IKW} still holds for $\Pi$, where the probability
in~(\ref{eq:IKW}) is over both the choice of $A$ and $B$, \emph{and
over the internal randomness of $\Pi$}. We will rely on this fact
in a crucial way in this work.

\subsection{\label{sec:Sampling-tools}Sampling tools}

The following is the standard definition of a sampler, stated in the
terminology of graphs, see e.g. \cite{IJKW08}.
\begin{definition}
[Sampler Graph] A bipartite graph $G=(L,R,E)$ is said to be an
$\left(\varepsilon,\delta\right)$-sampler if, for every function
$f:L\to[0,1]$, there are at most $\delta\card R$ vertices $u\in R$
for which 
\[
\left|\E_{v\in N(u)}[f(v)]-\E_{v\in L}[f(v)]\right|>\varepsilon.
\]
Observe that if $G$ is an $(\varepsilon,\delta)$-sampler, and if
$F\subset L$, then by considering the function $f\equiv1_{F}$ we
get that there are at most $\delta\card R$ vertices $u\in R$ for
which 
\[
\left|\Pr_{v\in N(u)}[v\in F]-\Pr_{v\in L}[v\in F]\right|>\varepsilon.
\]
The following lemma is stated in \cite[Lemma 2.2]{IKW09} and is proved
implicitly in \cite[Lemma 2.9]{IJKW08}. For completeness, we include
its proof.\end{definition}
\begin{lemma}
[\label{lem:subspace-point-sampler}Subspace-point sampler \cite{IJKW08}]Let
$d'<d$ be natural numbers, let $V$ be a linear space over a finite
field $\F$, and let $W$ be a fixed $d'$-subspace of $V$. Let $G$
be the bipartite graph whose left vertices are all points of $V$
and whose right vertices are all $d$-subspaces of $V$ that contain
$W$. We place an edge between a $d$-subspace $X$ and $x\in V$
if and only if $x\in X$. Then $G$ is an $(\tau+\frac{1}{\left|\F\right|^{d-d'}},\frac{1}{\card{\F}^{d-d'-2}\cdot\tau^{2}})$-sampler
for every $\tau>0$.\end{lemma}
\begin{myproof}
Fix a function $f:V\to\left[0,1\right]$. We show that for a uniformly
distributed $d$-subspace $X\subseteq V$ that contains $W$ it holds
with probability at least $1-\frac{1}{\card{\F}^{d-d'-2}\cdot\tau^{2}}$
that
\[
\left|\E_{x\in X}\left[f(x)\right]-\E_{v\in V}\left[f(v)\right]\right|\le\tau+\frac{1}{\left|\F\right|^{d-d'}}.
\]
Let $\overline{W}$ be a fixed subspace of $V$ for which $V=W\oplus\overline{W}$.
Let $f_{W}:\overline{W}\to\left[0,1\right]$ be the function that
maps each vector $\overline{w}$ of $\overline{W}$ to $\E_{v\in\overline{w}+W}\left[f(v)\right]$,
and observe that $\E_{v\in V}\left[f(v)\right]=\E_{\overline{w}\in\overline{W}}\left[f_{W}(\overline{w})\right]$.
Furthermore, observe that every $d$-subspace $X$ that contains $W$
can be written as $X=W\oplus U$ where $U$ is a $\left(d-d'\right)$-subspace
of $\overline{W}$, and moreover that $\E_{x\in X}\left[f(x)\right]=\E_{u\in U}\left[f_{W}(u)\right]$.
Thus, it suffices to prove that for a uniformly distributed $\left(d-d'\right)$-subspace
$U$ of $\overline{W}$ it holds with probability at least $1-\frac{1}{\card{\F}^{d-d'-2}\cdot\tau^{2}}$
that 
\begin{equation}
\left|\E_{u\in U}\left[f_{W}(u)\right]-\E_{\overline{w}\in\overline{W}}\left[f_{W}(\overline{w})\right]\right|\le\tau+\frac{1}{\left|\F\right|^{d-d'}}.\label{eq:subspace-point-sampler}
\end{equation}
To that end, let $U$ be a uniformly distributed $\left(d-d'\right)$-subspace
of $\overline{W}$. Let $S_{1}$ be a uniformly distributed set of
$Q\eqdef\frac{\left|\F\right|^{d-d'}-1}{\left|\F\right|-1}$ vectors
of $U$ such that every two vectors in $S_{1}$ are linearly independent%
\footnote{Such a set can be sampled, for example, by iteratively choosing a
uniformly distributed vector of $U$ that is linearly independent
from each of the previously chosen vectors individually. It is not
hard to see that such a process will halt after choosing~$Q\eqdef\frac{\left|\F\right|^{d-d'}-1}{\left|\F\right|-1}$
vectors.%
}. For every $\alpha\in\F^{*}$ let $S_{\alpha}$ be the set obtained
by multiplying every vector in $S_{1}$ by $\alpha$. Observe that
all the sets $S_{\alpha}$ have the property that every two vectors
in $S_{\alpha}$ are linearly independent, and that the sets $S_{\alpha}$
form a partition of $U\backslash\left\{ {0}\right\} $. We will show
that for every $\alpha\in\F^{*}$ it holds with probability at least
$1-\frac{1}{\card{\F}^{d-d'-1}\cdot\tau^{2}}$ that 
\[
\left|\E_{u\in S_{\alpha}}\left[f_{W}(u)\right]-\E_{\overline{w}\in\overline{W}}\left[f_{W}(\overline{w})\right]\right|\le\tau,
\]
and the required result will follow by taking the union bound over
all $\alpha\in\F^{*}$, and by noting that the vector ${0}$ contributes
at most $\frac{1}{\left|\F\right|^{d-d'}}$ to the difference in Inequality~\ref{eq:subspace-point-sampler}.

Fix $\alpha\in\F^{*}$, and let $s_{1},\ldots,s_{Q}$ be the vectors
in $S_{\alpha}$. It is a known fact that $s_{1},\ldots,s_{Q}$ are
pair-wise independent and uniformly distributed vectors of $\overline{W}$
(over the random choice of $U$). This implies that $f_{W}(s_{1}),\ldots,f_{W}(s_{Q})$
are pair-wise independent random variables with expectation $\E_{\overline{w}\in\overline{W}}\left[f_{W}(\overline{w})\right]$,
and therefore by the Chebyshev inequality it follows that
\[
\Pr\left[\left|\frac{1}{Q}\sum_{i=1}^{Q}f_{W}(s_{i})-\E_{\overline{w}\in\overline{W}}\left[f_{W}(\overline{w})\right]\right|>\tau\right]\le\frac{1}{Q\cdot\tau^{2}}\le\frac{1}{\card{\F}^{d-d'-1}\cdot\tau^{2}},
\]
as required.
\end{myproof}

\subsection{\label{subsec:PCP}Constraint graphs and PCPs}

As discussed in the introduction, the focus of this work is on claims
that can be verified by reading a small number of symbols of the proof.
A PCP system for a language $L$ is an oracle machine $M$, called
a verifier, that has oracle access to a proof $\pi$ over an alphabet
$\Sigma$. The verifier $M$ reads the input $x$, tosses $r$ coins,
makes at most $q$ ``oracle'' queries into $\pi$, and then accepts
or rejects. If $x$ is in the language then it is required that $M$
accepts with probability~$1$ for some $\pi$, and otherwise it is
required that $M$ accepts with probability at most $\varepsilon$
for every $\pi$. More formally:
\begin{definition}
\label{def:PCP-verifier}Let $r,q:\N\to\N$, and let $\Sigma$ be
a function that maps the natural numbers to finite alphabets. A $\left(r,q\right)_{\Sigma}$\emph{-PCP
verifier} $M$ is a probabilistic polynomial time oracle machine that
when given input $x\in\B^{*}$, tosses at most $r(\left|x\right|)$
coins, makes at most $q\left(\left|x\right|\right)$ \emph{non-adaptive}
queries to an oracle that is a string over $\Sigma(\left|x\right|)$,
and outputs either ``accept'' or ``reject''. We refer to $r$,
$q$, and $\Sigma$ as the \emph{randomness complexity}, \emph{query
complexity}, and \emph{proof alphabet} of the verifier respectively.\end{definition}
\begin{remark}
\label{rem:randomness-to-proof-length}Note that for an $\left(r,q\right)_{\Sigma}$-PCP
verifier $M$ and an input $x$, we can assume without loss of generality
that the oracle is a string of length at most $2^{r(\left|x\right|)}\cdot q(\left|x\right|)$,
since this is the maximal number of different queries that $M$ can
make. Hence, it is unnecessary to keep track of the proof length of
the verifier.\end{remark}
\begin{definition}
\label{def:PCP}Let $r$, $q$ and $\Sigma$ be as in Definition~\ref{def:PCP-verifier},
let $L\subseteq\B^{*}$ and let $\varepsilon:\N\to[0,1)$. We say
that $L\in\mathbf{PCP}_{\varepsilon,\Sigma}\left[r,q\right]$ if there
exists an $\left(r,q\right)_{\Sigma}$-PCP verifier $M$ that satisfies
the following requirements:
\begin{itemize}
\item \textbf{Completeness:} For every $x\in L$, there exists $\pi\in\Sigma\left(\left|x\right|\right)^{*}$
such that $\Pr\left[M^{\pi}(x)\mbox{ accepts}\right]=1$.
\item \textbf{Soundness:} For every $x\notin L$ and for every $\pi\in\Sigma\left(\left|x\right|\right)^{*}$
it holds that $\Pr\left[M^{\pi}(x)\mbox{ accepts}\right]\nolinebreak\le\nolinebreak\varepsilon\left(\left|x\right|\right)$. 
\end{itemize}
\end{definition}
One possible formulation of the PCP theorem is as follows.
\begin{theorem}
[\label{thm:PCP-thm}PCP Theorem~\cite{AS98,ALMSS98}]There exist
universal constant $\varepsilon\in\left(0,1\right)$ and a finite
alphabet $\Sigma$ such that $\NP\subseteq\mathbf{PCP}_{\varepsilon,\Sigma}\left[O(\log n),2\right]$.
\end{theorem}
PCPs that have query complexity $2$ correspond to graphs in a natural
way: Consider the action of an $\left(r,2\right)_{\Sigma}$-verifier
$M$ on some fixed string $x$, and let $r\eqdef r(\left|x\right|)$,$\Sigma\eqdef\Sigma(\left|x\right|)$.
The verifier $M$ is given access to some proof string $\pi$ of length
$\ell$, and may make $2^{r}$ possible tests on this string, where
each such test consists of making two queries to $\pi$ and deciding
according to the answers. We now view the action of $M$ as a graph
in the following way. We consider the graph $G$ whose vertices are
the coordinates in $\left[\ell\right]$, and that has an edge for
each possible test of the verifier $M$. The endpoints of an edge
$e$ of $G$ are the coordinates that are queried by $M$ in the test
that corresponds to $e$. We also associate an edge $e$ with a constraint
$c_{e}\in\Sigma\times\Sigma$, which contains all the pairs of answers
that make $M$ accept when performing the test that corresponds to
$e$. We think of $\pi$ as an assignment that assigns the vertices
of $G$ values in $\Sigma$, and say that $\pi$ \emph{satisfies}
an edge $\left(u,v\right)$ if $\left(\pi(u),\pi(v)\right)\in c_{\left(u,v\right)}$.
If $x\in L$, then it is required that there exists some assignment
$\pi$ that satisfies all the edges of $G$, and otherwise it is required
that every assignment satisfies at most $\varepsilon$~fraction of
the edges. This correspondence is called the FGLSS correspondence~\cite{FGLSS96}.
We turn to state it formally:
\begin{definition}
[\label{def:cgraph}Constraint graph] A \emph{(directed) constraint
graph} is a directed graph $G=(V,E)$ together with an alphabet $\Sigma$,
and, for each edge $(u,v)\in E$, a binary constraint $c_{u,v}\subseteq\Sigma\times\Sigma$.
The \emph{size} of $G$ is the number of edges of $G$. The graph
is said to have \emph{projection constraints} if it is bipartite with
all the edges directed from the left to the right, and every constraint
$c_{u,v}$ has an associated function $f_{u,v}:\Sigma\to\Sigma$ such
that $c_{u,v}$ is satisfied by $(a,b)$ if and only if $f_{u,v}(a)=b$.\\
Given an assignment $\pi:V\to\Sigma$, we define 
\[
\sat(G,\pi)=\Pr_{(u,v)\in E}[(\pi(u),\pi(v))\in c_{u,v}]\quad\aand\quad\sat(G)=\max_{\pi}(\sat(G,\pi)).
\]
 We also denote $\unsat(G,\pi)=1-\sat(G,\pi)$ and similarly $\unsat(G)=1-\sat(G)$.\end{definition}
\begin{remark}
Note that Definition~\ref{def:cgraph} uses \emph{directed graphs},
while the common definition of constraint graphs refers to undirected
graphs.
\end{remark}

\begin{remark}
Note that if the graph $G$ has projection constraints, then this
is simply a label cover instance with projection constraints~\cite{AL96}.\end{remark}
\begin{proposition}
[\label{prop:corr}FGLSS correspondence~\cite{FGLSS96}]The following
two statements are equivalent:
\begin{itemize}
\item $L\in\mathbf{PCP}_{\varepsilon,\Sigma}\left[r,2\right]$.
\item There exists a polynomial-time algorithm that transforms strings $x\in\B^{*}$
to constraint graphs $G_{x}$ of size $2^{r(\left|x\right|)}$ with
alphabet $\Sigma\left(\left|x\right|\right)$ such that: (1)~if $x\in L$
then $\sat(G_{x})=1$, and (2)~if $x\not\in L$ then $\sat(G_{x})\le\eps$.
\end{itemize}
Given a PCP system for $L$, we refer to the corresponding family
of graphs $\set{G_{x}}$ where $x$ ranges over all possible instances
as its \emph{underlying graph family}. If the graphs $\left\{ G_{x}\right\} $
have projection constraints then we say that the PCP system has the
\emph{projection property}.
\end{proposition}
Using the \cite{FGLSS96} correspondence, we can rephrase the PCP
theorem in the terminology of constraint graphs:
\begin{theorem}
[\label{thm:PCP-thm-using-graphs}PCP Theorem for constraint graphs]There
exist universal constant $\varepsilon\in\left(0,1\right)$ and a finite
alphabet $\Sigma$ such that for every language $L\in\NP$ the following
holds: There exists a polynomial time reduction that on input $x\in\B^{*}$,
outputs a constraint graph $G_{x}$ such that if $x\in L$ then $\sat(G_{x})=1$
and otherwise $\sat(G_{x})\le\varepsilon$.\end{theorem}
\begin{remark}
The connection between PCPs and approximation problems (such as Proposition~\ref{prop:corr})
was discovered by~\cite{FGLSS96}. However, the precise correspondence
between PCPs and constraint graphs that is given in Proposition~\ref{prop:corr}
was only stated for the first time by~\cite{ALMSS98}. Still, in
the rest of this paper we refer to Proposition~\ref{prop:corr} as
the \cite{FGLSS96} correspondence.
\end{remark}

\begin{remark}
\label{rem:randomness-vs-size}Note the tight relationship between
the randomness complexity of the PCP and the size of the corresponding
constraint graphs. In particular, observe that PCP verifiers with
randomness complexity $O(\log n)$ correspond to constraint graphs
of polynomial size. This relationship is one of the main reasons for
the study of the randomness complexity of PCP verifiers.

Moreover, recall that the work of~\cite{MR08} constructs PCPs that
are very randomness efficient, i.e., have randomness complexity~$\left(1+o(1)\right)\log n$
(see also Remark~\ref{rem:MR-short}). This randomness efficiency
is translated into constraints graphs of almost-linear size, namely~$n^{1+o(1)}$.
\end{remark}

\subsection{Basic facts about random subspaces}

In this section we present two useful propositions about random subspaces.
The following proposition says that a uniformly distributed subspace
is independent from every fixed subspace with high probability.
\begin{proposition}
\label{pro:random-subspaces-disjoint}Let $d,d'\in\N$ such that $d>2d'$,
and let $V$ be a $d$-dimensional space. Let $W_{1}$ be a uniformly
distributed $d'$-subspace of $V$, and let $W_{2}$ be a fixed $d'$-subspace
of $V$. Then, 
\[
\Pr[W_{1}\cap W_{2}=\set 0]\ge1-2\cdot d'/\left|\F\right|^{d-2\cdot d'}.
\]
\end{proposition}
\begin{myproof}
Suppose that $W_{1}$ is chosen by choosing random basis vectors $v_{1},\ldots,v_{d'}$
one after the other. It is easy to see that $W_{1}\cap W_{2}\ne\left\{ 0\right\} $
only if $v_{i}\in\mbox{span}\left(W_{2}\cup\left\{ v_{1},\ldots,v_{i-1}\right\} \right)$
for some $i\in\left[d'\right]$. For each fixed $i$, the vector $v_{i}$
is uniformly distributed in $V\backslash\mbox{span}\left\{ v_{1},\ldots,v_{i-1}\right\} $,
and therefore the probability that $v_{i}\in\mbox{span}\left(W_{2}\cup\left\{ v_{1},\ldots,v_{i-1}\right\} \right)$
for a fixed $i$ is at most
\begin{eqnarray}
\frac{\left|\mbox{span}\left(W_{2}\cup\left\{ v_{1},\ldots,v_{i-1}\right\} \right)\right|}{\left|V\backslash\mbox{span}\left\{ v_{1},\ldots,v_{i-1}\right\} \right|} & = & \frac{\left|\F\right|^{d'+i-1}}{\left|\F\right|^{d}-\left|\F\right|^{i-1}}\nonumber \\
 & \le & \frac{2\cdot\left|\F\right|^{d'+i-1}}{\left|\F\right|^{d}}\label{eq:analysis-event-dimension-B-bound}\\
 & \le & \frac{2\cdot\left|\F\right|^{2\cdot d'-1}}{\left|\F\right|^{d}}\nonumber \\
 & \le & \frac{2}{\left|\F\right|^{d-2\cdot d'}},\nonumber 
\end{eqnarray}
where Inequality~\ref{eq:analysis-event-dimension-B-bound} can be
observed by noting that $\left|\F\right|^{i-1}\le\left|\F\right|^{d-1}\le\frac{1}{2}\cdot\left|\F\right|^{d}$.
By the union bound, the probability that this event occurs for some
$i\in\left[d'\right]$ is at most $\frac{2\cdot d'}{\left|\F\right|^{d-2\cdot d'}}$.
It follows that the probability that $W_{1}\cap W_{2}\ne\left\{ {0}\right\} $
is at most $\frac{2\cdot d'}{\left|\F\right|^{d-2\cdot d'}}$ as required. 
\end{myproof}
The following proposition says that the span of $d'$ uniformly distributed
vectors is with high probability a uniformly distributed $d'$-subspace. 
\begin{proposition}
\label{pro:random-vectors-have-full-dim}Let $V$ be a $d$-dimensional
space over a finite field~$\F$, let $w_{1},\ldots,w_{d'}$ be independent
and uniformly distributed vectors of $V$, and let $W=\span\left\{ w_{1},\ldots,w_{d'}\right\} $.
Then, with probability at least $1-d'/\left|\F\right|^{d-d'}$ it
holds that $\dim W=d'$. Furthermore, conditioned on the latter event,
$W$ is a uniformly distributed $d'$-subspace of $V$.\end{proposition}
\begin{myproof}
The fact that $\dim W=d'$ with probability at least $1-d'/\left|\F\right|^{d-d'}$
can be proved in essentially the same way as Proposition~\ref{pro:random-subspaces-disjoint}.
To see that conditioned on the latter event it holds that the subspace
$W$ is uniformly distributed, observe that since $w_{1},\ldots,w_{d'}$
were originally chosen to be uniformly distributed, all the possible
$d'$-sets of linearly independent vectors have the same probability
to occur.
\end{myproof}
Finally, the following proposition shows the equivalence of two different
ways of choosing subspaces $A_{1},A_{2}\subseteq B$ where $A_{1}$
and $A_{2}$ are independent.
\begin{proposition*}
\label{pro:triplet-distributions-equivalent}Let $V$ be a linear
space over a finite field $\F$, and let $d_{0},d_{1}\in\N$ be such
that $d_{0}<d_{1}<\dim V$. The following two distributions over $d_{0}$-subspaces
$A_{1}$, $A_{2}$ and a $d_{1}$-subspace $B$ are the same: 
\begin{enumerate}
\item Choose $B$ to be a uniformly distributed $d_{1}$-subspace of $V$,
and then choose $A_{1}$ and $A_{2}$ to be two uniformly distributed
and independent $d_{0}$-subspaces of $B$. 
\item Choose $A_{1}$ and $A_{2}$ to be two uniformly distributed and independent
$d_{0}$-subspaces of $V$, and then choose $B$ to be a uniformly
distributed $d_{1}$-subspace of $V$ that contains $A_{1}$ and $A_{2}$. 
\end{enumerate}
\end{proposition*}
\begin{myproof}
Observe that choosing $A_{1}$, $A_{2}$, $B$ under the first distribution
amounts to choosing $d_{1}$ uniformly distributed and linearly independent
vectors in $V$ (those vectors will serve as the basis of $B$), and
then choosing two disjoint subsets of those vectors to serve as the
basis of $A_{1}$ and as the basis of $A_{2}$. On the other hand,
choosing $A_{1}$, $A_{2}$ and $B$ under the second distribution
amounts to choosing $d_{0}$ uniformly distributed and linearly independent
vectors in $V$ to serve as the basis of $A_{1}$, then choosing another
$d_{0}$ uniformly distributed and linearly independent vectors in
$V$ to serve as the basis of $A_{2}$ while making sure that this
basis is also linearly independent from the basis of $A_{1}$, and
then completing the basis of $A_{1}$ and the basis of $A_{2}$ to
a basis of $B$. It is easy to see that those two distributions over
a set of $d_{1}$ vectors and its two disjoint subsets are identical.
\end{myproof}

\subsection{\label{sub:similarity-of-distributions}Similarity of distributions}

In this section we introduce a notion of ``similarity of distributions'',
which we will use in the second part of the paper. Let $X_{1}$ and
$X_{2}$ be two random variables that take values from a set $\mathcal{X}$,
and let $\smooth\in(0,1]$. We say that $X_{1}$ and $X_{2}$ are
\textsf{$\smooth$-similar} if for every $x\in\mathcal{X}$ it holds
that
\[
\smooth\cdot\Pr\left[X_{1}=x\right]\le\Pr\left[X_{2}=x\right]\le\frac{1}{\smooth}\cdot\Pr\left[X_{1}=x\right].
\]
Note that if $X_{1}$ and $X_{2}$ are $\gamma$-similar then actually
it holds for every $S\subseteq\mathcal{X}$ that
\[
\smooth\cdot\Pr\left[X_{1}\in S\right]\le\Pr\left[X_{2}\in S\right]\le\frac{1}{\smooth}\cdot\Pr\left[X_{1}\in S\right],
\]
The following claim says roughly that if $f$ is a randomized function,
then the random variable $f(X_{1})$ is $\gamma$-similar to $f(X_{2})$.
\begin{claim}
\label{cla:similarity of derived variables}Let $X_{1}$ and $X_{2}$
be two random variables that take values from a set $\mathcal{X}$
that are $\smooth$-similar. Let $Y_{1}$ and $Y_{2}$ be two random
variables that take values from a set $\mathcal{Y}$ such that for
every $x\in\mathcal{X}$, $y\in\mathcal{Y}$ it holds that
\[
\Pr\left[Y_{1}=y|X_{1}=x\right]=\Pr\left[Y_{2}=y|X_{2}=x\right].
\]
Then, the variables $Y_{1}$, $Y_{2}$ are $\smooth$-similar.\end{claim}
\begin{myproof}
It holds that
\begin{eqnarray*}
\Pr\left[Y_{1}=y\right] & = & \sum_{x\in\mathcal{X}}\Pr\left[Y_{1}=y|X_{1}=x\right]\cdot\Pr\left[X_{1}=x\right]\\
 & = & \sum_{x\in\mathcal{X}}\Pr\left[Y_{2}=y|X_{2}=x\right]\cdot\Pr\left[X_{1}=x\right]\\
 & \ge & \sum_{x\in\mathcal{X}}\Pr\left[Y_{2}=y|X_{2}=x\right]\cdot\smooth\cdot\Pr\left[X_{2}=x\right]\\
 & = & \smooth\cdot\Pr\left[Y_{2}=y\right].
\end{eqnarray*}
Similarly it can be proved that $\Pr\left[Y_{1}=y\right]\le\frac{1}{\smooth}\cdot\Pr\left[Y_{2}=y\right]$.
\end{myproof}

\subsection{Expanders}

Expanders are graphs with certain properties that make them extremely
useful for many applications in theoretical computer science. Below
we give a definition of expanders that suits our needs.
\begin{definition}
\label{def:expansion}Let $G=\left(V,E\right)$ be a $d$-regular
graph. Let $E\left(S,\overline{S}\right)$ be the set of edges from
a subset $S\subseteq V$ to its complement. We say that $G$ has edge
expansion $h$ if for every $S\subseteq V$ such that $\left|S\right|\le\left|V\right|/2$
it holds that 
\[
\left|E(S,\overline{S})\right|\ge h\cdot d_{0}\cdot\left|S\right|.
\]

\end{definition}
A useful fact is that there exist constant degree expanders over any
number of vertices:
\begin{fact}
\label{fac:expanders-exist}There exist $d_{0}\in\N$ and $h_{0}>0$
such that there exists a polynomial-time constructable family $\left\{ G_{n}\right\} _{n\in\N}$
of $d_{0}$-regular graphs $G_{n}$ on $n$ vertices that have edge
expansion $h_{0}$ (such graphs are called expanders).
\end{fact}

\section{\label{sec:main}Main theorem}

In this section we prove our main PCP theorem (Theorem~\ref{thm:main}),
which asserts the existence of two-query PCPs with soundness error~$\varepsilon(n)$
for any function $1/n^{\kappa}\le\varepsilon(n)\le1/\poly\log n$.
To that end, we use the PCP theorem for graphs (Theorem~\ref{thm:PCP-thm-using-graphs})
to reduce the problem of deciding membership of a string $x$ in the
language $L$ to the problem of checking the satisfiability of a constraint
graph with constant soundness error. We then show that every constraint
graph can be transformed into one that has ``linear structure'',
defined shortly below. This is done in Lemma~\ref{lem:embedding-linear-structure},
which directly proves Theorem~\ref{thm:structure} (the existence
of PCPs with linear structure). Finally, in Lemma~\ref{lem:derandomized-parallel-repetition}
we prove a derandomized parallel repetition theorem for constraint
graphs with linear structure. Theorem~\ref{thm:main} follows by
combining the two lemmas. We begin by defining the notion of a graph
with linear structure.
\begin{definition}
[\label{def:linear-structure}Linear Structure]We say that a directed
graph $G$ has a \em{linear structure} if it satisfies the following
conditions: 
\begin{enumerate}
\item \label{enu:linear-structure-vertices}The vertices of $G$ can be
identified with the linear space $\F^{m}$, where $\F$ is a finite
field and $m\in\N$. 
\item \label{enu:linear-structure-edges}We identify the set of pairs of
vertices $\left(\F^{m}\right)^{2}$ with the linear space $\F^{2m}$.
Using this identification, the edges $E$ of $G$ are required to
form a linear subspace of $\F^{2m}$. 
\item \label{enu:linear-structure-sides-have-full-dim}We require that $\l\left(E\right)=\r\left(E\right)=\F^{m}$.
In other words, this means that every vertex of $G$ is both the left
endpoint of some edge and the right point of some edge. 
\end{enumerate}
\end{definition}
\begin{remark}
We mention that although it is not required by Definition~\ref{def:linear-structure},
a graph with linear structure must be regular, i.e., all the vertices
in the graph have the same in-degree and out-degree. This is a straightforward
corollary of Items~\ref{enu:linear-structure-edges} and~\ref{enu:linear-structure-sides-have-full-dim}
of the definition.
\end{remark}
The following lemmas are proved in Sections~\ref{sec:embedding-linear-structure}
and~\ref{sec:derandomized-parallel-repetition} respectively.
\begin{lemma}
[\label{lem:embedding-linear-structure}Linear Structure Embedding]There
exists a polynomial time procedure that satisfies the following requirements: 
\begin{itemize}
\item \textbf{Input:}

\begin{itemize}
\item A constraint graph $G$ of size $n$ over alphabet $\Sigma$. 
\item A finite field $\F$ of size $q$. 
\end{itemize}
\item \textbf{Output:} A constraint graph $G'=\left(\F^{m},E'\right)$ such
that the following holds:

\begin{itemize}
\item $G'$ has a linear structure. 
\item The size of $G'$ is at most $O\left(q^{2}\cdot n\right)$. 
\item $G'$ has alphabet $\Sigma^{O(\log_{q}(n))}$. 
\item If $G$ is satisfiable then $G'$ is satisfiable. 
\item If $\unsat\left(G\right)\ge\rho$ then $\unsat\left(G'\right)\ge\Omega\left(\frac{1}{q\cdot\log_{q}(n)}\cdot\rho\right)$. 
\end{itemize}
\end{itemize}
\end{lemma}

\begin{lemma}
[\label{lem:derandomized-parallel-repetition}Derandomized Parallel
Repetition]There exist a universal constant $h$ and a polynomial
time procedure that satisfy the following requirements: 
\begin{itemize}
\item \textbf{Input:}

\begin{itemize}
\item A finite field $\F$ of size $q$ 
\item A constraint graph $G=\left(\F^{m},E\right)$ over alphabet $\Sigma$
that has a linear structure. 
\item A parameter $d_{0}\in\N$ such that $d_{0}<m/h^{2}$. This parameter
will determine the dimension of linear subspaces used in the derandomized
parallel repetition, and thus together with $q$ will determine the
number of repetitions used in the derandomized parallel repetition.
\item A parameter $\rho\in\left(0,1\right)$ such that $\rho\ge h\cdot d_{0}\cdot q^{-d_{0}/h}$.
Intuitively, the parameter~$\rho$ should be chosen such that~$1-\rho$
is an upper bound on the soundness error of $G$.
\end{itemize}
\item \textbf{Output:} A constraint graph $G'$ such that the following
holds:

\begin{itemize}
\item $G'$ has size $n^{O\left(d_{0}\right)}$. 
\item $G'$ has alphabet $\Sigma^{q^{O(d_{0})}}$. 
\item If $G$ is satisfiable then $G'$ is satisfiable. 
\item If $\sat\left(G\right)<1-\rho$ then $\sat\left(G'\right)<h\cdot d_{0}\cdot q^{-d_{0}/h}$.
\item $G'$ has the projection property.
\end{itemize}
\end{itemize}
\end{lemma}
We turn to prove the main theorem from the above lemmas.
\begin{theorem*}
[\ref{thm:main}, restated]There exists a constant $\kappa>0$ such
that for every function $\varepsilon:\N\to\left(0,1\right)$ satisfying
$1/n^{\kappa}\le\varepsilon(n)\le1/\poly\log n$ the following holds:
Every language $L\in\NP$ has a two-query PCP system with perfect
completeness, soundness error $1/\poly\log n$, alphabet size $2^{1/\poly\left(\varepsilon\right)}$,
proof length $\poly\left(n\right)$, and randomness complexity $O(\log n)$.
Furthermore, the verifier in this PCP system makes only `projection'
queries.\end{theorem*}
\begin{myproof}
Let $\kappa>0$ be a constant to be chosen later, and let $\varepsilon:\N\to\left(0,1\right)$
be a function satisfying $1/n^{\kappa}\le\varepsilon(n)\le1/\poly\log n$.
Fix a language $L\in\NP$. We show that $L$ has a two-query PCP system
with perfect completeness, soundness error $\varepsilon(n)$ and alphabet
size $2^{1/\poly\left(\varepsilon\right)}$, which has the projection
property. By the \cite{FGLSS96} correspondence (Proposition~\ref{prop:corr}),
it suffices to show a polynomial time procedure that on input $x\in\B^{*}$,
outputs a constraint graph $G'$ of size $\poly\left(n\right)$ such
that the following holds: If $x\in L$ then $G'$ is satisfiable (i.e.
$\sat(G')=1$), and if $x\not\in L$ then $\sat(G')\le\varepsilon(n)$.
The procedure begins by transforming $x$, using the PCP theorem for
constraint graphs (Theorem~\ref{thm:PCP-thm-using-graphs}), to a
constraint graph $G$ of size $n=\poly\left|x\right|$ such that if
$x\in L$ then $\sat\left(G\right)=1$ and if $x\not\in L$ then $\sat\left(G\right)\le\varepsilon_{0}$,
where $\varepsilon_{0}\in(0,1)$ is a universal constant that does
not depend on $x$. Let $n=\poly\left(\left|x\right|\right)$ be the
size of $G$, and let $\rho_{0}=1-\varepsilon_{0}$.

Next, the procedure sets $\F$ to be the smallest field of size at
least $1/\left(\varepsilon(n)\right)^{c}$ for some constant $c>1$
to be determined later, and sets $q=\left|\F\right|$. Note that $q\ge\poly\log n$.
The procedure now invokes Lemma~\ref{lem:embedding-linear-structure}
(linear structure embedding) on input $G$ and $\F$, thus obtaining
a new constraint graph $G_{1}$. Note that by Lemma~\ref{lem:embedding-linear-structure}
if $\unsat\left(G\right)\ge\rho_{0}$, then $\rho_{1}\eqdef\unsat\left(G_{1}\right)\ge\Omega\left(\frac{1}{q\cdot\log_{q}\left(n\right)}\cdot\rho_{0}\right)$.

Finally, the procedure sets $d_{0}$ to be an arbitrary constant such
that $\rho_{1}\ge h\cdot d_{0}\cdot q^{-d_{0}/h}$ . Note that this
is indeed possible, since $\log_{q}\left(1/\rho_{1}\right)$ is a
constant that depends only on $\rho$ (here we use the fact that~$q\ge\poly\log n$).
Finally, the procedure invokes Lemma~\ref{lem:derandomized-parallel-repetition}
(derandomized parallel repetition) on input $G_{1}$, $\F$, $\rho_{1}$,
and $d_{0}$, and outputs the resulting constraint graph $G'$. We
note that we use here the assumption that $\varepsilon(n)\ge n^{\kappa}$,
and choose~$\kappa$ to be sufficiently small, in order to guarantee
that $G_{1}$ satisfies the requirements of Lemma~\ref{lem:derandomized-parallel-repetition}. 

It remains to analyze the parameters of $G'$. It is not hard to see
that $G'$ has size~$n^{O(d_{0})}$ and alphabet $\Sigma^{q^{O(d_{0})}}=\Sigma^{1/\poly\left(\varepsilon\right)}$.
Furthermore, if $\unsat\left(G\right)\ge\rho$, then $\unsat\left(G_{1}\right)\ge\rho_{1}$.
Therefore, by Lemma~\ref{lem:derandomized-parallel-repetition} and
by the choice of~$d_{0}$, it holds that $\sat(G')\le O(1/q^{\Omega(1)})$.
Since $q=1/\left(\varepsilon(n)\right)^{c}$, it holds for sufficiently
large $c$ that $\sat(G')\le\varepsilon(n)$, as required.\end{myproof}
\begin{remark}
Recall that \cite{MR08} prove a stronger version of the main theorem,
saying that for every soundness error $\varepsilon(n)>n^{\kappa}$,
not necessarily upper bounded by $1/\poly\log n$, it holds that $\NP$
has a PCP system with soundness $\varepsilon$ and alphabet size $\exp\left(1/\poly(\varepsilon)\right)$
(Theorem~\ref{thm:MR}). If one could prove a stronger version of
Lemma~\ref{lem:embedding-linear-structure} (Linear Structure Embedding)
in which the soundness of $G'$ is $\rho/\poly\left(q\right)$ and
the alphabet size is $\left|\Sigma\right|^{\poly\left(q\right)}$
then the stronger Theorem~\ref{thm:MR} would follow using the same
proof as above, without using a composition technique as in~\cite{MR08,DH09},
by choosing $q$ to be sufficiently small.
\end{remark}

\begin{remark}
The reduction described in Theorem~\ref{thm:main} yields graphs
of polynomial size, but not of nearly-linear size as in~\cite{MR08}
(see Remark~\ref{rem:randomness-to-proof-length}). In fact, the
construction of graphs with linear structure (Lemma~\ref{lem:embedding-linear-structure})
is nearly linear size (taking an instance of size $n$ to an instance
of size $q^{2}\cdot n$). The part that incurs a polynomial and not
nearly-linear blow-up is the derandomized parallel repetition (Lemma~\ref{lem:derandomized-parallel-repetition})
that relies on the derandomized direct product. It is possible that
a more efficient derandomized direct product may lead to a nearly-linear
size construction in total.
\end{remark}

\section{\label{sec:embedding-linear-structure}PCPs with Linear Structure}

In this section we prove Lemma~\ref{lem:embedding-linear-structure}
(linear structure embedding), which implies Theorem~\ref{thm:structure}
(the existence of PCPs with linear structure) by combining it with
the PCP theorem (Theorem~\ref{thm:PCP-thm-using-graphs}). The lemma
which says that every constraint graph can be transformed into one
that has linear structure. To this end, we use a family of structured
graphs called de-Bruijn graphs. We show that de-Bruijn graphs have
linear structure, and that every constraint graph can be embedded
in some sense on a de-Bruijn graph. This embedding technique is a
variant of a technique introduced by Babai et. al. \cite{BFLS91}
and Polishchuk and Spielman~\cite{PS94} for embedding circuits on
de-Bruijn graphs. We begin by defining de-Bruijn graphs.
\begin{definition}
\label{def:deBruijn}Let $\Lambda$ be a finite alphabet and let $m\in\N$.
The \emph{de~Bruijn graph} $\DBs$ is the directed graph whose vertices
set is $\Lambda^{m}$ such that each vertex $\left(\alpha_{1},\ldots,\alpha_{m}\right)\in\Lambda^{m}$
has outgoing edges to all the vertices of the form $\left(\alpha_{2},\ldots,\alpha_{m},\beta\right)$
for $\beta\in\Lambda$.\end{definition}
\begin{remark}
We note that previous works used a slightly different notion, the
``wrapped de~Bruijn graph'', which is a layered graph in which
the edges between layers are connected as in the de~Bruijn graph.
Also, we note that previous works fixed $\Lambda$ to be the binary
alphabet, while we we use a general alphabet.
\end{remark}
Lemma~\ref{lem:embedding-linear-structure} follows easily from the
following two propositions. Proposition~\ref{pro:de-bruijn-linear-structure}
says that de~Bruijn graphs have linear structure. Proposition~\ref{pro:de-bruijn-embedding}
says that any constraint graph can be embedded on a de~Bruijn graph.
\begin{proposition}
\label{pro:de-bruijn-linear-structure}Let $\F$ be a finite field
and let $m\in\N$. Then, the de~Bruijn graph $\DBf$ has linear structure.\end{proposition}
\begin{myproof}
Items~\ref{enu:linear-structure-vertices} and~\ref{enu:linear-structure-sides-have-full-dim}
of the definition of linear structure~(Definition~\ref{def:linear-structure})
follow immediately from the definition of de~Bruijn graphs. To see
that Item~\ref{enu:linear-structure-edges} holds, observe that in
order for a tuple in $\F^{2m}$ to be an edge of~$\DBf$, it only
needs to satisfy equality constraints, which are in turn linear constraints.
Thus, the set of edges of~$\DBf$ form a linear subspace of $\F^{2m}$.\end{myproof}
\begin{proposition}
[\label{pro:de-bruijn-embedding}Embedding on de-Bruijn graphs]There
exists a polynomial time procedure that satisfies the following requirements: 
\begin{itemize}
\item \textbf{Input:}

\begin{itemize}
\item A constraint graph $G$ of size $n$ over alphabet $\Sigma$. 
\item A finite alphabet $\Lambda$.
\item A natural number $m$ such that $\left|\Lambda\right|^{m}\ge2\cdot n$ 
\end{itemize}
\item \textbf{Output:} A constraint graph $G'$ such that the following
holds:

\begin{itemize}
\item The underlying graph of $G'$ is the de~Bruijn graph $\DBs$. 
\item The size of $G'$ is $\left|\Lambda\right|^{m+1}$. 
\item $G'$ has alphabet $\Sigma^{O(m)}$. 
\item If $G$ is satisfiable then $G'$ is satisfiable. 
\item If $\unsat\left(G\right)\ge\rho$ then $\unsat\left(G'\right)\ge\Omega\left(\frac{n}{\left|\Lambda\right|^{m+1}\cdot m}\cdot\rho\right)$. 
\end{itemize}
\end{itemize}
\end{proposition}
Lemma~\ref{lem:embedding-linear-structure} (linear structure embedding)
is obtained by invoking Proposition~\ref{pro:de-bruijn-embedding}
with $\Lambda=\F$, $m=\left\lceil \log_{q}\left(2\cdot n\right)\right\rceil $
and combining it with Proposition~\ref{pro:de-bruijn-linear-structure}.
The rest of this section is devoted to proving Proposition~\ref{pro:de-bruijn-embedding},
and is organized as follows: In Section~\ref{sub:de-Bruijn-routing}
we give the required background on the routing properties of de~Bruijn
graphs. Then, in Section~\ref{sub:de-Bruijn-overview}, we give an
outline of the proof of Proposition~\ref{pro:de-bruijn-embedding}.
Finally, we give the full proof of the proposition in Section~\ref{sub:de-Bruijn-details}.

\subsection{\label{sub:de-Bruijn-routing}de Bruijn graphs as routing networks}

The crucial property of the de~Bruijn graphs that we use is that
the de~Bruijn graph is a \textsf{permutation routing network}. To
explain the intuition that underlies this notion, let us think of
the vertices of the de~Bruijn graph as computers in a network, such
that two computers can communicate if and only if they are connected
by an edge. Furthermore, sending a message from a computer to its
neighbor takes one unit of time. Suppose that each computer in the
network wishes to send a message to some other computer in the network,
and furthermore each computer needs to receive a message from exactly
one computer (that is, the mapping from source computers to target
computers is a permutation). Then, the routing property of the de~Bruijn
network says that we can find paths in the network that have the following
properties:
\begin{enumerate}
\item Each path corresponds to a message that needs to be sent, and goes
from the message's source computer to its target computer.
\item If all the messages are sent simultaneously along their corresponding
paths, then at each unit of time, each computer processes exactly
one message. By ``processing'' we mean that the computer receives
the message from one of its neighbors and sends it to one of its neighbors.
\item The paths are of length exactly $2\cdot m$. This means that if all
the messages are sent simultaneously along their corresponding paths,
then after $2\cdot m$ units of time all the messages will reach their
destination.
\end{enumerate}
Formally, this property can be stated as follows.
\begin{fact}
\label{fac:routing}Let $\DBs$ be a de-Brujin graph. Then, given
a permutation $\mu$ on the vertices of $\DBs$ one can find a set
of undirected paths of length $l=2m$ which connect each vertex $v$
to $\mu(v)$ and which have the following property: For every $j\in\left[l\right]$,
each vertex $v$ is the $j$-th vertex of exactly one path. Furthermore,
finding the paths can be done in time that is polynomial in the size
of $\DBs$.
\end{fact}
Fact~\ref{fac:routing} is proved in~\cite{L92} for the special
case of $\Lambda=\B$. The proof of the general case essentially follows
the original proof, except that the looping algorithm of Bene\v{s}
is replaced with the decomposition of $d$-regular graphs to $d$
perfect matchings. For completeness, we give the proof of the general
case in Appendix~\ref{sec:Routing-on-deBruijn}.
\begin{remark}
Note that the paths mentioned in Fact~\ref{fac:routing} are undirected.
That is, if a vertex~$u$ appears immediately after a vertex~$v$
in path, then either $\left(u,v\right)$ or $\left(v,u\right)$ are
edges of $\DBs$.
\end{remark}

\subsection{\label{sub:de-Bruijn-overview}Proof overview}

Suppose we are given as input a constraint graph $G$ which we want
to embed on $\DB=\DBs$. Recall that the size of $G$ is at most~$\left|\Lambda\right|^{m}$,
so we may identify the vertices of $G$ with some of the vertices
of $\DB$.

\paragraph*{Handling degree $1$}

As a warm up, assume that $G$ has degree $1$, i.e., $G$ is a perfect
matching. In this case, we construct $G'$ as follows. We choose the
alphabet of $G'$ to be $\Sigma^{l}$ for $l\eqdef2m$. Fix any assignment
$\pi$ to $G$. We describe how to construct a corresponding assignment
$\pi'$ to $G'$. We think of the vertices of $G$ as computers, such
that each vertex $v$ wants to send the value $\pi(v)$ as a message
to its unique neighbor in $G$. Using the routing property of the
de~Bruijn graph, we find paths for routing those messages along the
edges of $G'$. Recall that if all the messages are sent simultaneously
along those paths, then every computer has to deal with one packet
at each unit of time, for $l$ units of time. We now define the assignment
$\pi'$ to assign each vertex $v$ of $G'$ a tuple in $\Sigma^{l}$
whose $j$-th element is the message with which $v$ deals at the
$j$-th unit of time.

We define the constraints of $G'$ such that they verify that the
routing is done correctly. That is, if the computer $u$ is supposed
to send a message to a vertex $v$ between the $j$-th unit of time
and the $\left(j+1\right)$-th unit of time, then the constraint of
the edge between $u$ and $v$ checks that $\pi'\left(u\right)_{j}=\pi'(v)_{j+1}$.
Furthermore, for each edge $\left(u,v\right)$ of $G$, the constraints
of $G'$ check that the values $\pi'\left(v\right)_{l}$ and $\pi'\left(v\right)_{1}$
satisfy the edge $\left(u,v\right)$. This condition should hold because
if $\pi'$ was constructed correctly according to $\pi$ then $\pi'\left(v\right)_{l}=\pi(u)$
and $\pi'\left(v\right)_{1}=\pi(v)$. It should be clear that the
constraints of $G'$ ``simulate'' the constraints of $G$. We discuss
the exact behavior of the soundness error in the detailed proof.

\paragraph*{Handling arbitrary degree graphs}

Using the expander replacement technique of Papadimitriou and Yannakakis
\cite{PY91}, we may assume that $G$ is $d$-regular for some universal
constant $d$. The $d$-regularity of $G$ implies that the edges
of $G$ can be partitioned to $d$ disjoint perfect matchings $\mu_{1},\ldots,\mu_{d}$
in polynomial time (see, e.g., \cite[Proposition 18.1.2]{C98}). Now,
we set the alphabet of $G'$ to be $\left(\Sigma^{l}\right)^{d}$,
and handle each of the matchings $\mu_{i}$ as before, each time using
a ``different part'' of the alphabet symbols. In other words, the
alphabet of $G'$ consists of $d$-tuples of $\Sigma^{l}$, and so
the constraints used to handle each matching $\mu_{i}$ will refer
to the $i$-th coordinates in those tuples. Finally, for vertex $v$,
its constraints will also check that the message it sends in each
of the $d$ paths is the same. In other words, if $\pi'\left(v\right)=\left(\sigma_{1},\ldots,\sigma_{d}\right)\in\left(\Sigma^{l}\right)^{d}$
then the constraints will check that~$\left(\sigma_{1}\right)_{1}=\ldots=\left(\sigma_{d}\right)_{1}$.
As before, the constraints of resulting graph $G'$ ``simulate''
the constraints of the original graph $G$.
\begin{remark}
Observe that the foregoing proof used only the routing property of
de~Bruijn graphs, and will work for any graph that satisfies this
property. In other words, Proposition~\ref{pro:de-bruijn-embedding}
(embedding on de-Bruijn graphs) holds for any graph for which Fact~\ref{fac:routing}
holds.
\end{remark}

\subsection{\label{sub:de-Bruijn-details}Detailed proof}

We use the following version of the expander-replacement technique
of \cite{PY91}.
\begin{lemma}
[{\label{lem:degree-reduction}\cite[Lemma 3.2]{D07}}]There exist
universal constants $c,d\in\N$ and a polynomial time procedure that
when given as input a constraint graph $G$ of size $n$ outputs a
constraint graph $G'$ of size $2\cdot d\cdot n$ over alphabet $\Sigma$
such that the following holds: 
\begin{itemize}
\item $G'$ has $2\cdot n$ vertices and is $d$-regular. 
\item If $G$ is satisfiable then so is $G'$. 
\item If $\unsat\left(G\right)\ge\rho$ then $\unsat\left(G'\right)\ge\rho/c$. 
\end{itemize}
\end{lemma}
We turn to proving Proposition~\ref{pro:de-bruijn-embedding} (embedding
on de-Bruijn graphs). When given as input a constraint graph $G$,
a finite alphabet~$\Lambda$ and a natural number $m$ such that
$\left|\Lambda^{m}\right|\ge2\cdot n$, the procedure of Proposition~\ref{pro:de-bruijn-embedding}
acts as follows. The procedure begins by invoking Lemma~\ref{lem:degree-reduction}
on $G$, resulting in a $d$-regular constraint graph $G_{1}$ over
$2\cdot n$ vertices. Then, the vertices of $G_{1}$ are identified
with a subset of the vertices of $\DB=\DBs$ (note that this is possible
since $\left|\Lambda^{m}\right|\ge2\cdot n$).

Next, the procedure partitions the edges of $G_{1}$ to $d$~disjoint
perfect matchings, and views those matchings as permutations $\mu_{1},\ldots,\mu_{d}$
on the vertices of~$\DB$ in the following way: Given a vertex $v$
of~$\DB$, if~$v$ is identified with a vertex of~$G_{1}$ then
$\mu_{i}$ maps $v$ to its unique neighbor in $G$ via the $i$-th
matching, and otherwise $\mu_{i}$ maps $v$ to itself. The procedure
then applies Fact~\ref{fac:routing} to each permutation~$\mu_{i}$
resulting in a set of paths $\mathcal{P}_{i}$ of length $l\eqdef2m$.
Let $\mathcal{P}=\bigcup\mathcal{P}_{i}$.

Finally, the procedure constructs $G'$ in the following way. We set
the alphabet of $G'$ to be $\Sigma^{l\cdot d}$, viewed as $\left(\Sigma^{l}\right)^{d}$.
If $\sigma\in\left(\Sigma^{l}\right)^{d}$, and we denote $\sigma=\left(\sigma_{1},\ldots,\sigma_{d}\right)$,
then we denote by $\sigma_{i,j}$ the element $\left(\sigma_{i}\right)_{j}\in\Sigma$.
To define the constraints of $G'$, let us consider their action on
an assignment $\pi'$ of $G'$. An edge $\left(u,v\right)$ of $\DB'$
is associated with the constraint that accepts if and only if all
the following conditions hold: 
\begin{enumerate}
\item For every $i\in\left[d\right]$, the values $\left(\pi'\left(u\right)_{i,l},\pi'\left(u\right)_{i,1}\right)$
satisfy the edge $\left(\mu_{i}^{-1}(u),u\right)$ of $G$. 
\item It holds that $\pi'\left(u\right)_{1,1}=\ldots=\pi'\left(u\right)_{d,1}$
and that $\pi'\left(v\right)_{1,1}=\ldots=\pi'\left(v\right)_{d,1}$.
\item \label{enu:de-bruijn-routing-consistency}For every $i\in\left[d\right]$
and $j\in\left[l-1\right]$ such that $u$ and $v$ are the $j$-th
and $\left(j+1\right)$-th vertices of a path in $p\in\mathcal{P}_{i}$
respectively, it holds that $\pi'\left(u\right)_{i,j}\ne\pi'\left(v\right)_{i,j+1}$.
\item Same as Condition~\ref{enu:de-bruijn-routing-consistency}, but when
$v$ is the $j$-th vertex of $p$ and $u$ is its $\left(j+1\right)$-th
vertex.
\end{enumerate}
The size of $G'$ is indeed $\left|\Lambda\right|^{m+1}$, since the
graph is $\left|\Lambda\right|$-regular and contains $\left|\Lambda\right|^{m}$
vertices. Furthermore, if $G$ is satisfiable, then so is $G'$: The
satisfiability of $G$ implies the satisfiability of $G_{1}$, so
there exists a satisfying assignment $\pi_{1}$ for $G_{1}$. We construct
a satisfying assignment $\pi'$ from $\pi_{1}$ by assigning each
vertex $v$ of $G'$ a value $\pi'\left(v\right)$, such that for
each $i\in\left[d\right]$, if $v$ is the $j$-th vertex of a path
$p\in\mathcal{P}_{i}$ that connects the vertices $u$ and $\mu_{i}(u)$,
then we set $\pi'\left(v\right)_{i,j}=\pi_{1}(u)$. Note that this
is well defined, since every vertex is the $j$-th vertex of exactly
one path in $\mathcal{P}_{i}$.

It remains to analyze the soundness of $G'$. Suppose that $\unsat\left(G\right)\ge\rho$.
Then, by Lemma~\ref{lem:degree-reduction} it holds that $\unsat\left(G_{1}\right)\ge\rho/c$.
Let $\pi'$ be an assignment to $G'$ that minimizes the fraction
of violated edges of $G'$. Without loss of generality, we may assume
that for every vertex $v$ of the $\DB$ it holds that $\pi'\left(v\right)_{1,1}=\ldots=\pi'\left(v\right)_{d,1}$:
If there is a vertex $v$ that does not match this condition, all
of the edges attached to $v$ are violated and therefore we can modify
the $\pi'(v)$ to match this condition without increasing the fraction
of violated edges of $\pi'$. Define an assignment $\pi_{1}$ to $G_{1}$
by setting $\pi_{1}(v)=\pi'\left(v\right)_{1,1}$ (when $v$ is viewed
as a vertex of $\DB$).

Since $\unsat\left(G_{1}\right)\ge\rho/c$, it holds that $\pi_{1}$
violates at least $\rho/c$~fraction of the edges of $G_{1}$, or
in other words $\pi_{1}$ violates at least $\rho\cdot2\cdot n\cdot d/c$
edges of $G_{1}$. Thus, there must exist a permutation~$\mu_{i}$
such that $\pi_{1}$ violates at least $\rho\cdot2\cdot n/c$ edges
of $G_{1}$ of the form $\left(u,\mu_{i}(u)\right)$. Fix such an
edge $\left(u,\mu_{i}(u)\right)$ and consider the corresponding path
$p\in\mathcal{P}_{i}$. Observe that $\pi'$ must violate at least
one of the edges of $p$: To see it, note that if $\pi'$ would satisfy
all the edges on $p$, then it would imply that $\pi'\left(\mu_{i}(u)\right)_{i,l}=\pi_{1}(u)$
and that $\pi'\left(\mu_{i}(u)\right)_{i,1}=\pi_{1}(\mu_{i}(u))$,
but the last two values violate the edge $\left(u,\mu_{i}(u)\right)$
of $G_{1}$, and therefore $\pi'$ must violate the last edge of $p$
- contradiction. It follows that for each of the $\rho\cdot2\cdot n/c$
edges of the matching $\mu_{i}$ that are violated by $\pi_{1}$ it
holds that $\pi'$ violates at least one edge of their corresponding
path. By averaging there must exist $j\in\left[l\right]$ such that
for at least $\rho\cdot2\cdot n/c\cdot l$ edges of the matching $\mu_{i}$
it holds that $\pi'$ violates the $j$-th edge of their corresponding
path.

Now, by the definition of the paths in $\mathcal{P}_{i}$, no edge
of $G'$ can be the $j$-th edge of two distinct paths in $\mathcal{P}_{i}$,
and therefore it follows that there at least $\rho\cdot2\cdot n/c\cdot l$
edges of $G'$ are violated by $\pi'$. Finally, there are $\left|\Lambda\right|^{m+1}$
edges in $G'$, and this implies that $\pi'$ violates a fraction
of the edges of $G'$ that is at least
\[
\frac{\rho\cdot2\cdot n/c\cdot l}{\left|\Lambda\right|^{m+1}}=\Omega\left(\frac{n}{\left|\Lambda\right|^{m+1}\cdot l}\cdot\rho\right),
\]
as required.\qed

\section{\label{sec:derandomized-parallel-repetition}Derandomized Parallel
Repetition of Constraint Graphs with Linear Structure}

\global\long\def\soundness{\varepsilon}
In this section we prove Lemma~\ref{lem:derandomized-parallel-repetition},
restated below, by implementing a form of derandomized parallel repetition
on graphs that have linear structure.
\begin{lemma}
[\ref{lem:derandomized-parallel-repetition}, restated]There exist
a universal constant $h$ and a polynomial time procedure that satisfy
the following requirements: 
\begin{itemize}
\item \textbf{Input:}

\begin{itemize}
\item A finite field $\F$ of size $q$ 
\item A constraint graph $G=\left(\F^{m},E\right)$ over alphabet $\Sigma$
that has a linear structure. 
\item A parameter $d_{0}\in\N$ such that $d_{0}<m/h^{2}$. This parameter
will determine the dimension of linear subspaces used in the derandomized
parallel repetition, and thus together with $q$ will determine the
number of repetitions used in the derandomized parallel repetition.
\item A parameter $\rho\in\left(0,1\right)$ such that $\rho\ge h\cdot d_{0}\cdot q^{-d_{0}/h}$.
Intuitively, the parameter~$\rho$ should be chosen such that~$1-\rho$
is an upper bound on the soundness error of $G$.
\end{itemize}
\item \textbf{Output:} A constraint graph $G'$ such that the following
holds:

\begin{itemize}
\item $G'$ has size $n^{O\left(d_{0}\right)}$. 
\item $G'$ has alphabet $\Sigma^{q^{O(d_{0})}}$. 
\item If $G$ is satisfiable then $G'$ is satisfiable. 
\item If $\sat\left(G\right)<1-\rho$ then $\sat\left(G'\right)<h\cdot d_{0}\cdot q^{-d_{0}/h}$.
\item $G'$ has the projection property
\end{itemize}
\end{itemize}
\end{lemma}
\global\long\def\EE{E}
\global\long\def\EF{F}
The basic idea of the proof is as follows. $G'$ contains two kinds
of vertices: the first kind corresponds to small subspaces of the
vertices space $\F^{m}$, and of the other kind corresponds to small
subspaces of the edges space $E$, where in both cases ``small subspaces''
means $O\left(d_{0}\right)$-dimensional subspaces. A satisfying assignment
$\Pi$ to $G'$ is expected to be constructed in the following way:
Take a satisfying assignment~$\pi$ to~$G$. For each vertex of
$G'$ which is a subspace $A$ of vertices, the assignment $\Pi$
should assign $A$ to $\pi_{|A}$. For each vertex of $G'$ which
is a subspace $\EF$ of edges, the assignment~$\Pi$ should assign~$\EF$
to~$\pi_{|\l\left(\EF\right)\cup\r\left(\EF\right)}$.

The edges of $G'$ are constructed so as to simulate a test on $\Pi$
to which we refer as the ``E-test'', and acts roughly as follows
(see Figure~\ref{fig:E-test} for the actual test): Choose a random
subspace $\EF$ of edges and a random subspace $A$ of endpoints of
$\EF$, and accept if and only if the labeling of the endpoints of
the edges in $\EF$ by $\Pi\left(\EF\right)$ satisfies the edges
and is consistent with the labeling of the vertices of $A$ by $\Pi\left(A\right)$.

The intuition that underlies the soundness analysis of $G'$ is the
following: The E-test performs some form of a ``derandomized direct
product test'' on $\Pi$ - if we compare it to the $P$-test (Figure~\ref{fig:P-test}),
then the pair $\left(A,F\right)$ here is analogous to the pair $\left(A,B\right)$
there. Therefore, if $\Pi\left(\EF\right)$ is consistent with~$\Pi\left(A\right)$,
the labeling~$\Pi\left(\EF\right)$ should be roughly consistent
with some assignment $\pi$ to $G$. Therefore, by checking that the
labeling $\Pi\left(F\right)$ satisfies the edges in $F$, the E-test
checks that $\pi$ satisfies many edges of $\pi$ in parallel. In
this sense, the E-test can be thought as a form of ``derandomized
parallel repetition''.

The rest of this section is organized as follows. In Section~\ref{sub:derand-parallel-repetition-construction}
we provide a formal description of the construction of $G'$ and analyze
all its parameters except for the soundness. In order to analyze the
soundness of $G'$, we introduce in Section~\ref{sub:derandomized-parallel-repetition-specialized-test}
a specialized direct product test. Finally, in Section~\ref{sub:derandomized-parallel-repetition-soundness},
we analyze the soundness of $G'$ by reducing it to the analysis of
the specialized direct product test.
\begin{notation}
Given a function $f:U\to\Sigma$ and two subsets $S,T\subseteq U$
we denote by $f_{|\left(S,T\right)}$ the pair of functions $\left(f_{|S},f_{|T}\right)$.
\end{notation}

\begin{notation}
Recall that in Notation~\ref{not:function-distance} we denoted the
notation $f\apx{\alpha}g$ ($f\napx{\alpha}g$) to mean that $f$
and $g$ differ on at most (more than) $\alpha$~fraction of the
elements of $U$. We now extend this notation to pairs of functions.
Given two pairs of functions $f_{1},f_{2}:U\to\Sigma$ and $g_{1},g_{2}:V\to\Sigma$,
we denote by $\left(f_{1},g_{1}\right)\apx{\alpha}\left(f_{2},g_{2}\right)$
the fact that both $f_{1}\apx{\alpha}f_{2}$ and $g_{1}\apx{\alpha}g_{2}$,
and otherwise we denote $\left(f_{1},g_{1}\right)\napx{\alpha}\left(f_{2},g_{2}\right)$.
\end{notation}

\subsection{\label{sub:derand-parallel-repetition-construction}The construction
of $G'$}

\begin{figure}[h!t]
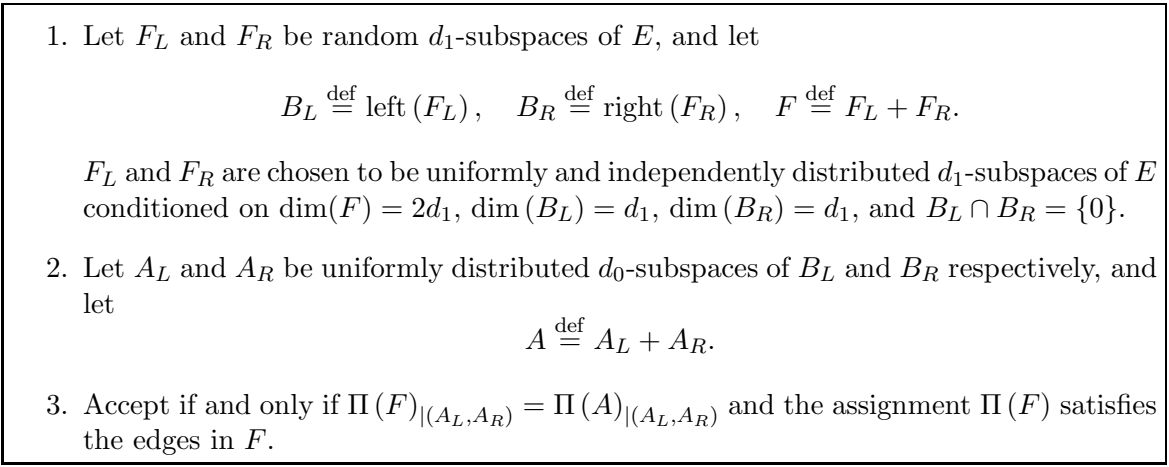

\centering \fbox{%
\begin{minipage}[c]{6in}%
\centering 
\begin{enumerate}
\item Let ${\EF}_{L}$ and ${\EF}_{R}$ be random $d_{1}$-subspaces of
$\EE$, and let 
\[
B_{L}\eqdef\l\left({\EF}_{L}\right),\quad B_{R}\eqdef\r\left({\EF}_{R}\right),\quad{\EF}\eqdef{\EF}_{L}+{\EF}_{R}.
\]
${\EF}_{L}$ and ${\EF}_{R}$ are chosen to be uniformly and independently
distributed $d_{1}$-subspaces of~$E$ conditioned on $\dim({\EF})=2d_{1}$,
$\dim\left(B_{L}\right)=d_{1}$, $\dim\left(B_{R}\right)=d_{1}$,
and $B_{L}\nolinebreak\cap\nolinebreak B_{R}\nolinebreak=\nolinebreak\set 0$.
\item Let $A_{L}$ and $A_{R}$ be uniformly distributed $d_{0}$-subspaces
of $B_{L}$ and $B_{R}$ respectively, and let 
\[
A\eqdef A_{L}+A_{R}.
\]

\item Accept if and only if $\Pi\left({\EF}\right)_{|\left(A_{L},A_{R}\right)}=\Pi\left(A\right)_{|\left(A_{L},A_{R}\right)}$
and the assignment $\Pi\left({\EF}\right)$ satisfies the edges in
${\EF}$. \end{enumerate}
\end{minipage}} \caption{\label{fig:E-test}The E-test}
\end{figure}

We begin by describing the construction of $G'$. Let $G=(\F^{m},\EE)$
be the given constraint graph, let $d_{0}$ be the parameter from
Lemma~\ref{lem:derandomized-parallel-repetition}, and let $d_{1}=h\cdot d_{0}$
where $h$ is the universal constant from Lemma~\ref{lem:derandomized-parallel-repetition}
to be chosen later. The graph $G'$ is bipartite. The right vertices
of $G'$ are identified with all the $2d_{0}$-subspaces of $\F^{m}$
(the vertex space of $G$). The left vertices of $G'$ are identified
with all the $2d_{1}$-subspaces of the edge space $E$ of $G$. An
assignment $\Pi$ to $G'$ should label each $2d_{0}$-subspace $A$
of~$\F^{m}$ with a function from $A$ to $\Sigma$, and each $2d_{1}$-subspace
$F$ of~$E$ with a function that maps the endpoints of the edges
in $F$ to $\Sigma$. The edges of $G'$ are constructed such that
they simulate the action of the ``E-test'' described in Figure~\ref{fig:E-test}.

The completeness of $G'$ is clear. It is also clear that $G'$ has
projection constraints. Let us verify the size and alphabet-size of
$G'$. The size of $G'$ is at most the number of $2d_{1}$-subspaces
of $\EE$ multiplied by the number of $2d_{0}$-subspaces of $\F^{m}$,
which is $\left|\EE\right|^{2d_{1}}\cdot\left|\F^{m}\right|^{2d_{0}}$.
It holds that $d_{0}<d_{1}$, and furthermore the linear structure
of $G'$ implies that $\dim\EE\ge m$ (by Item~\ref{enu:linear-structure-sides-have-full-dim}
of Definition~\ref{def:linear-structure}), so it follows that $\left|\F^{m}\right|^{2d_{0}}\le\left|\EE\right|^{2d_{1}}$
and thus $\left|\EE\right|^{2d_{1}}\cdot\left|\F^{m}\right|^{2d_{0}}\le\left|\EE\right|^{4d_{1}}$.
Finally, observe that the size of $G$ is $n=\left|\EE\right|$, so
it follows that the size of $G'$ is at most $n^{4d_{1}}=n^{O(d_{0})}$,
as required.

For the alphabet size, recall that an edges subspace $\EF$ is labeled
by a function that maps the endpoints of the edges to $\Sigma$. Such
a function can be represented by a string in $\Sigma^{2\cdot q^{2\cdot d_{1}}}$,
since each $2d_{1}$-subspace $\EF$ contains $q^{2d_{1}}$ edges
and each has two endpoints. It can be observed similarly that the
labels assigned by $\Pi$ to $2d_{0}$-subspaces $A$ of $\F^{m}$
can be represented by strings in~$\Sigma^{2\cdot q^{2\cdot d_{1}}}$.
The alphabet of $G'$ is therefore $\Sigma^{2\cdot q^{2\cdot d_{1}}}=\Sigma^{q^{O(d_{0})}}$,
as required.

\subsection{\label{sub:derandomized-parallel-repetition-specialized-test}The
specialized direct product test}

In order to analyze the soundness of the E-test, we introduce a variant
of the direct product test of~\cite{IKW09} that is specialized to
our needs. We refer to this variant as the \emph{specialized direct
product test}, abbreviated the ``S-test''.

Given an string $\pi:\F^{m}\to\Sigma$, we define its \emph{S-direct
product} $\Pi$ (with respect to $d_{0},d_{1}\in\N$) as follows:
$\Pi$ assigns each $2d_{0}$-subspace $A\subseteq\F^{m}$ the function~$\pi_{|A}$,
and assigns each pair of independent $d_{1}$-subspaces $\left(B_{1},B_{2}\right)$
the pair of functions $\pi_{|\left(B_{1},B_{2}\right)}$.

We turn to consider the task of testing whether a given assignment
$\Pi$ is the S-direct product of some string $\pi:\F^{m}\to\Sigma$.
In our settings, we are given an assignment $\Pi$ that assigns each
$2d_{0}$-subspace $A$ to a function $a:A\to\Sigma$ and each pair
of independent $d_{1}$-subspaces $\left(B_{1},B_{2}\right)$ to a
pair of functions $b_{1}:B_{1}\to\Sigma$, $b_{2}:B_{2}\to\Sigma$.
We wish to check whether $\Pi$ is a S-direct product of some $\pi:\F^{m}\to\Sigma$.
To this end we invoke the S-test, described in Figure~\ref{fig:S-test}.

\begin{figure}
\centering\fbox{%
\begin{minipage}[c]{6in}%
\begin{enumerate}
\item Choose uniformly distributed pair of independent $d_{1}$-subspaces
$B_{1},B_{2}$ of $\F^{m}$.
\item Choose uniformly distributed pair of $d_{0}$-subspaces $A_{1}\subseteq B_{1}$,
$A_{2}\subseteq B_{2}$.
\item Accept if and only if $\Pi\left(B_{1},B_{2}\right)_{|\left(A_{1},A_{2}\right)}=\Pi\left(A_{1}+A_{2}\right)_{|\left(A_{1},A_{2}\right)}$. \end{enumerate}
\end{minipage}} \caption{\label{fig:S-test}The S-test}
\end{figure}

It is easy to see that if $\Pi$ is a S-direct product then the S-test
always accepts. Furthermore, it can be shown that if $\Pi$ is ``far''
from being a S-direct product, then the S-test rejects with high probability.
As in the P-test, this holds even if $\Pi$ is a randomized assignment.
Formally, we have the following result.
\begin{theorem}
[\label{thm:S-test}the soundness of the S-test]There exist universal
constants $\hs',\cs\in\N$ such that the following holds: Let $d_{0}\in\N$,
$d_{1}\ge\hs'\cdot d_{0}$, and $m\ge\hs'\cdot d_{1}$, and let $\varepsilon\ge\hs'\cdot d_{0}\cdot q^{-d_{0}/\hs'}$,
$\as\eqdef\hs'\cdot d_{0}\cdot q^{-d_{0}/\hs'}$. Suppose that a (possibly
randomized) assignment $\Pi$ passes the S-test with probability at
least~$\varepsilon$. Then there exists an assignment $\pi:\F^{m}\to\Sigma$
for which the following holds. Let $B_{1}$, $B_{2}$ be uniformly
distributed and independent $d_{1}$-subspaces of $\F^{m}$, let $A_{1}$
and $A_{2}$ be uniformly distributed $d_{0}$-subspaces of $B_{1}$
and $B_{2}$ respectively, and denote $A=A_{1}+A_{2}$. Then: 
\begin{equation}
\Pr\left[\Pi\left(B_{1},B_{2}\right)_{|\left(A_{1},A_{2}\right)}=\Pi\left(A\right)_{|\left(A_{1},A_{2}\right)}\aand\Pi\left(B_{1},B_{2}\right)\apx{\as}\pi_{|\left(B_{1},B_{2}\right)}\right]=\Omega\left(\varepsilon^{\cs}\right).\label{eq:S-test}
\end{equation}

\end{theorem}
We defer the proof of Theorem~\ref{thm:S-test} to Section~\ref{sec:S-test-analysis}.
\begin{remark}
Note that Equation~\ref{eq:S-test} only says that $\Pi$ is close
to the S-direct product of $\pi$ on pairs $\left(B_{1},B_{2}\right)$,
and not necessarily on $2d_{0}$-subspaces~$A$. In fact, it could
be also proved that~$\Pi$ is close to the S-direct product of $\pi$
on the $2d_{0}$-subspaces, but this is unnecessary for our purposes.
\end{remark}

\subsection{\label{sub:derandomized-parallel-repetition-soundness}The soundness
of the derandomized parallel repetition}

In this section we prove the soundness of $G'$: namely, that if $\sat\left(G\right)<1-\rho$,
then 
\[
\sat(G')\le\soundness\eqdef h\cdot d_{0}\cdot q^{-d_{0}/h},
\]
where $h$ is the universal constant from Lemma~\ref{lem:derandomized-parallel-repetition}
(derandomized parallel repetition). We will choose $h$ to be sufficiently
large such that the various inequalities in the following proof will
hold. To this end, we note that throughout all the following proof,
increasing the choice of $h$ does not break any of our assumptions
on $h$, so we can always choose a larger $h$ to satisfy the required
inequalities. 

Let $h'$ and $c$ be the universal constants whose existence is guaranteed
by Theorem~\ref{thm:S-test} (the soundness of the S-test), and let
$\alpha$ denote the corresponding value from Theorem~\ref{thm:S-test}.
We will choose the constant $h$ to be at least~$h'$.

Let $\Pi$ be an assignment to $G'$. Let us denote by $\mathcal{T}$
the event in which the E-test accepts $\Pi$. With a slight abuse
of notation, for a subspace ${\EF}\subseteq\EE$ and an assignment
$\pi:\F^{m}\to\Sigma$, we denote by $\Pi\left({\EF}\right)\apx{\alpha}\pi$
the claim that for at least $1-\alpha$~fraction of the edges $e$
of ${\EF}$ it holds that $\Pi\left({\EF}\right)$ is consistent with
$\pi$ on both the endpoints of $e$, and otherwise we denote $\Pi\left({\EF}\right)\napx{\alpha}\pi$.
Our proof is based on two steps: 
\begin{itemize}
\item We will show (in Proposition~\ref{pro:test-passes-consistently}
below) that if the test accepts with probability $\soundness$, then
it is ``because'' $\Pi$ is consistent with some underlying assignment
$\pi:\F^{m}\to\Sigma$. This is done essentially by observing that
the E-test ``contains'' an S-test, and reducing to the analysis
of the S-test. 
\item On the other hand, we will show (in Proposition~\ref{pro:inconsistency-with-fixed-assignment}
below) that for every assignment $\pi:\F^{m}\to\nolinebreak\Sigma$
the probability that the test accepts while being consistent with
$\pi$ is negligible. This is done roughly as follows: Any fixed assignment
$\pi$ is rejected by at least $\rho$ fraction of $G$'s edges. Furthermore,
the subspace ${\EF}$ queried by the test is approximately a uniformly
distributed subspace of $E$, and hence a good sampler of $\EE$.
It follows ${\EF}$ must contain $\approx\nolinebreak\rho$~fraction
of edges of $G$ that reject $\pi$, and therefore $\Pi\left({\EF}\right)$
must be inconsistent with~$\pi$.
\end{itemize}
The conclusions of each of the foregoing two steps clearly contradict
each other, we therefore conclude that the E-test accepts with probability
less than $\soundness$. We now state the two said propositions, which
formalize the foregoing two steps, and which are proved in Sections~\ref{sub:test-passes-consistently}
and~\ref{sub:inconsistency-with-fixed-assignment} respectively.
\begin{proposition}
\label{pro:test-passes-consistently}There exists $\varepsilon_{0}=\Omega\left(\soundness^{c}\right)$
such that the following holds: If $\Pr\left[\mathcal{T}\right]\ge\soundness$,
then there exists an assignment $\pi:\F^{m}\to\Sigma$ such that $\Pr\left[\mathcal{T}\aand\Pi\left({\EF}\right)\apx{4\cdot\alpha}\pi\right]\ge\nolinebreak\varepsilon_{0}$.
\end{proposition}

\begin{proposition}
\label{pro:inconsistency-with-fixed-assignment}Let $\varepsilon$
be as in Proposition~\ref{pro:test-passes-consistently}. Then, for
every assignment $\pi:\F^{m}\to\Sigma$ it holds that $\Pr\left[\mathcal{T}\aand\Pi\left({\EF}\right)\apx{4\cdot\alpha}\pi\right]<\varepsilon_{0}$.
\end{proposition}
\noindent Clearly, the two propositions together imply that $\Pr[\mathcal{T}]\le\soundness$,
as required.\medskip{}

Before turning to the proofs of Propositions~\ref{pro:test-passes-consistently}
and \ref{pro:inconsistency-with-fixed-assignment} let us state a
useful claim that says that if we take a random $d$-subspace of edges
and project it to its left endpoints (respectively, right endpoints),
we get a random $d$-subspace of vertices with high probability.
\begin{claim}
\label{cla:projection-to-sides}Let $d\in\N$ and let $E_{a}$ be
a uniformly distributed $d$-subspace of $\EE$. Then, $\Pr\left[\dim\left(\l\left(E_{a}\right)\right)=d\right]\ge1-d/q^{m-d}$,
and conditioned on $\dim\left(\l\left(E_{a}\right)\right)=d$, it
holds that $\l\left(E_{a}\right)$ is a uniformly distributed $d$-subspace
of $\F^{m}$. The same holds for $\r\left(E_{a}\right)$.

More generally, let $E_{b}$ be a fixed subspace of $\EE$ such that
$\dim\left(E_{b}\right)>d$ and $\dim\left(\l\left(E_{b}\right)\right)=D\nolinebreak>\nolinebreak d$.
Let $E_{a}$ be a uniformly distributed $d$-subspace of $E_{b}$.
Then, $\Pr\left[\dim\left(\l\left(E_{a}\right)\right)=d\right]\nolinebreak\ge\nolinebreak1-d/q^{D-d}$,
and conditioned on $\dim\left(\l\left(E_{a}\right)\right)=d$, it
holds that $\l\left(E_{a}\right)$ is a uniformly distributed $d$-subspace
of $\l\left(E_{b}\right)$. Again, the same holds for $\r\left(E_{a}\right)$.
\end{claim}
We defer the proof of to Appendix~\ref{sec:proof-of-projection-to-sides}

\subsubsection{\label{sub:test-passes-consistently}Proof of Proposition~\ref{pro:test-passes-consistently}}

Suppose that $\Pr\left[\mathcal{T}\right]\ge\soundness$. We prove
Proposition~\ref{pro:test-passes-consistently} by arguing that the
E-test contains an ``implicit S-test'' and applying Theorem~\ref{thm:S-test}
(the soundness of the S-test).

Observe that, without loss of generality, we may assume that for every
edge-subspace $F$ such that $\Pi\left(F\right)$ violates one of
the edges in $F$, it holds that $\Pi\left(F\right)_{\left(A_{L},A_{R}\right)}\ne\Pi\left(A\right)_{\left(A_{L},A_{R}\right)}$
for any choice of $A_{L}$ and $A_{R}$. The reason is that for every
such $F$, we can modify $\Pi\left(F\right)$ such that it assigns
symbols outside of the alphabet~$\Sigma$ of~$G$, so~$\Pi\left(F\right)$
will always disagree with~$\Pi\left(A\right)$. Note that this modification
indeed does not change the acceptance probability of $\Pi$. This
assumption that we make on $\Pi$ implies in particular that the event
$\mathcal{T}$ is equivalent to the event $\Pi\left(F\right)_{\left(A_{L},A_{R}\right)}\ne\Pi\left(A\right)_{\left(A_{L},A_{R}\right)}$,
and this equivalence is used in the following analysis. 

We turn back to the proof of Proposition~\ref{pro:test-passes-consistently}.
We begin the proof by extending $\Pi$ to pairs of independent $d_{1}$-subspaces
of $\F^{m}$ in a randomized manner as follows: Given a pair of independent
$d_{1}$-subspaces $B_{1}$ and $B_{2}$, we choose ${\EF}_{1}$ and
${\EF}_{2}$ to be uniformly distributed and independent $d_{1}$-subspaces
of $\EE$ such that $\l\left({\EF}_{1}\right)=B_{1}$ and $\r\left({\EF}_{2}\right)=B_{2}$,
and set $\Pi\left(B_{1},B_{2}\right)=\Pi\left({\EF}_{1}+{\EF}_{2}\right)_{|\left(B_{1},B_{2}\right)}$.

Now, observe that the probability that the E-test accepts equals to
the probability that the S-test accepts the extended $\Pi$. The reason
is that the subspaces $B_{L}$, $B_{R}$, $A_{L}$, $A_{R}$ of the
E-test are distributed like the subspaces $B_{1}$, $B_{2}$, $A_{1}$,
$A_{2}$ of the S-test. It thus follows the E-test performs in a way
an S-test on the extended assignment $\Pi$.

Next, we note that by choosing $h$ to be sufficiently large, the
foregoing ``implicit S-test'' matches the requirements of Theorem~\ref{thm:S-test}
(the soundness of the S-test), and we can thus apply this theorem.
It follows that there exists an assignment $\pi:\F^{m}\to\Sigma$
such that 
\begin{equation}
\Pr\left[\Pi\left(B_{L},B_{R}\right)_{\left(A_{L},A_{R}\right)}=\Pi\left(A\right)_{|\left(A_{L},A_{R}\right)}\and\Pi\left(B_{L},B_{R}\right)\apx{\alpha}\pi_{\left(B_{L},B_{R}\right)}\right]\ge\Omega\left(\soundness^{c}\right).\label{eq:derand-par-rep-S-test-conclusion}
\end{equation}
By using the equivalence between the event \emph{$\mathcal{T}$} and
the event $\Pi\left(F\right)_{\left(A_{L},A_{R}\right)}\ne\Pi\left(A\right)_{\left(A_{L},A_{R}\right)}$,
it follows that Inequality~\ref{eq:derand-par-rep-S-test-conclusion}
is equivalent to the inequality 
\begin{equation}
\Pr\left[\mathcal{T}\and\Pi\left({\EF}\right)_{|\left(B_{L},B_{R}\right)}\apx{\alpha}\pi_{|\left(B_{L},B_{R}\right)}\right]\ge\Omega\left(\soundness^{c}\right).\label{eq:E_0-consistent-on-B_L_R}
\end{equation}
We turn to show that 
\[
\Pr\left[\mathcal{T}\and\Pi\left({\EF}\right)\apx{4\alpha}\pi\right]\ge\Omega\left(\soundness^{c}\right).
\]
 We will prove that if ${\EF}$ is such that $\Pi\left({\EF}\right)\napx{4\alpha}\pi$,
then for a random choice of $B_{L},B_{R}$ conditioned on ${\EF}$,
it is highly unlikely that Inequality \ref{eq:E_0-consistent-on-B_L_R}
still holds. Formally, we will prove the following.
\begin{claim}
\label{cla:inconsistent-edges-spaces-inconsistent-on-S-test}For every
fixed $2d_{0}$-subspace $F_{0}$ of $E$ such that $\Pi\left(F_{0}\right)\napx{4\alpha}\pi$,
it holds that
\[
\Pr\left[\left.\Pi\left({\EF}\right)_{|\left(B_{L},B_{R}\right)}\apx{\alpha}\pi_{|\left(B_{L},B_{R}\right)}\right|{F}=F_{0}\right]\le1/\left(q^{d_{1}-2}\cdot\alpha^{2}\right).
\]

\end{claim}
\noindent We defer the proof of Claim~\ref{cla:inconsistent-edges-spaces-inconsistent-on-S-test}
to the end of this section. Claim~\ref{cla:inconsistent-edges-spaces-inconsistent-on-S-test}
immediately implies the following.
\begin{corollary}
\label{cor:inconsistent-edges-spaces-inconsistent-on-S-test}It holds
that
\[
\Pr\left[\left.\Pi\left({\EF}\right)_{|\left(B_{L},B_{R}\right)}\apx{\alpha}\pi_{|\left(B_{L},B_{R}\right)}\right|\Pi\left({\EF}\right)\napx{4\alpha}\pi\right]\le1/\left(q^{d_{1}-2}\cdot\left(\alpha/2\right)^{2}\right).
\]

\end{corollary}
\noindent By combining Corollary~\ref{cor:inconsistent-edges-spaces-inconsistent-on-S-test}
with Inequality~\ref{eq:E_0-consistent-on-B_L_R}, and by choosing
$h$ to be sufficiently large, it follows that 
\[
\Pr\left[\mathcal{T}\and\Pi\left({\EF}\right)_{|\left(B_{L},B_{R}\right)}\apx{\alpha}\pi_{|\left(B_{L},B_{R}\right)}\and\Pi\left({\EF}\right)\apx{4\alpha}\pi\right]\ge\Omega\left(\varepsilon^{c}\right).
\]
This implies that
\[
\Pr\left[\mathcal{T}\and\Pi\left({\EF}\right)\apx{4\alpha}\pi\right]\ge\Omega\left(\varepsilon^{c}\right).
\]
Setting $\varepsilon_{0}$ to be the latter lower bound finishes the
proof.\qed
\begin{myproof}
[Proof of Claim~\ref{cla:inconsistent-edges-spaces-inconsistent-on-S-test}.]Observe
that the assumption $\Pi\left({\EF_{0}}\right)\napx{4\alpha}\pi$
implies that one of the following holds 
\[
\Pi\left({\EF_{0}}\right)_{|\l\left({\EF}_{0}\right)}\napx{2\alpha}\pi_{|\l\left({\EF}_{0}\right)},
\]
\[
\Pi\left({\EF}_{0}\right)_{|\r\left({\EF}_{0}\right)}\napx{2\alpha}\pi_{|\r\left({\EF}_{0}\right)}.
\]
Without loss of generality, assume that the first holds. Now, when
conditioning on ${\EF}={\EF}_{0}$, it holds that $\EF_{L}$ is a
uniformly distributed $d_{1}$-subspace of ${F}_{0}$ satisfying $\dim\left(\l\left(\EF_{L}\right)\right)=d_{1}$.
By Claim~\ref{cla:projection-to-sides} (with $E_{b}={\EF}_{0}$
and $E_{a}=\EF_{L}$), under the conditioning on $\dim\left(\l\left(\EF_{L}\right)\right)=d_{1}$,
it holds that $B_{L}\eqdef\l\left(\EF_{L}\right)$ is a uniformly
distributed $d_{1}$-subspace of $\l\left({\EF}_{0}\right)$. Therefore,
by Lemma~\ref{lem:subspace-point-sampler} (subspace-point sampler),
the event $\Pi\left({\EF}\right)_{|B_{L}}\napx{\alpha}\pi_{|B_{L}}$
occurs with probability at least 
\[
1-1/\left(q^{d_{1}-2}\cdot\left(\alpha-q^{-d_{1}}\right)^{2}\right)\ge1-1/\left(q^{d_{1}-2}\cdot\left(\alpha/2\right)^{2}\right),
\]
as required.
\end{myproof}

\subsubsection{\label{sub:inconsistency-with-fixed-assignment}Proof of Proposition~\ref{pro:inconsistency-with-fixed-assignment}}

Fix an assignment $\pi:\F^{m}\to\Sigma$. By assumption it holds that
$\sat\left(G\right)<1-\rho$, and therefore $\pi$ must violate a
set $E^{*}$ of edges of $G$ of density at least $\rho$. Below we
will show that at least $\rho/2$~fraction of the edges in ${\EF}$
are in $E^{*}$ with probability greater than $1-\varepsilon_{0}$.
Now, observe that $\Pi\left({\EF}\right)$ cannot satisfy the edges
of $F$ and at the same time be consistent with $\pi$ on the edges
in $E^{*}$, and hence whenever the latter event occurs it either
holds that the E-test fails or that $\Pi\left({\EF}\right)\napx{\rho/2}\pi$.
However, for sufficiently large choice of $h$, it holds that $\rho/2>4\cdot\alpha$,
and therefore the probability that the E-test passes and at the same
time it holds that $\Pi\left({\EF}\right)\apx{4\cdot\alpha}\pi$ is
less than $\varepsilon_{0}$, as required.

It remains to show that
\[
\Pr\left[\frac{\left|{\EF}\cap E^{*}\right|}{\left|{\EF}\right|}\ge\rho/2\right]>1-\varepsilon_{0}.
\]
We prove the above inequality by showing that ${\EF}$ is close to
being a uniformly distributed $2d_{1}$-subspace of $\EE$, and then
applying Lemma~\ref{lem:subspace-point-sampler} (subspace-point
sampler). To this end, let $F_{L}'$ and $F_{R}'$ be uniformly distributed
$d_{1}$-subspaces of $F$, and let $F'=F_{L}'+F_{R}'$. Let us denote
by $\mathcal{E}_{1}$ the event in which $\dim\left(F'\right)=2d_{1}$,
and by $\mathcal{E}_{2}$ the event in which $\l\left(F_{L}'\right)$
and $\r\left(F_{R}'\right)$ are independent and are of dimension~$d_{1}$.
Observe that conditioned on $\mathcal{E}_{1}$ and $\mathcal{E}_{2}$
the subspace $F'$ is distributed exactly like the subspace ${\EF}$.
It therefore holds that
\begin{eqnarray*}
\Pr\left[\frac{\left|{\EF}\cap E^{*}\right|}{\left|{\EF}\right|}\ge\rho/2\right] & = & \Pr\left[\left.\frac{\left|F'\cap E^{*}\right|}{\left|F'\right|}\ge\rho/2\right|\mathcal{E}_{1}\aand\mathcal{E}_{2}\right]\\
 & \ge & \Pr\left[\left.\frac{\left|F'\cap E^{*}\right|}{\left|F'\right|}\ge\rho/2\aand\mathcal{E}_{2}\right|\mathcal{E}_{1}\right]\\
 & \ge & \Pr\left[\left.\frac{\left|F'\cap E^{*}\right|}{\left|F'\right|}\ge\rho/2\right|\mathcal{E}_{1}\right]-\Pr\left[\neg\mathcal{E}_{2}|\mathcal{E}_{1}\right]\\
 & \ge & \Pr\left[\left.\frac{\left|F'\cap E^{*}\right|}{\left|F'\right|}\ge\rho/2\right|\mathcal{E}_{1}\right]-\frac{\Pr\left[\neg\mathcal{E}_{2}\right]}{\Pr\left[\mathcal{E}_{1}\right]}.
\end{eqnarray*}
Now, observe that conditioned on $\mathcal{E}_{1}$, the subspace
$F'$ is a uniformly distributed $2d_{1}$-subspace of $\EE$. Thus,
by Lemma~\ref{lem:subspace-point-sampler} (subspace-point sampler)
it holds that
\[
\Pr\left[\left.\frac{\left|F'\cap E^{*}\right|}{\left|F'\right|}\ge\rho/2\right|\mathcal{E}_{1}\right]\ge1-1/q^{2d_{1}-2}\cdot\left(\rho/2-q^{-2d_{1}}\right)^{2}\ge1-1/q^{2d_{1}-2}\cdot\left(\rho/3\right)^{2}.
\]
Moreover, by Proposition~\ref{pro:random-subspaces-disjoint} it
holds that 
\begin{eqnarray*}
\Pr\left[\mathcal{E}_{1}\right] & \ge & 1-2d_{1}/q^{\dim\EE-2d_{1}}\\
 & \ge & 1-2d_{1}/q^{m-2d_{1}}\\
 & \ge & \frac{1}{2},
\end{eqnarray*}

Finally, we upper bound $\Pr\left[\neg\mathcal{E}_{2}\right]$ by
showing that $\Pr\left[\mathcal{E}_{2}\right]\ge1-4d_{1}/q^{m-2\cdot d_{1}}$.
By Claim~\ref{cla:projection-to-sides} (with $E_{b}=\EE$ and $E_{a}=F_{L}',F_{R}'$)
it holds that $\dim\left(\l\left(F_{L}'\right)\right)=\dim\left(\r\left(F_{R}'\right)\right)=d_{1}$
with probability at least $1-2\cdot d_{1}/q^{m-d_{1}}$. Furthermore,
conditioned on the latter event, it holds that $\l\left(F_{L}'\right)$
and $\r\left(F_{R}'\right)$ are uniformly distributed $d_{1}$-subspaces
of $\F^{m}$, and it is also easy to see that those subspaces are
independent. By Proposition~\ref{pro:random-subspaces-disjoint},
this implies that conditioned on $\dim\left(\l\left(F_{L}'\right)\right)=\dim\left(\r\left(F_{R}'\right)\right)=d_{1}$
the subspaces $\l\left(F_{L}'\right)$ and $\r\left(F_{R}'\right)$
are independent with probability at least $1-2d_{1}/q^{m-2\cdot d_{1}}$,
and hence $\Pr\left[\mathcal{E}_{2}\right]\ge1-4d_{1}/q^{m-2\cdot d_{1}}$
as required.

We conclude that that
\begin{eqnarray*}
\Pr\left[\frac{\left|{\EF}\cap E^{*}\right|}{\left|{\EF}\right|}\ge\rho/2\right] & \ge & \Pr\left[\left.\frac{\left|F'\cap E^{*}\right|}{\left|F'\right|}\ge\rho/2\right|\mathcal{E}_{1}\right]-\frac{\Pr\left[\neg\mathcal{E}_{2}\right]}{\Pr\left[\mathcal{E}_{1}\right]}\\
 & \ge & 1-1/q^{2\cdot d_{1}-2}\cdot\left(\rho/3\right)^{2}-\frac{4\cdot d_{1}/q^{m-2\cdot d_{1}}}{1/2}\\
 & = & 1-1/q^{2\cdot d_{1}-2}\cdot\left(\rho/3\right)^{2}-8\cdot d_{1}/q^{m-2\cdot d_{1}}\\
 & > & 1-\varepsilon_{0},
\end{eqnarray*}
where the last inequality holds for sufficiently large choice of $h$.
This concludes the proof.\qed

\section{\label{sec:dPCP}Decodable PCPs}

The PCP theorem says that $\csat$ has a proof system in which the
(randomized) verifier reads only $O(1)$ bits from the proof. In known
constructions this proof is invariably an \emph{encoding} of a satisfying
assignment to the input circuit. Although this is not stipulated by
the classical definition of a PCP, the fact that a PCP is really an
encoding of a `standard' NP witness is sometimes useful. Various attempts
to capture this behavior gave rise to such objects as PCPs of Proximity
(PCPPs) \cite{BGHSV06} or assignment testers~\cite{DR06}, and more
recently to decodable PCPs (dPCPs) \cite{DH09}.

\paragraph{Application: alphabet reduction through composition.}

The notion of dPCPs is useful for reducing the alphabet size of PCPs
with small soundness error via composition. They were introduced in
\cite{DH09} in an attempt to simplify and modularize the construction
of \cite{MR08}. Indeed this notion is a refinement of \cite{MR08}'s
so-called ``locally decode or reject codes (LDRCs)'' which allowed
\cite{DH09} prove a generic two-query composition theorem. This theorem
allows one to improve parameters of a PCP using any dPCP. The only
known construction of a dPCP (until this work) is the so-called ``manifold
vs. point'' construction. In the next sections we give a new construction
of a dPCP by adapting the work of the previous sections to a dPCP.
Our dPCP can then be plugged into the composition scheme of \cite{DH09}
to reprove the result of \cite{MR08}. We sketch this in Section~\ref{sub:MR}.

\paragraph{Decodable PCPs and PCPs of Proximity (PCPPs).}

We can define dPCPs for any NP language but we focus on the language
$\csat$ since it suffices for our purposes. A dPCP system for $\csat$
is a proof system in which the satisfying assignments of the input
circuit are \emph{encoded} into a special ``dPCP'' format. These
encodings can then be both locally verified and locally decoded in
a probabilistic manner. In other words, the verifier is given an input
circuit as well as oracle access to a proof string, and is able to
simultaneously check that the given string is a valid encoding of
a \emph{satisfying} assignment, as well as to decode a random symbol
in that assignment. The formal definition is given below in Section~\ref{sub:dPCP}.

dPCPs are closely related to PCPs of proximity \cite{BGHSV06} or
assignment testers \cite{DR06} (to be defined shortly below). In
fact dPCPs were first defined in the context of low soundness error
to overcome inherent limitations of PCPPs in this parameter range.
In this work we extend the definition of a dPCP also to the high soundness
error range (i.e. matching the parameter range of PCPPs). We call
these uniquely decodable PCPs (udPCPs) as opposed to list decodable
dPCPs. It is natural to consider such an object in our context since
our approach is to reduce the error by parallel repetition. Thus we
start with a dPCP with relatively high error and then reduce the error.
Uniquely decodable PCPs turn out to be roughly equivalent to PCPPs
in the sense that any PCPP can be used to construct a udPCP and vice
versa. In retrospect, we find the notion of udPCPs (and dPCPs) just
as natural as that of PCPPs. In fact, many known constructions of
PCPPs work by implicitly constructing a udPCP and then adding comparison
checks. \\

As mentioned above, our main goal in Sections~\ref{sec:dPCP}, \ref{sec:dPCP-embedding-linear-structure},
and \ref{sec:dPCP-derandomized-parallel-repetition} is to give a
new construction of dPCPs with low soundness error (Theorem~\ref{thm:main-dPCP}).
Our construction of dPCPs with low soundness error follows the same
steps as our construction of PCPs with low soundness error: In the
first step, we construct a dPCP with high soundness error (that is,
a udPCP). In the second step, we apply derandomized parallel repetition
to the foregoing udPCP to reduce its soundness error to a sub-constant
function.

In the following subsections we recall the definitions of PCPPs (Section~\ref{sec:PCPP})
and define udPCPs (Section~\ref{sub:dPCP}). We then prove the equivalence
of PCPPs and udPCPs. Next we state two lemmas that capture the two
main steps in constructing dPCPs. This is followed by a proof of Theorem~\ref{thm:main-dPCP}
(construction of dPCPs). Finally, we sketch a proof of Theorem~\ref{thm:MR}
(the \cite{MR08} result) based on Theorem~\ref{thm:main-dPCP}.

\subsection{\label{sec:PCPP}Recalling the definition of PCPPs}

PCPs of Proximity (PCPPs) were defined simultaneously in \cite{BGHSV06}
and in \cite{DR06} under the name assignment testers. PCPPs allow
the verifier to check not only that a given circuit is satisfiable,
but also that a given assignment is (close to being) satisfying. They
were introduced for various motivations, and in particular, they facilitate
composition of PCPs which is important for constructing PCPs with
reasonable parameters.

Intuitively, a PCP verifier for $\csat$ is an oracle machine $V$
that is given as input a circuit $\varphi:\B^{t}\to\B$, and is also
given oracle access to an assignment $x$ to $\varphi$ and a proof
$\pi$. The verifier $V$ is required to verify that $x$ is close
to a satisfying assignment of $\varphi$, and to do so by making only
few queries to $x$ and $\pi$. For technical reasons, it is often
preferable to define $V$ in a different way. In this definition,
instead of requiring that $V$ makes few queries to its a oracle and
decides according to the answers it gets, we require that $V$ outputs
explicitly the queries it intends to make and the predicate $\psi$
it intends to apply to the answers it gets. The advantage of this
definition is that it allows us to measure the complexity of the predicate~$\psi$.
The formal definitions of PCPP are given below.
\begin{definition}
[\label{def:PCPP-verifier}PCPP verifier]A \emph{PCPP verifier}
for $\csat$ is a probabilistic polynomial-time algorithm $V$ that
on input circuit $\varphi:\B^{t}\to\B$ of size $n$ tosses $r(n)$
coins and generates 
\begin{enumerate}
\item $q=q(n)$ queries $I=\left(i_{1},\ldots,i_{q}\right)$ in $\left[t+\ell\right]$
(where $\ell=\ell\left(n\right)$ and the queries are viewed as coordinates
of a string in $\B^{t+\ell}$). 
\item A circuit $\psi:\B^{q}\to\left\{ 0,1\right\} $ of size at most $s(n)$. 
\end{enumerate}

We shall refer to $r(n)$, $q(n)$, $\ell(n)$, and $s(n)$ as the
\emph{randomness complexity}, \emph{query complexity}, \emph{proof
length}, and \emph{decision complexity} respectively.

\end{definition}

\begin{definition}
[\label{def:PCPP}PCPPs]Let $V$, $r(n)$, $q(n)$, $\ell(n)$,
and $s(n)$, be as in Definition~\ref{def:PCPP-verifier}, and let
$\rho:\N\to(0,1]$. We say that $V$ is a \emph{PCPP system} for $\csat_{\B}$
with \emph{rejection ratio} $\rho$ if the following holds for every
circuit $\varphi:\B^{t}\to\B$ of size $n$: 
\begin{itemize}
\item \textbf{Completeness:} For every satisfying assignment $x$ for $\varphi$
there exists a proof string $\pi_{x}\in\B^{\ell}$ such that
\[
\Pr_{I,\psi}\left[\psi\left(\left(x\circ\pi_{x}\right)_{|I}\right)=1\right]=1,
\]
where $I$ and $\psi$ are the (random) output of $V\left(\varphi\right)$. 
\item \textbf{Soundness:} For every $x\in\B^{t}$ that is $\varepsilon$-far
from a satisfying assignment to $\varphi$ and every proof string
$\pi\in\B^{\ell}$ the following holds:
\[
\Pr_{I,\psi}\left[\psi\left(\left(x\circ\pi\right)_{|I}\right)=0\right]\ge\rho\cdot\varepsilon.
\]

\end{itemize}
\end{definition}
The starting point for our construction of a dPCP is the fact that
NP has PCPPs with reasonable parameters:
\begin{theorem}
[\cite{BGHSV06,DR06}]\label{thm:PCPP}$\csat_{\B}$ has a PCPP
system with randomness complexity $O(\log n)$, query complexity $O(1)$,
proof length $\poly(n)$, decision complexity $O(1)$, and rejection
ratio $\Omega(1)$.\end{theorem}
\begin{remark}
\label{rem:strong-vs-weak-PCPP}The PCPPs described in Definition~\ref{def:PCPP}
are known in the literature as ``strong PCPPs''. Here, the term
``strong'' means that the rejection probability is linearly related
to to the distance $\varepsilon$ of $x$ from a satisfying assignment.
In particular, this implies that even if $\varepsilon$ is small (but
non-zero), then the PCPP rejects with non-zero probability.

An alternative definition of PCPPs, known as ``weak PCPPs'', requires
only that every assignment $x\in\B^{t}$ that is very far from a satisfying
assignment will be rejected with high probability, while $x$'s that
are close to a satisfying assignment may be accepted with probability~$1$.
\end{remark}

\subsection{\label{sub:dPCP}The definition of decodable PCPs}

Decodable PCPs (dPCPs) were defined in the work of~\cite{DH09} in
order to overcome certain limitations of PCPPs%
\footnote{In particular, using arguments in the spirit of \cite{BHLM09}, it
is easy to prove that a PCPP that has low soundness error must make
at least three queries. Hence, PCPPs can not be used to construct
two-query PCPs with low soundness error.%
}. As mentioned above, the definition of~\cite{DH09} is only useful
if the soundness error is indeed very low. Below, we recall the definition
of~\cite{DH09} and suggest an alternative definition for the case
where the soundness error is high. This alternative definition will
be useful later in the construction of decodable PCPs with low soundness
error.

\subsubsection{Recalling the definition of \cite{DH09}}

Intuitively, a PCP decoder for $\csat$ is an oracle machine $D$
that is given as input a circuit $\varphi$, and is also given oracle
access to a ``proof'' $\pi$ that is supposed to be the encoding
of some \emph{satisfying} assignment $x$ to~$\varphi$. The PCP
decoder $D$ is required to decode a uniformly distributed coordinate
$k$ of the assignment $x$ by making only few queries to~$\pi$.
It could also be the case that the proof $\pi$ is too corrupted for
the decoding to be possible, in which case $D$ is allowed to output
a special failure symbol $\bot$. Thus, we say that $D$ has made
an error only if it outputs a symbol other than $x_{k}$ and $\bot$.
We refer to the probability of the latter event as the ``decoding
error of $D$'', and would like it to be minimal. We do note, however,
that if~$\pi$ is not corrupted, then~$D$ is not allowed to output~$\bot$.

It turns out that if we wish the decoding error of $D$ to be very
small, we need to relax the foregoing definition, and allow the PCP
decoder $D$ to perform ``list decoding''. That is, instead of requiring
that there would be a single assignment $x$ that is decoded by $D$,
we only require that there exists a \emph{short} list of assignments
$x^{1},\ldots,x^{L}$ such that the decoder outputs either $\bot$
or one of the symbols $x_{k}^{1},\ldots,x_{k}^{L}$ with very high
probability. Of course, this is meaningless if the assignments are
binary strings, and therefore we extend the definition of $\csat$
to circuits whose inputs are symbols from some large alphabet $\Gamma$.

We turn to give the formal definitions of (list-)decodable PCPs. As
in the case of PCPPs, instead of letting the decoder make the queries
and process the answers directly, we require the decoder to output
the queries and a circuit~$\psi$ that given the answers to the queries
outputs the decoded value.
\begin{notation}
Let $\Sigma$ and $\Gamma$ be finite alphabets, and let $f:\Gamma^{k}\to\Sigma^{n}$
be a function. We say that a circuit $C$ computes $f$ if it takes
as input a binary string of length $k\cdot\left\lceil \log\left|\Gamma\right|\right\rceil $
and outputs a binary string of length $n\cdot\left\lceil \log\left|\Sigma\right|\right\rceil $
that represent the input in $\Gamma^{k}$ and the output in $\Gamma^{n}$
in the natural way. We will usually omit the function $f$ and simply
refer to the circuit $C:\Gamma^{k}\to\Sigma^{n}$. We will also view
the circuit $C$ as taking as input $k$ symbols in $\Gamma$ and
outputs $n$ symbols in~$\Sigma$. Given a circuit $\varphi:\Gamma^{t}\to\B$,
an assignment $x\in\Gamma^{t}$ for $\varphi$ is said to \emph{satisfy}
$\varphi$ if $\varphi(x)$, and otherwise it is said to be \emph{unsatisfying}.\end{notation}
\begin{definition}
[{\label{def:PCP-decoders}PCP decoders, similar to~\cite[Definition 3.1]{DH09}}]Let
$r,q,s,\ell:\N\to\N$, and let $\Gamma$, $\Sigma$ be functions that
map each $n\in\N$ to some finite alphabet. A \emph{PCP decoder} for
$\csat_{\iab}$ over \emph{proof alphabet~}$\oab$ is a probabilistic
polynomial-time algorithm $D$ that for every $n\in\N$ acts as follows.
Let $\Gamma=\Gamma(n)$, $\Sigma=\Sigma(n)$, $\ell=\ell(n)$. When
given as input an input circuit $\varphi:\iab^{t}\to\B$ of size $n$
and an index $k\in\left[t\right]$, the PCP decoder~$D$ tosses $r(n)$
coins and generates 
\begin{enumerate}
\item A sequence of queries $I=\left(i_{1},\ldots,i_{q(n)}\right)$ in $\left[\ell\right]$
(where the queries are viewed as coordinates of a proof string in
$\Gamma^{\ell}$). 
\item A circuit $\psi:\oab^{q(n)}\to\Gamma\cup\left\{ \bot\right\} $ of
size at most $s(n)$. 
\end{enumerate}

We shall refer to the functions $r(n)$, $q(n)$, $\ell(n)$, and
$s(n)$ as the \emph{randomness complexity}, \emph{query complexity},
\emph{proof length}, and \emph{decoding complexity} respectively.
Without loss of generality we have $\ell\left(n\right)=2^{r(n)}\cdot q(n)\cdot t$.

\end{definition}

\begin{definition}
[{\label{def:List-Decodable-PCPs}List Decodable PCPs, similar to
\cite[Definition 3.2]{DH09}}]Let $D$, $\iab$, $\oab$, and $\ell$
be as in Definition~\ref{def:PCP-decoders}, and $L:\N\to\N$ and
$\varepsilon:\N\to\left[0,1\right]$. We say that a PCP decoder $D$
with the foregoing parameters is a \emph{(list) decodable PCP} \emph{system}
for $\csat_{\iab}$ (abbreviated ldPCP) with\textsf{ }\emph{list size}~$L=L(n)$,\emph{
soundness error }$\varepsilon=\varepsilon(n)$ if the following holds
for every circuit $\varphi:\iab^{t}\to\B$ of size $n$: 
\begin{itemize}
\item \textbf{Completeness:} For every $x\in\iab^{t}$ such that $\varphi(x)=1$
there exists a proof string $\pi_{x}\in\oab^{\ell}$ such that
\[
\Pr_{k;I,\psi}\left[\psi\left(\pi_{x|I}\right)=x_{k}\right]=1,
\]
 where $k$ is uniformly distributed in $\left[t\right]$ and $I$
and $\psi$ are the (random) output of $D\left(\varphi,k\right)$. 
\item \textbf{Soundness:} For every proof string $\pi\in\oab^{\ell}$, there
exist a (possibly empty) list of satisfying assignments $x^{1},\ldots,x^{L}\in\Gamma^{t}$
for $\varphi$ such that
\[
\Pr_{k;I,\psi}\left[\psi\left(\pi_{|I}\right)\notin\left\{ x_{k}^{1},\ldots,x_{k}^{L},\bot\right\} \right]\le\varepsilon,
\]
where $k$, $I$, $\psi$ are as before.
\end{itemize}
\end{definition}

\subsubsection{Uniquely-decodable PCPs}

We turn to discuss our suggested definition for dPCPs for the case
of high soundness error. If the soundness error is high, then we can
actually require the PCP decoder to decode a unique assignment, instead
of decoding a list of assignments. Thus, we refer to dPCPs with high
soundness error as ``uniquely decodable PCPs'' (udPCPs).

The straightforward definition for udPCPs would be to take the foregoing
definition of ldPCPs, and set~$\varepsilon$ to be large and~$L$
to be~$1$. However, this definition turns out to be useless for
our purposes. To see why, recall that our ultimate goal is to construct
dPCPs with \emph{low} \emph{error} by first constructing dPCPs with
\emph{high error} and then decreasing their error using derandomized
parallel repetition. However, if we define udPCPs using the above
straightforward definition, then it is not even clear that \emph{sequential
repetition} decreases their error%
\footnote{The problem in performing sequential repetition for such definition
of udPCPs is that we must invoke the PCP decoder on a \emph{uniformly
distributed and independent index $k$} in each invocation, and it
is not clear how to use invocations for different indices~$k$ in
order to decrease the error.%
}.

We therefore use the following alternative definition for udPCP. We
now require that if the proof $\pi$ is such that the PCP decoder
$D$ errs with high probability, then $D$ detects that there is an
error with at least proportional probability. In other words, we require
that the probability that~$D$ outputs $\bot$ is related to the
probability that~$D$ errs. Observe that such PCP decoders can indeed
be improved by sequential repetition: If the proof $\pi$ is erroneous
and we invoke the PCP decoder $D$ many times, then the probability
that $D$ detects the error and outputs $\bot$ improves. Below we
give the formal definition.
\begin{definition}
\label{def:decoding-error}Let $D$, $\iab$, $\oab$, and $\ell$
be as in Definition~\ref{def:PCP-decoders}. Let $\varphi:\Gamma^{t}\to\B$
be a circuit of size $n$, let $x$ be an assignment to~$\varphi$,
and let $\pi\in\Sigma^{\ell(n)}$ be a proof for $D$. We define the
\emph{decoding error of $D$ on $\pi$ with respect to }$x$ as the
probability
\[
\Pr_{k;I,\psi}\left[\psi\left(\pi_{|I}\right)\notin\left\{ x_{k},\bot\right\} \right],
\]
where $k$, $I$, $\psi$ are as in Definition~\ref{def:List-Decodable-PCPs}.
We define the \emph{decoding error}\textsf{ }\emph{of $D$ on}\textsf{
$\pi$} as the minimal decoding error of $D$ on $\pi$ with respect
to an assignment~$x'$ for $\varphi$, over all possible assignments~$x'$
to~$\varphi$.
\end{definition}

\begin{definition}
[\label{def:Unique-Decodable-PCPs}Uniquely Decodable PCPs]Let $D$,
$\iab$, $\oab$, and $\ell$ be as in Definition~\ref{def:PCP-decoders},
and let $\rho:\N\to\left[0,1\right]$. We say that the PCP decoder
$D$ is a \emph{(uniquely) decodable PCP system} for $\csat_{\iab}$
(abbreviated udPCP) with\textsf{ }\emph{rejection ratio}\textsf{ }$\rho$
if for every circuit $\varphi:\iab^{t}\to\B$ of size $n$ the PCP
decoder $D$ satisfies the completeness requirement of Definition~\ref{def:List-Decodable-PCPs},
and furthermore satisfies the following requirement: 
\begin{itemize}
\item \textbf{Soundness:} For every proof string $\pi\in\oab^{\ell}$, if
$D$ has decoding error $\varepsilon$ on $\pi$ then 
\[
\Pr_{k;I,\psi}\left[\psi\left(\pi_{|I}\right)=\bot\right]\ge\rho(n)\cdot\varepsilon,
\]
where $k$, $I$, $\psi$ are as in Definition~\ref{def:List-Decodable-PCPs}.
\end{itemize}
\end{definition}
\begin{remark}
We could have also defined the decoding error of $D$ on $\pi$ with
respect to $x$ as the probability $\Pr_{k;I,\psi}\left[\psi\left(\pi_{|I}\right)\ne x_{k}\right]$.
This definition may be more natural, but it is more convenient to
work with the current definition.
\end{remark}

\begin{remark}
\label{rem:strong-vs-weak-udPCP}Note that the soundness requirement
in our definition of udPCPs is similar to the soundness requirement
of PCPPs, and in particular to definition of soundness of \emph{strong}
PCPPs (see Remark~\ref{rem:strong-vs-weak-PCPP}). We could also
use a definition that is analogous to the definition of a \emph{weak}
PCPP. Specifically, we could have required only that when the decoding
error is very large, the decoder rejects with high probability. However,
our definition is stronger, and since we can satisfy it, we prefer
to work with it. It is also more convenient to work with this definition
throughout this work.
\end{remark}
We next argue that every PCPP implies a udPCP.
\begin{proposition}
\label{pro:PCPP-implies-udPCP}Let $V$ be a PCPP system for $\csat_{\B}$
with randomness complexity $r(n)$, query complexity $q(n)$, proof
length $\ell(n)$, decision complexity $s(n)$, and rejection ratio
$\rho(n)$. Then, for every $u:\N\to\N$ there exists a udPCP for
$\csat_{\B^{u(n)}}$ with proof alphabet $\B$, randomness complexity
$r(n)$, query complexity $q(n)+u(n)$, proof length $n+\ell(n)$,
decoding complexity $s(n)+O\left(u(n)\right)$, and rejection ratio
$\rho(n)/u(n)$.\end{proposition}
\begin{myproof}
Let $u:\N\to\N$ and denote $u=u(n)$. For every circuit $\varphi:\left(\B^{u}\right)^{t}\to\B$
of size $n$ and satisfying assignment $x$ for $\varphi$, we define
the corresponding proof string for $D$ to be $x\circ\pi_{x}$, where
$\pi_{x}$ is the proof string of $V$ for $x$ when $x$ is treated
as a binary string.

Fix a circuit $\varphi:\left(\B^{u}\right)^{t}\to\B$ and $k\in\left[t\right]$,
and let $x'\in\B^{u\cdot t}$, $\pi\in\B^{\ell}$. On input $\left(\varphi,k\right)$
and oracle access to a proof $x'\circ\pi$, the decoder $D$ first
emulates the verifier $V$ on $\varphi$ with oracle access to $x'\circ\pi_{x}$.
If $V$ rejects, then~$D$ outputs~$\bot$. Otherwise, $D$ queries
the coordinates
\[
u\cdot\left(k-1\right)+1,\ldots,u\cdot k
\]
of~$x$ and outputs the tuple of answers as the symbol in $\B^{u}$
that it is ought to decode.

It should be clear that $D$ satisfies the completeness requirement,
and has the correct randomness complexity, query complexity, proof
length, and decoding complexity.

It remains to analyze the rejection ratio of $D$. Let $\pi'$ be
a proof string for $D$ and assume that $\pi'=x\circ\pi$ where $x\in\B^{u\cdot t}$
and $\pi\in\B^{\ell}$. Let $x_{0}$ be the satisfying assignment
of $\varphi$ that is nearest to $x$ when viewed as a binary string.
Let $\varepsilon$ be the relative distance between $x$ and $x_{0}$
when viewed as strings over the alphabet $\B^{u}$. Clearly, the decoding
error of $D$ on $x\circ\pi$ with respect to $x_{0}$ is $\varepsilon$,
and is an upper bound on the decoding error of $D$. Furthermore,
the relative distance between $x$ and $x_{0}$ as binary strings
is at least $\varepsilon/u$. Thus, the emulation of $V$ rejects
$x\circ\pi$ with probability at least $\rho(n)\cdot\varepsilon/u$,
and this is also the rejection probability of $D$, as required.\end{myproof}
\begin{remark}
One could also prove Proposition~\ref{pro:PCPP-implies-udPCP} without
a loss of a factor of $u$ in the rejection ratio $\rho$ using error
correcting codes.
\end{remark}

\begin{remark}
It is not hard to see that the converse of Proposition~\ref{pro:PCPP-implies-udPCP}
also holds. Namely, given a udPCP it is easy to construct from it
a PCPP. Roughly, given a udPCP~$D$, construct a PCPP verifier that
when given oracle access to~$x\circ\pi$, invokes $D$ with oracle
access to~$\pi$ on a uniformly distributed~$k$, and verifies that
the output of~$D$ equals~$x_{k}$.
\end{remark}

\begin{remark}
Our definition of udPCPs (Definition~\ref{def:Unique-Decodable-PCPs})
bears some similarities to the notion of relaxed locally decodable
codes~\cite{BGHSV06}, which are also constructed using PCPPs. However,
the notions are fundamentally different. The most important difference
between the notions is that while the decoder of a relaxed LDC should
decode any possible message, the decoder of a udPCP is required to
decode only satisfying assignments of a given circuit. This makes
udPCPs significantly more powerful, and in fact makes them equivalent
to PCPPs. A secondary difference is that when a udPCP is given oracle
access to a corrupted oracle then it can output $\bot$ with any probability,
while a relaxed LDC is required to output $x_{k}$ (instead of $\bot$)
with some given probability.
\end{remark}

\subsection{Decoding graphs}

\subsubsection{The definition of decoding graphs}

Recall that in the first part of the paper, we often found it more
convenient to work with constraint graphs instead of working with
PCPs. We now define the notion of ``decoding graphs'', which will
serve as the graph analogue of decoding PCPs just as constraint graphs
serve as the graph analogue of PCPs.
\begin{definition}
[\label{def:decoding-graphs}Decoding graphs]A \emph{(directed)
decoding graph} is a directed graph $G=(V,E)$ that is augmented with
the following objects: 
\begin{enumerate}
\item A circuit $\varphi:\Gamma^{t}\to\B$, to which we refer as the \emph{input
circuit}. Here $\Gamma$ denotes some finite alphabet.
\item A finite alphabet $\Sigma$, to which we refer as the \emph{alphabet
of $G$}. 
\item For each edge $e\in E$, an index $k_{e}\in\left[t\right]$, and a
circuit $\psi_{e}:\Sigma\times\Sigma\to\Gamma\cup\left\{ \bot\right\} $.
We say that $e$ is \emph{associated with} $k_{e}$ and $\psi_{e}$.
For $k\in\left[t\right]$, we denote by $E_{k}$ the set of edges
associated with $k$. 
\end{enumerate}
The \emph{size} of $G$ is the number of edges of $G$. We say that
$G$ has \emph{decoding complexity} $s$ if all the circuits are of
size at most~$s$. It is required that $G$ satisfies the following
property:
\begin{itemize}
\item \textbf{Completeness:} For every satisfying assignment $x\in\Gamma^{t}$
to $\varphi$, there exists an assignment $\pi_{x}:V\to\Sigma$ to
$G$ such that the following holds. For every edge $\left(u,v\right)$
that is associated with an index $k=k_{\left(u,v\right)}$ and a circuit
$\psi=\psi_{\left(u,v\right)}$, it holds that $\psi\left(\pi(u),\pi(v)\right)=x_{k}$.
\end{itemize}
\end{definition}
\begin{notation}
\label{not:decoding-graphs}We will use the following terminology
regarding constraint graphs: Let $G=\left(V,E\right)$ be a decoding
graph with input circuit $\varphi:\Gamma^{t}\to\B$ alphabet $\Sigma$.
\begin{enumerate}
\item Let $\left(u,v\right)\in E$ and $\psi=\psi_{\left(u,v\right)}$ be
and edge its associated circuit, and let $\pi:V\to\Sigma$ be an assignment
to $G$. If $\psi$ outputs $\bot$ on input $\left(\pi(u),\pi(v)\right)$
then we say that $\left(u,v\right)$ \emph{rejects $\pi$} (or that
$\pi$ violates $\left(u,v\right)$), and otherwise we say that $\left(u,v\right)$
\emph{accepts }$\pi$ (or that $\pi$ satisfies $\left(u,v\right)$).
\item Let $\left(u,v\right)$, $\psi$, and $\pi$ be as before, let $k=k_{\left(u,v\right)}$
be the index associated with $\left(u,v\right)$, and let $x$ be
an assignment to $\varphi$. We say that $\left(u,v\right)$ \emph{fails
to decode~}$x$ if $\psi\left(\pi(u),\pi(v)\right)\notin\left\{ x_{k},\bot\right\} $.
When $x$ is clear from the context we will omit it, and we will also
say that $\left(u,v\right)$ \emph{errs}, or that $\left(u,v\right)$
\emph{decodes correctly} (if $\left(u,v\right)$ does not err). Note
that outputting\emph{ }$\bot$ is \emph{not considered to be failure}.
\item \label{enu:decoding-projection-property}We say that $G$ has the
\emph{projection property} if for every circuit $\psi_{\left(u,v\right)}$
has an associated function $f_{\left(u,v\right)}:\Sigma\to\Sigma$
such that $\psi_{\left(u,v\right)}\left(a,b\right)\ne\bot$ if and
only if $f_{\left(u,v\right)}(a)=b$.
\item We refer to the quantity $\log\left(\max_{k\in\left[t\right]}\left|E_{k}\right|\right)$
as the \emph{randomness complexity} of $G$, since it upper bounds
the number of bits required to choose a uniformly distributed edge
that is associated with a particular index.
\end{enumerate}
\end{notation}
We turn to define soundness properties of decoding graphs. As in the
case of decodable PCPs, we have two definitions, one for the case
of high soundness error (unique decoding) and one for the case of
low soundness error (list decoding).
\begin{definition}
\label{def:decoding-graphs-soundness}Let $G=\left(V,E\right)$, $\Sigma$,
$\Gamma$, $\varphi$ be as before, and let $\pi:V\to\Sigma$ be an
assignment to~$G$. 
\begin{itemize}
\item \textbf{Unique decoding soundness:} For every satisfying assignment
$x\in\Gamma^{t}$ to $\varphi$, we define the \emph{decoding error
of $G$ on $\pi$ with respect to $x$} as the probability
\[
\Pr_{k\in\left[t\right],\left(u,v\right)\in E_{k}}\left[\psi_{\left(u,v\right)}\left(\pi\left(u\right),\pi\left(v\right)\right)\notin\left\{ x_{k},\bot\right\} \right],
\]
where $k$ is uniformly distributed in $\left[t\right]$ and $\left(u,v\right)$
is uniformly distributed in $E_{k}$. Note that the edge $\left(u,v\right)$
is chosen according to the decoding distribution of $G$.\\
We define the \emph{decoding error of $G$ on $\pi$} as the minimal
decoding error of $G$ on $\pi$ with respect to any satisfying assignment
of~$\varphi$. Now, we say that $G$ has rejection ratio $\rho$
if for every assignment $\pi$ to $G$, if $G$ has decoding error
$\varepsilon$ on $\pi$ then it holds that
\[
\Pr_{k\in\left[t\right],\left(u,v\right)\in E_{k}}\left[\psi_{\left(u,v\right)}\left(\pi\left(u\right),\pi\left(v\right)\right)=\bot\right]\ge\rho\cdot\varepsilon,
\]
 where $k$ and $\left(u,v\right)$ are chosen as before. 
\item \textbf{List decoding soundness:} We say that $G$ is \emph{list-decoding}
with \emph{list size} $L$ and \emph{soundness error}~$\varepsilon$
if for every assignment $\pi$ to $G$ there exists a (possibly empty)
list of satisfying assignments $x^{1},\ldots,x^{L}\in\Gamma^{k}$
for $\varphi$ such that
\[
\Pr_{k\in\left[t\right],\left(u,v\right)\in E_{k}}\left[\psi_{\left(u,v\right)}\left(\pi\left(u\right),\pi\left(v\right)\right)\notin\left\{ x_{k}^{1},\ldots,x_{k}^{L},\bot\right\} \right]\le\varepsilon,
\]
where $k$ and $\left(u,v\right)$ are chosen as before
\end{itemize}
\end{definition}
The following proposition gives the correspondence between decoding
PCPs and decoding graphs, in analogy to the correspondence between
PCPs and constraint graphs.
\begin{proposition}
\label{pro:decoding-correspondence}Let $r,s,\ell,\rho,\Gamma,\Sigma$
be as in Definition~\ref{def:Unique-Decodable-PCPs}. The following
two statements are equivalent: 
\begin{itemize}
\item $\csat_{\Gamma}$ has a udPCP with query complexity~$2$, randomness
complexity $r$, decoding complexity $s$, proof length $\ell$, proof
alphabet $\Sigma$, and rejection ratio $\rho$. 
\item There exists a polynomial-time transformation that transforms a circuit
$\varphi:\Gamma^{t}\to\B$ of size $n$ to a decoding graph $G=\left(V,E\right)$
with $\ell(n)$ vertices, randomness complexity $r(n)$, decoding
complexity $s(n)$, proof alphabet $\Sigma\left(n\right)$, and rejection
ratio $\rho(n)$. 
\end{itemize}
A similar equivalence holds for ldPCPs and list-decoding graphs.
\end{proposition}

\subsubsection{Additional properties of decoding graphs}

Recall that when discussing constraint graphs, we were interested
in the probability that a uniformly distributed edge of the graph
is satisfied by a given assignment. As can be seen in Definition~\ref{def:decoding-graphs-soundness},
when discussing decoding graphs we are interested in a different distribution
over the edges, defined below.
\begin{definition}
The \emph{decoding distribution} $\mathcal{D}_{G}$ of a decoding
graph $G=\left(V,E\right)$ is the distribution over the edges of
$G$ that is corresponds to the following way for picking a random
edge of $G$: Choose $k\in\left[t\right]$ uniformly at random, and
then choose an edge uniformly at random from $E_{k}$.
\end{definition}
It is usually inconvenient to analyze the decoding distribution of
the graphs we work with. However, we will work only with graphs whose
decoding distribution is similar to the uniform distribution over
the edges (where similarity is defined as in Section~\ref{sub:similarity-of-distributions}).
The following definition aims to capture this property, which allows
us to analyze the uniform distribution instead of the decoding distribution.
\begin{definition}
We say that a decoding graph $G=\left(V,E\right)$ has \emph{smoothness}
$\smooth$ if its decoding distribution is $\smooth$-similar to the
uniform distribution over $E$. 
\end{definition}
The following proposition gives a comfortable way for calculating
the smoothness of a decoding graph. Intuitively, observe that if all
the sets $E_{k}$ are of the same size then the decoding distribution
is identical to the uniform distribution. We now observe that if the
sizes of the sets $E_{k}$ are close to each other then the decoding
distribution is similar to the uniform distribution.
\begin{proposition}
[\label{pro:smoothness-criterion}Smoothness criterion]A decoding
graph $G$ with edge-set $E$ has smoothness $\smooth$ if and only
if for every $k\in\left[t\right]$, the number of edges that are associated
with $k$ is between $\smooth\cdot\frac{\left|E\right|}{t}$ and $\frac{1}{\smooth}\cdot\frac{\left|E\right|}{t}$.\end{proposition}
\begin{myproof}
Observe that if there are $m_{k}$ edges associated with $k\in\left[t\right]$
then the probability for such an edge to be chosen under the decoding
distribution is $\frac{1}{t}\cdot\frac{1}{m_{k}}$ while the corresponding
probability under the uniform distribution is $\frac{1}{\left|E\right|}$.
Now apply the definition of similarity of distributions.
\end{myproof}
We will often want our decoding graphs to be regular, or at least
have bounded degree. The precise definition follows.
\begin{definition}
We say that a decoding graph $G$ has \emph{degree bound} $d\in\N$
if all the in-degrees and all out-degrees of the vertices in $G$
are bounded by $d$. We say that it is \textsf{$d$}\emph{-regular}
if every vertex has \emph{exactly} $d$ incoming edges and \emph{exactly}
$d$ outgoing edges.
\end{definition}

\subsubsection{General udPCPs and decoding graphs}

Proposition~\ref{pro:decoding-correspondence} gave us only a correspondence
between decoding graphs and udPCPs that makes exactly two queries.
The next proposition shows that in fact any udPCP, even if it uses
more than two queries, gives rise to a procedure that transforms circuits
to decoding graphs with related parameters and unique decoding soundness.
A nice property of this procedure is that it generates decoding graphs
that are regular and have smoothness~$1$, which will be useful later
in this work.
\begin{proposition}
\label{pro:any-PCP-decoder-implies-decoding-graphs}Let $\iab$, $\oab$,
$r(n)$, $q(n)$, $\ell(n)$, $s(n)$, and $\rho(n)$ be as in Definition~\ref{def:Unique-Decodable-PCPs},
and let $h_{0}$ and $d_{0}$ be the constants from Fact~\ref{fac:expanders-exist}.
If there exists a udPCP $D$ for $\csat_{\iab}$ with the foregoing
parameters, then there exists a polynomial time procedure that acts
as follows. When given a circuit $\varphi:\Gamma^{t}\to\B$ of size
$n$, the procedure outputs a corresponding vertex-decoding graph
$G=\left(V,E\right)$ with randomness complexity $r(n)+\log\left(d_{0}\cdot q(n)\right)$,
alphabet $\oab^{q(n)}$, decoding complexity $s(n)+\poly\log\left|\Sigma(n)\right|$,
and rejection ratio $\Omega\left(\rho(n)/\left(q(n)\right)^{2}\right)$.
Furthermore, $G$ is~$\left(q(n)\cdot d_{0}\right)$-regular, and
has $t\cdot2^{r(n)}$ vertices and smoothness~$1$.\end{proposition}
\begin{myproof}
[Proof sketch]The proof is a variant of a well known technique for
reducing the query complexity of a PCP verifier to $2$, and its full
details are provided in Appendix~\ref{sec:any-PCP-decoder-implies-decoding-graphs}.
The graph $G$ is constructed roughly as follows: The graph $G$ has
a vertex for every possible invocation of the decoder $D$. Each such
vertex $v$ is expected to be labeled with the answers that $D$ receives
to its queries on the corresponding invocation, and the edges that
are connected to $v$ check that those answers are not rejected by
$D$. The edges of $G$ also verify that the labels of the different
vertices are consistent with each other, and in order to save in the
number of edges we choose the consistency checks according to an expander.

Observe that since a vertex should be labeled with all the answers
that $D$ gets to its queries on this particular invocation, we can
use those labels to perform decoding. In particular, given that an
edge $\left(u,v\right)$ accepts, the value that it decodes can be
decided based only on the label of $u$. This property will be useful
in Section~\ref{sec:dPCP-embedding-linear-structure} (see Definition~\ref{def:vertex-decoding-graphs}
for details).
\end{myproof}

\subsection{\label{sub:dPCP-construction}Our construction of dPCPs, Theorem~\ref{thm:main-dPCP}}

In this section we state and prove Theorem~\ref{thm:main-dPCP}.
\begin{theorem*}
[\ref{thm:main-dPCP}, dPCP, restated formally]For every function
$\Gamma$ that maps natural numbers to finite alphabets such that
$\left|\Gamma(n)\right|\le2^{\poly\log n}$ the following holds. There
exists an ldPCP $D$ for $\csat_{\Gamma}$with query complexity $2$,
proof alphabet $2^{\poly\log n}$, randomness complexity $O(\log n)$,
soundness error $1/\log^{\Omega(1)}n$, and list size $\poly\log n$.
Furthermore, $D$ has the projection property (see Notation~\ref{not:decoding-graphs},
Item~\ref{enu:decoding-projection-property}).
\end{theorem*}
We prove this theorem analogously to the proof of Theorem~\ref{thm:main},
which asserts the existence of two-query PCPs with soundness error~$1/\poly\log n$.
Our starting point is a known construction of a PCPP, stated here
as Theorem~\ref{thm:PCPP} which is then reduced to a transformation
mapping circuits to decoding graphs. We then have two main steps.
The first is to equip the decoding graphs with linear structure, as
formulated in Lemma~\ref{lem:decoding-linear-structure}. The second
step is to reduce the error by derandomized parallel repetition, as
stated in Lemma~\ref{lem:decoding-derand-parallel-repetition}. Theorem~\ref{thm:main-dPCP}
follows by combining the two lemmas which we state next,
\begin{lemma}
[\label{lem:decoding-linear-structure}Linear Structure Embedding
for udPCPs] There exists a polynomial time procedure that satisfies
the following requirements: 
\begin{itemize}
\item \textbf{Input:}

\begin{itemize}
\item A decoding graph $G$ of size $n$ for input circuit $\varphi:\Gamma^{t}\to\B$
with alphabet $\Sigma$, rejection ratio $\rho$, decoding complexity
$s$, and smoothness $\smooth$. 
\item A finite field $\F$ of size $q$ such that $q\ge4\cdot d_{0}^{2}$,
where $d_{0}$ is the constant from Fact~\ref{fac:expanders-exist}. 
\end{itemize}
\item \textbf{Output:} A decoding graph $G'=\left(\F^{m},E'\right)$ for
$\varphi$ such that the following holds:

\begin{itemize}
\item $G'$ has a linear structure. 
\item The size of $G'$ is at most $O\left(q\cdot n/\gamma\right)$. 
\item $G'$ has alphabet $\Sigma^{O(\log_{q}(n/\gamma))}$. 
\item $G'$ has rejection ratio $\Omega\left(\rho/q^{2}\cdot\log_{q}(n/\gamma)\right)$ 
\item $G'$ has decision complexity $s+\poly\left(\log_{q}\left(n/\gamma\right),\log\left|\Gamma\right|\right)$ 
\item $G'$ has smoothness $\Omega\left(1/q\right)$. 
\end{itemize}
\end{itemize}
\end{lemma}

\begin{lemma}
[\label{lem:decoding-derand-parallel-repetition}Derandomized Parallel
Repetition for dPCPs]There exist a universal constant $h$ and a
polynomial time procedure that satisfy the following requirements: 
\begin{itemize}
\item \textbf{Input:}

\begin{itemize}
\item A finite field $\F$ of size $q$.
\item A decoding graph $G=\left(\F^{m},E\right)$ of size $n$ for input
circuit $\varphi:\Gamma^{t}\to\B$ with linear structure, alphabet
$\Sigma$, rejection ratio $\rho$, decision complexity $s$, and
smoothness $\smooth$.
\item The rejection ratio $\rho$ of $G$.
\item A parameter $d_{0}\in\N$ such that $d_{0}<m/h^{2}$ and $\rho\ge h\cdot d_{0}\cdot q^{-d_{0}/h}/\gamma$.
\end{itemize}
\item \textbf{Output:} A decoding graph $G'$ for $\varphi$ such that the
following holds:

\begin{itemize}
\item $G'$ has size $n^{O\left(d_{0}\right)}$. 
\item $G'$ has alphabet $\Sigma^{q^{O(d_{0})}}$. 
\item $G'$ is list-decoding with soundness error $\varepsilon\eqdef h\cdot d_{0}\cdot q^{-d_{0}/h}/\gamma$
and list size $L\eqdef q^{O(d_{0})}$.
\item $G'$ has the projection property.
\item $G'$ has decoding complexity $q^{O(d_{0})}\cdot\left(s+\poly\log\left|\Sigma\right|\right)$.
\end{itemize}
\end{itemize}
\end{lemma}
We now turn to prove Theorem~\ref{thm:main-dPCP}.
\begin{myproof}
Let $V$ be a PCPP verifier for $\csat$ as in Theorem~\ref{thm:PCPP}.
By Proposition~\ref{pro:PCPP-implies-udPCP} this implies a udPCP
for $\csat$ with similar parameters. Next, by Proposition~\ref{pro:any-PCP-decoder-implies-decoding-graphs}
we get a polynomial time transformation taking a circuit $\varphi:\B^{n}\to\B$
into a vertex-decoding graph. The graph $G$ has the following parameters.
The randomness complexity is $r(n)=O(\log n)$, the decoding complexity,
rejection ratio, and constant proof alphabet are constant, and the
smoothness is $1$.

We choose $\F$ to be the smallest finite field of size at least $\log n$,
and set $\F$ to be the finite field of size $q$. We now invoke Lemma~\ref{lem:decoding-linear-structure}
(linear structure embedding for udPCPs) on input $G$ and $\F$, and
obtain a new vertex-decoding graph $G_{1}$ with linear structure
and parameters: 
\begin{itemize}
\item The size of $G_{1}$ is at most $O(q\cdot n)$. 
\item $G_{1}$ has alphabet size $2^{O(\log_{q}(n))}$. 
\item $G_{1}$ has rejection ratio $\rho_{1}\eqdef\Omega\left(\rho/q^{2}\cdot\log_{q}(n)\right)$ 
\item $G_{1}$ has decision complexity $\poly(\log_{q}n)$ 
\item $G_{1}$ has smoothness $\gamma_{1}=\Omega\left(\frac{1}{q}\right)$. 
\end{itemize}
Finally, we set $d_{0}$ to be an arbitrary constant such that $\rho_{1}\ge h\cdot d_{0}\cdot q^{-d_{0}/h}/\gamma_{1}$
. Note that this is indeed possible, since $\log_{q}\left(1/\rho_{1}\right)$
is a constant that depends only on $\rho$. Finally, we invoke Lemma~\ref{lem:decoding-derand-parallel-repetition}
(derandomized parallel repetition for dPCPs) on input $G_{1}$, $\F$,
$\rho_{1}$, and $d_{0}$, and denote by $G'$ the output decoding
graph. The transformation taking the initial input $\varphi$ into
$G'$ (via intermediate steps $G$ and $G_{1}$) is equivalent, by
Proposition~\ref{pro:decoding-correspondence}, to a dPCP with the
claimed parameters.
\end{myproof}

\subsection{\label{sub:MR}Proof of the result of \cite{MR08}, Theorem~\ref{thm:MR}}

Our Theorem~\ref{thm:main} asserts the existence of a two query
PCP with soundness error $1/\poly\log n$ and alphabet size $\card{\Sigma}=2^{\poly\log n}$.
In this section we will sketch a proof of Theorem~\ref{thm:MR} in
which the alphabet size $\card{\Sigma}$ can be any value smaller
than $2^{\poly\log n}$ while maintaining the relation of $\eps\le1/\poly(\log\card{\Sigma})$.
\begin{theorem*}
[\textbf{\ref{thm:MR}}, restated,~\cite{MR08}]For any function
$\varepsilon(n)\ge1/\poly\log n$ the class $\NP$ has a two-query
PCP verifier with perfect completeness, soundness error at most $\varepsilon$
over alphabet $\Sigma$ of size at most $\card{\Sigma}\le2^{1/\poly(\varepsilon)}$.
\end{theorem*}
\noindent Our proof of Theorem~\ref{thm:MR} relies on the scheme
of \cite{DH09} who showed a generic way to compose a PCP with a dPCP,
and then proved Theorem~\ref{thm:MR} by repeating the composition
step, assuming the existence of two building blocks: a PCP and a dPCP.
We plug in our constructions of a PCP (Theorem~\ref{thm:main}) and
of a dPCP (Theorem~\ref{thm:main-dPCP}) into the composition scheme
of \cite{DH09} and obtain a new construction of the verifier of Theorem~\ref{thm:MR}
that does not rely on low degree polynomials.
\begin{remark}
\label{rem:MR-short} An important feature of the theorem of \cite{MR08}
asserts that the verifier is randomness-efficient, i.e. it uses only
$(1+o(1))\log n$ random bits rather than $O(\log n)$ random bits.
This is equivalent to constructing constraint graphs of almost-linear
size rather than polynomial size (see Remark~\ref{rem:randomness-vs-size}).
Using the composition scheme of \cite{DH09}, the outcome will be
randomness efficient as long as the PCP verifier at the outermost
level of composition is randomness-efficient. It does not, for example,
depend on whether the dPCP is randomness-efficient.

However, since our PCP verifier from Theorem~\ref{thm:main} is not
randomness-efficient, we can only get this additional feature by relying
at the outermost level on a PCP verifier as in \cite{MR08}. The dPCP
can still be based on our Theorem~\ref{thm:main-dPCP}. Alternatively,
if we also base the outermost PCP on theorem~\ref{thm:main} we get
a polynomial-size construction, but not a ``randomness-efficient''
one. It is also conceivable that the construction of Theorem~\ref{thm:main}
can be improved to yield a randomness-efficient PCP, and we leave
this for future work.
\end{remark}
In order to state the generic composition theorem of~\cite{DH09}
let us first define the \emph{decision complexity} of a PCP verifier.
Roughly speaking, a PCP verifier has decision complexity $s(n)$ if
every constraint in the underlying constraint graph can be computed
by a circuit of size at most $s(n)$%
\footnote{More precisely, the verifier should be able to compute this circuit
based on its input and its randomness.%
}. This definition is analogous to the definition of the decoding complexity
of a PCP decoder. It is easy to see that the PCP verifier (from Theorem~\ref{thm:main})
has decision complexity $\poly\log n$ in the same way that the dPCP
decoder (from Theorem~\ref{thm:main-dPCP}) was shown to have decoding
complexity $\poly\log n$.

We turn to state the composition theorem of~\cite{DH09}. As in all
composition theorems in the literature, the goal of this theorem is
to take an ``outer verifier'' (in this case, a PCP verifier), which
has a large alphabet, and reduce its alphabet size by composing it
with an ``inner verifier'' (in this case, a PCP decoder). The gain
is obtained from the fact that the inner verifier is invoked on a
claim of size $s(n)\ll n$, and thus can have a much smaller alphabet
than the outer verifier. The result of the composition is a verifier
that has the alphabet size roughly as of the inner verifier, and can
still be invoked on a claim of size~$n$. However, the composed verifier
accumulates soundness error from the invocations of both the outer
verifier and the inner verifier, and thus the composition does not
come ``for free''.
\begin{theorem}
[\label{thm:DH}Paraphrasing \cite{DH09}]Let $V$ and $D$ be a
PCP verifier and a PCP decoder as follows: 
\begin{enumerate}
\item Let $V$ be a two-query PCP verifier for $\NP$ with perfect completeness,
soundness error $\Delta(n)$, alphabet size $\card{\Sigma(n)}$, and
decision complexity $s(n)$. Assume further that the PCP verifier
makes projection queries. 
\item Let $D$ be a two-query PCP decoder for $\csat_{\Gamma}$ for some
$\Gamma(n)$. Assume $D$ has perfect completeness, soundness error
$\delta(n)$, list size $L(n)$, and alphabet size $\card{\sigma(n)}$.
\end{enumerate}
If both $V$ and $D$ have the projection property then there is a
PCP verifier $V\circledast D$ with the following properties. $V\circledast D$
invokes $D$ on inputs of length at most $s(n)$. $V\circledast D$
has perfect completeness, soundness error $O(\delta(s(n))+L(s(n))\Delta(n))$,
alphabet size $\card{\sigma(s(n))}^{\poly(L(s(n))/\delta(s(n)))}$,
and $V\circledast D$ has the projection property.
\end{theorem}
As discussed above, the main gain from this theorem is that the alphabet
size of $V\circledast D$ is much smaller than that of $V$. Let us
see how this is useful. Suppose we take $V,D$ from Theorems \ref{thm:main}
and \ref{thm:main-dPCP}. We have $\Sigma(n)\le2^{\poly\log n},s(n)=\poly\log n$,
and $\sigma(n)\le2^{\poly\log n}$. Thus, $\sigma(s(n))=2^{\poly\log\log(n)}$.
Similarly $L(s(n))\le\poly\log\log n$ and $\delta(s(n))=1/\poly\log\log n$.
This results in alphabet size of $2^{\poly\log\log(n)}$ and soundness
error of $1/\poly\log\log n$. By composing this verifier again with
$D$ (yielding $(V\circledast D)\circledast D$) one can inductively
obtain a PCP verifier with soundness error $1/\poly\log^{(i)}n$ for
any $i$ and corresponding alphabet size $\card{\Sigma}=2^{1/\poly(\epsilon)}$.
To get \emph{any} alphabet size $\card{\Sigma}$ one must do careful
padding and we do not go into these details.\medskip{}

The composition theorem (Theorem~\ref{thm:DH}) is stated here in
the two-query terminology (rather than in the terminology of ``robust''
PCPs). Let us now give a brief outline of how to obtain this version
from the version of \cite{DH09}:
\begin{enumerate}
\item \emph{From two-query to robust:} Use Lemma 2.5 of \cite{DH09} to
deduce existence of a robust PCP $rV$ and a robust dPCP $rD$ with
parameters related to $V$ and $D$. In particular, the number of
accepting views for $rD$ is bounded by $\card{\sigma}$. 
\item \emph{Composition:} Apply Theorem 4.2 of \cite{DH09} with parameter
$\eps=\delta/L\ge\card{\sigma}^{\Omega(1)}$. Deduce a new robust
PCP $rV\circledast rD$ with parameters as follows. The soundness
error is $\delta+L\Delta+4L\eps=O(\delta+L\Delta)$. The number of
accepting views is at most $\card{\sigma}^{4/\eps^{4}}$ (this follows
from inspecting the proof, but not directly from the theorem statement). 
\item \emph{Back to two queries:} Again use Lemma 2.5 to move back to a
two query PCP. The new alphabet size is at most the number of accepting
views of $rV\circledast rD$ which is at most $\card{\sigma(s(n))}^{4/\eps^{4}}=\card{\sigma}^{(L/\delta)^{O(1)}}$
as claimed.\qed
\end{enumerate}

\section{\label{sec:dPCP-embedding-linear-structure}Decoding PCPs with Linear
Structure}

In this section we prove Lemma~\ref{lem:decoding-linear-structure},
i.e., that every decoding graph $G$ can be embedded on a graph that
has linear structure. The heart of the proof is very similar to the
proof of the corresponding lemma for constraint graphs (Lemma~\ref{lem:embedding-linear-structure})
with few adaptations to the setting of decoding graphs. Two important
differences are the following:
\begin{enumerate}
\item \label{enu:udPCPs-linear-structure-not-enough-vertices}Recall that
we prove Lemma~\ref{lem:embedding-linear-structure} by embedding
the constraint graph~$G$ on a de~Bruijn graph $\DB$, and that
this is done by identifying the vertices of $G$ with the vertices
of~$\DB$. Furthermore, recall that if $\DB$ has more vertices than
$G$, then some of the vertices of $\DB$ are not identified with
vertices of $G$, and thus we place only trivial constraints on those
vertices.\\
This construction does not work for decoding graphs. The reason is
that in the setting of decoding graphs every edge needs to be able
to decode some index~$k\in\left[t\right]$. Furthermore, every edge
that fails to decode must contribute to the fraction of rejecting
edges. Thus, we can not have many trivial edges.\\
In order to resolve this issue, we prove a proposition that allows
us to ensure that $G$ has exactly the same number of vertices as
in $\DB$, see Proposition~\ref{pro:decoding-increasing-vertices-number}
below.\\
We note that Item~\ref{enu:udPCPs-linear-structure-not-enough-vertices}
is \emph{not} caused by the fact we chose a strong definition of udPCP
and not a weak one (see Remark~\ref{rem:strong-vs-weak-udPCP}).
Even if we used a weak definition of udPCP, requiring edges to reject
only if the decoding error is above some threshold, we still could
not use dummy vertices and edges in the embedding, as this would cause
the aforementioned threshold to be too large for our purposes.
\item Recall that in the embedding of constraint graphs on de~Bruijn graphs
we used the expander-replacement technique (Lemma~\ref{lem:degree-reduction})
to make sure that the graph $G$ has small degree. Since such a lemma
was not proved for decoding graphs in previous works, we have to prove
it on our own. This is done in Proposition~\ref{pro:decoding-degree-reduction}
below.
\end{enumerate}
The rest of this section is organized as follows. In Section~\ref{sub:decoding-linear-structure-auxiliary-propositions}
we prove the aforementioned Propositions~\ref{pro:decoding-degree-reduction}
and~\ref{pro:decoding-increasing-vertices-number}. Then, in Section~\ref{sub:embedding-decoding-graphs-on-de-bruijn-graphs},
we prove Lemma~\ref{lem:decoding-linear-structure}.

\subsection{\label{sub:decoding-linear-structure-auxiliary-propositions}Auxiliary
propositions}

In this section we prove Propositions~\ref{pro:decoding-degree-reduction}
and~\ref{pro:decoding-increasing-vertices-number} mentioned above.
In order to state those two propositions, we need to define a special
kind of decoding graphs, called ``vertex-decoding graphs''. The
reason is that we only know how to prove Proposition~~\ref{pro:decoding-increasing-vertices-number}
for vertex-decoding graphs. Fortunately, we can convert any decoding
graph to a vertex-decoding one using Proposition~\ref{pro:decoding-degree-reduction}.

We move to define the notion of vertex-decoding graphs. Intuitively,
a decoding graph is vertex-decoding if the value that an edge $\left(u,v\right)$
decodes depends only on the labeling of $u$, while the labeling of
$v$ only affects on whether the edge accepts or rejects. The formal
definition follows.
\begin{definition}
[\label{def:vertex-decoding-graphs}Vertex-decoding graphs]We say
that a decoding graph $G$ is a \emph{vertex-decoding graph} if it
has the following properties: 
\begin{enumerate}
\item \label{enu:vertex-decoding-depends-on-source}For every edge $\left(u,v\right)$
of $G$ and its associated circuit $\psi=\psi_{\left(u,v\right)}$,
there exists a function $f:\Sigma\to\Gamma$ that satisfies the following:
For every assignment $\pi$ to the vertices of~$G$ for which $\psi\left(\pi(u),\pi(v)\right)\ne\bot$
it holds that $\psi\left(\pi(u),\pi(v)\right)=f\left(\pi(u)\right)$. 
\item \label{enu:vertex-decoding-every-vertex-decodes}Every vertex has
at least one outgoing edge. In other words, every vertex is capable
of decoding at least one index~$k\in\left[t\right]$. 
\end{enumerate}
\end{definition}
\begin{remark}
While the property of a graph being vertex-decoding is reminiscent
of the projection property, there are two important differences. First,
note that Item~\ref{enu:vertex-decoding-depends-on-source} in Definition~\ref{def:vertex-decoding-graphs}
is weaker than the projection property, since it only requires that~$\pi(u)$
determines the decoded value, and not necessarily~$\pi(v)$. Second,
note that Item~\ref{enu:vertex-decoding-every-vertex-decodes} is
not required by the projection property, and is actually violated
by the known constructions of graphs that have the projection property.
\end{remark}
We turn to prove Propositions~\ref{pro:decoding-degree-reduction}
and~\ref{pro:decoding-increasing-vertices-number}. We begin with
Proposition~\ref{pro:decoding-degree-reduction}, which says that
we can always reduce the degree of decoding graphs while paying only
a moderate cost in the parameters. As mentioned above, the proposition
also transforms the decoding graph into a vertex-decoding graph.
\begin{proposition}
\label{pro:decoding-degree-reduction}Let $d_{0}$ be the constant
from Fact~\ref{fac:expanders-exist}, and let $d=2d_{0}$. There
exists a polynomial time procedure that acts as follows:
\begin{itemize}
\item \textbf{Input}: A decoding graph $G$ of size $n$ for input circuit
$\varphi:\Gamma^{t}\to\B$ with alphabet $\Sigma$, rejection ratio
$\rho$, decoding complexity $s$, and smoothness~$\smooth$.
\item \textbf{Output:} A $d$-regular vertex-decoding graph $G'$ of size
at most $d\cdot n/\gamma$ for input circuit $\varphi$, alphabet
$\Sigma^{2}$, rejection ratio $\Omega\left(\rho\right)$, decoding
complexity $s+\poly\log\left|\Sigma\right|$, and smoothness $1$.
Furthermore, $G'$ has at most $n/\gamma$ vertices.
\end{itemize}
\end{proposition}
\begin{myproof}
[Proof sketch]We apply the same construction as in the proof of
Proposition~\ref{pro:any-PCP-decoder-implies-decoding-graphs}. Let
$\varphi:\Gamma^{t}\nolinebreak\to\nolinebreak\B$ be the input circuit
of $G$. The key observation is that $G$ corresponds to a decoder
$D$ that acts on $\varphi$ such that $D$ has query complexity $2$,
randomness complexity $\log\left(n/t\cdot\smooth\right)$, proof alphabet
$\Sigma$, rejection ratio $\rho$, and decoding complexity $s$.
The reason for the foregoing randomness complexity is that by the
smoothness of $G$ and by the smoothness criterion of Proposition~\ref{pro:smoothness-criterion},
it holds that for every $k\in\left[t\right]$ there are at most $n/t\cdot\smooth$
edges that are associated with $k$, and therefore choosing a uniformly
distributed edge that is associated with $G$ requires $\log\left(n/\left(t\cdot\smooth\right)\right)$
uniformly distributed bits. Now, by applying the construction of the
proof of Proposition~\ref{pro:any-PCP-decoder-implies-decoding-graphs}
to the decoder $D$, we obtain a graph $G'$ that satisfies the requirements.
The fact that $G'$ is vertex-decoding can be observed by examining
the construction of Proposition~\ref{pro:any-PCP-decoder-implies-decoding-graphs}
(see also the second paragraph in the above proof sketch of Proposition~\ref{pro:any-PCP-decoder-implies-decoding-graphs}).
\end{myproof}
We next prove Proposition~\ref{pro:decoding-increasing-vertices-number},
which says that we can increases the number of vertices of a vertex-decoding
graph to any size we wish, while paying only a small cost in the parameters.
This proposition will be used to ensure that the number of vertices
of a decoding graph $G$ is equal to the number of vertices of the
de~Bruijn graph on which we want to embed $G$.
\begin{proposition}
\label{pro:decoding-increasing-vertices-number}There exists a polynomial
time procedure that acts as follows:
\begin{itemize}
\item \textbf{Input:}

\begin{itemize}
\item A vertex-decoding graph $G$ of size $n$ for input circuit $\varphi:\Gamma^{t}\to\B$
with $\ell$ vertices, alphabet $\Sigma$, rejection ratio~$\rho$,
decoding complexity $s$, degree bound~$d$, and smoothness~$\smooth$.
\item A number $\ell'\in\N$ such that $\ell'\ge\ell$ (given in unary).
\end{itemize}
\item \textbf{Output: }Let $c\eqdef\left\lfloor \frac{\ell'}{\ell}\right\rfloor $
and let $d_{0}$ and $h_{0}$ be the constants from Fact~\ref{fac:expanders-exist}.
The procedure outputs a vertex-decoding graph $G'$ of size at most
$2\cdot(c+1)\cdot d_{0}\cdot n$ for input circuit~$\varphi$ that
has exactly $\ell'$ vertices and also has alphabet $\Sigma$, output
size $s+\poly\log\left|\Sigma\right|$, rejection ratio~$\Omega\left(\smooth^{2}\cdot\rho/d^{2}\right)$,
degree bound $2\cdot d_{0}\cdot d$, and smoothness $\frac{1}{2}\cdot\smooth$.
\end{itemize}
Furthermore, if $G$ is $d$-regular then $G'$ is $\left(2\cdot d_{0}\cdot d\right)$-regular
and has rejection ratio $\Omega\left(\smooth^{2}\cdot\rho\right)$.\end{proposition}
\begin{myproof}
[Proof sketch]The basic idea of the proof is as follows. Given the
graph $G$, we construct the graph $G'$ by replacing each vertex
$v$ of $G$ with multiple copies of $v$, such that the total number
of vertices becomes $\ell'$ as required. Each copy of $v$ will be
connected to the same edges as the original $v$. An assignment to
$G'$ will be required to assign the same value to all the copies
of $v$: Clearly, if an assignment $\pi'$ to $G'$ assigns the same
value to the copies of each vertex $v$ of $G$, then in a way $\pi'$
``behaves'' like an assignment to $G$, and we can use the soundness
of $G$ to establish the soundness of $G'$ with respect to $\pi'$.
In order to verify that the copies of a vertex $v$ are assigned the
same value, we will put equality constraints between the copies of
$v$. In order to save edges, the equality constraints are placed
according to the edges of an expander, and the analysis goes exactly
as in the proof of Proposition~\ref{pro:any-PCP-decoder-implies-decoding-graphs}.
We use the fact that $G$ is vertex~decoding in order to allow the
equality constraints to decode values even though they can use only
the labeling of a single vertex of~$G$. The rest of this proof consists
of the technical details of this construction, and is provided in
Appendix~\ref{sec:decoding-increasing-vertices-number}.
\end{myproof}

\subsection{\label{sub:embedding-decoding-graphs-on-de-bruijn-graphs}Embedding
decoding graphs on de~Bruijn graphs}

In this section we prove the following proposition, which implies
Lemma~\ref{lem:decoding-linear-structure} (linear structure embedding
for udPCPs) and is analogous to Proposition~\ref{pro:de-bruijn-embedding}
(embedding of constraint graphs on de-Bruijn graphs). The proof follows
the steps of Proposition~\ref{pro:de-bruijn-embedding} with the
few adaptations to the setting of decoding graphs. For intuition and
a high-level explanation of the proof, we refer the reader to Section~\ref{sec:embedding-linear-structure}
and in particular to Section~\ref{sub:de-Bruijn-overview}.
\begin{proposition}
[\label{pro:decoding-de-bruijn-embedding}Embedding Decoding Graphs
on de-Bruijn Graphs]Let $d_{0}$ be the constant of Fact~\ref{fac:expanders-exist}.
There exists a polynomial time procedure that satisfies the following
requirements: 
\begin{itemize}
\item \textbf{Input:}

\begin{itemize}
\item A decoding graph $G$ of size $n$ for an input circuit $\varphi:\Gamma^{t}\to\B$
with alphabet $\Sigma$, rejection ratio $\rho$, decoding complexity
$s$, and smoothness~$\smooth$.
\item A finite alphabet $\Lambda$ such that $\left|\Lambda\right|\ge4\cdot d_{0}^{2}$. 
\item A natural number $m$ such that $\left|\Lambda\right|^{m}\ge2\cdot d_{0}\cdot n/\gamma$.
\end{itemize}
\item \textbf{Output:} A decoding graph $G'$ for $\varphi$ such that the
following holds:

\begin{itemize}
\item The underlying graph of $G'$ is the de~Bruijn graph $\DBs$. 
\item The size of $G'$ is $\left|\Lambda\right|^{m+1}$. 
\item $G'$ has alphabet $\Sigma^{O(m)}$. 
\item $G'$ has rejection ratio $\Omega\left(\rho/\left|\Lambda\right|^{2}\cdot m\right)$.
\item $G'$ has smoothness at least $\smooth'\eqdef\Omega\left(\frac{1}{\left|\Lambda\right|}\right)$.
\item $G'$ has decision complexity $s+\poly\left(m,\log\left|\Sigma\right|\right)$
\end{itemize}
\end{itemize}
\end{proposition}
Let $G$, $\Lambda$, and $m$ be as in Proposition~\ref{pro:decoding-de-bruijn-embedding},
and let $\varphi:\Gamma^{t}\to\B$ be the input circuit of $G$. On
input $G$, $\Lambda$, and $m$, the procedure acts as follows. The
procedure first constructs a vertex-decoding graph $G_{1}$ by applying
to $G$ the procedure of Proposition~\ref{pro:decoding-degree-reduction},
and then applying to the resulting graph the procedure of Proposition~\ref{pro:decoding-increasing-vertices-number}
with $\ell'=\left|\Lambda\right|^{m}$. It can be verified that $G_{1}$
is a vertex-decoding graph for input circuit $\varphi$ with exactly
$\left|\Lambda\right|^{m}$ vertices, alphabet $\Sigma_{1}\eqdef\Sigma^{2}$,
rejection ratio $\rho_{1}=\Omega\left(\rho\right)$, decoding complexity
$s+\poly\log\left|\Sigma\right|$, and smoothness at least $\frac{1}{2}$.
Furthermore, $G_{1}$ is $d$-regular for $d=4\cdot d_{0}^{2}\le\left|\Lambda\right|$,
and is of size $d\cdot\left|\Lambda\right|^{m}$.

Then, the procedure identifies the vertices of $G_{1}$ with the vertices
of $\DB=\DBs$, partitions the the edges of $G_{1}$ to $d$ matchings
$\mu_{1},\ldots,\mu_{d}$, and views those matchings as permutations
on the vertices of $\DB$. We apply Fact~\ref{fac:routing} to each
permutation~$\mu_{i}$ resulting in a set of paths $\mathcal{P}_{i}$
of length $l\eqdef2m$. Let $\mathcal{P}=\bigcup\mathcal{P}_{i}$.

Next, the procedure constructs $G'$ in the following way. The alphabet
of $G'$ is set to be $\Sigma_{1}^{l\cdot d}$, viewed as $\left(\Sigma_{1}^{l}\right)^{d}$.
If $\sigma\in\left(\Sigma_{1}^{l}\right)^{d}$, and $\sigma=\left(\sigma_{1},\ldots,\sigma_{d}\right)$,
we denote by $\sigma_{i,j}$ the element $\left(\sigma_{i}\right)_{j}\in\Sigma_{1}$.
It remains to describe how to associate each edge $e$ of $G'$ with
an index $k_{e}\in\left[k\right]$ and with a circuit $\psi_{e}$.
To this end, we first describe in which cases a circuit $\psi_{e}$
accepts, and then describe how the index $k_{e}$ is chosen and what
is the output of $\psi_{e}$ when it accepts.

\paragraph*{The conditions in which $\psi_{e}$ accepts. }

Fix an edge $e'=\left(u,v\right)$ of $G'$, and let $\psi_{e}$ be
the circuit associated with $e$. The circuit $\psi_{e}$ accepts
in exactly the same cases in which the constraint that corresponds
to $e$ in the proof of Proposition~\ref{pro:de-bruijn-embedding}
(for constraint graphs) accepts. That is, the circuit $\psi_{e}$
accepts if and only if all of the following conditions hold:
\begin{enumerate}
\item For every $i\in\left[d\right]$, the values $\left(\pi'\left(u\right)_{i,l},\pi'\left(u\right)_{i,1}\right)$
satisfy the edge $\left(\mu_{i}^{-1}(u),u\right)$ of $G$. 
\item It holds that $\pi'\left(u\right)_{1,1}=\ldots=\pi'\left(u\right)_{d,1}$
and that $\pi'\left(v\right)_{1,1}=\ldots=\pi'\left(v\right)_{d,1}$.
\item \label{enu:decoding-de-bruijn-routing-consistency}For every $i\in\left[d\right]$
and $j\in\left[l-1\right]$ such that $u$ and $v$ are the $j$-th
and $\left(j+1\right)$-th vertices of a path in $p\in\mathcal{P}_{i}$
respectively, it holds that $\pi'\left(u\right)_{i,j}\ne\pi'\left(v\right)_{i,j+1}$.
\item Same as Condition~\ref{enu:decoding-de-bruijn-routing-consistency},
but when $v$ is the $j$-th vertex of $p$ and $u$ is its $\left(j+1\right)$-th
vertex.
\end{enumerate}

\paragraph*{The choice of $k_{e}$ and the output of $\psi_{e}$. }

Fix a vertex $u$ of $G'$. We describe the way we assign indices
$k_{e}$ to the outgoing edges of $u$, and the output of the circuits~$\psi_{e}$.
We begin by associating each of the $\left|\Lambda\right|$ outgoing
edges of $u$ in $G'$ with one of the $d$ outgoing edges of $u$
in $G_{1}$. This association is done in a ``balanced'' way - that
is, each outgoing edge of $u$ in $G_{1}$ is associated with either
$\left\lfloor \left|\Lambda\right|/d\right\rfloor $ or $\left\lceil \left|\Lambda\right|/d\right\rceil $
edges of $u$ in $G'$.

Now, let $e'$ be an outgoing edge of $u$ in $G'$, and suppose that
it is associated with an outgoing edge $e_{1}$ of $u$ in $G_{1}$,
and that $e_{1}$ belongs to the matching $\mu_{i}$. Let $k_{e_{1}}$
and $\psi_{e_{1}}$ be the index and circuit associated with $e_{1}$.
Recall that since $G_{1}$ is vertex-decoding, there exists a function
$f_{e_{1}}:\Sigma_{1}\to\Gamma$ such that whenever $\psi_{e_{1}}\left(a,b\right)\ne\bot$
it holds that $\psi_{e_{1}}\left(a,b\right)=f_{e_{1}}(a)$. We associate
$e'$ with the index $k_{e_{1}}$, and with the circuit $\psi_{e'}$
that is defined for every $a',b'\in\left(\Sigma_{1}^{l}\right)^{d}$
for which $\psi_{e'}\left(a,b\right)\ne\bot$ by 
\[
\psi_{e'}\left(a',b'\right)=f_{e_{1}}\left(\left(a'\right)_{1,1}\right).
\]
Note that $\psi_{e'}$ is indeed well defined, since the cases in
which $\psi_{e'}$ outputs $\bot$ were defined above.

\paragraph*{The parameters of $G'$. }

The size and alphabet of $G'$ are immediate, and the completeness
of $G'$ can be established in the same way as in Proposition~\ref{pro:de-bruijn-embedding}
(embedding of constraint graphs on de-Bruijn graphs). It can also
be verified that $G'$ has smoothness at least $\smooth'=\frac{1}{2\cdot\left|\Lambda\right|}$
using the smoothness criterion (Proposition~\ref{pro:smoothness-criterion})
and a straightforward calculation.

It remains to analyze the rejection ratio of $G'$. Let $\pi'$ be
an assignment to $G'$ that minimizes the ratio between the probability
that a random edge of $G'$ rejects $\pi'$ (under the decoding distribution)
to the decoding error of $G'$ on $\pi'$. As in the proof of Proposition~\ref{pro:de-bruijn-embedding},
we may assume that for every vertex $u$ of $\DB$ it holds that $\pi'\left(u\right)_{1,1}=\ldots=\pi'\left(u\right)_{d,1}$,
since otherwise we may modify $\pi'$ to such an assignment that satisfies
this property without increasing the rejection probability or decreasing
the decoding error. Let $\pi_{1}$ be the assignment to~$G_{1}$
defined by $\pi_{1}(u)=\pi'\left(u\right)_{1,1}$. Let $\varepsilon$
be the decoding error of $G_{1}$ on $\pi_{1}$, and let $x$ be the
assignment to $\varphi$ that achieves this decoding error. Let $\varepsilon'$
be the decoding error of $G'$ on $\pi'$ with respect to $x$. We
show that the rejection probability of $G'$ on $\pi'$ is at least
$\Omega\left(\smooth'\cdot\rho_{1}\cdot\varepsilon'/\left|\Lambda\right|\cdot m\right)$,
and this will yield the required rejection ratio. 

Observe that by the smoothness of $G_{1}$ (resp. $G'$), the fraction
of edges of $G_{1}$ (resp. $G'$) that fail to decode $x$ on $\pi_{1}$
(resp. $\pi'$) is at least $\varepsilon_{0}\eqdef\frac{1}{2}\cdot\varepsilon$
(resp. $\varepsilon_{0}'=\smooth'\cdot\varepsilon'$). Furthermore,
the fraction of edges of $G_{1}$ that reject $\pi_{1}$ is at least~$\rho_{1}\cdot\varepsilon_{0}$.
This implies, using the same argument as in the proof of Proposition~\ref{pro:de-bruijn-embedding},
that the fraction of edges of $G'$ that reject $\pi'$ is at least~$\Omega\left(\rho_{1}\cdot\varepsilon_{0}/\left|\Lambda\right|\cdot m\right)$.

We finish the proof by relating $\varepsilon_{0}'$ with $\varepsilon_{0}$.
To this end, observe that for every edge~$e'=\left(u,v\right)$ of
$G'$ and its associated edge~$e_{1}$ of $G_{1}$, the edge $e'$
fails to decode $x$ on $\pi'$ (i.e. $\psi_{e'}\left(\pi'\left(u\right)\right)\notin\left\{ x_{k_{e'}},\bot\right\} $)
only if $e_{1}$ fails to decode $x$ on $\pi_{1}$ (i.e. $\psi_{e_{1}}\left(\pi_{1}\left(u\right)\right)\notin\left\{ x_{k_{e_{1}}},\bot\right\} $).
Furthermore, each edge~$e_{1}$ of $G_{1}$ corresponds to either
$\left\lfloor \left|\Lambda\right|/d\right\rfloor $ or $\left\lceil \left|\Lambda\right|/d\right\rceil $
edges in $G'$. It can be verified by a straightforward calculation
that this implies that $\varepsilon_{0}'\le2\cdot\varepsilon_{0}$.
It now follows that the fraction of edges of $G'$ that reject $\pi'$
is at least
\begin{eqnarray*}
\Omega\left(\frac{\rho_{1}\cdot\varepsilon_{0}}{\left|\Lambda\right|\cdot m}\right) & \ge & \Omega\left(\frac{\rho_{1}\cdot\varepsilon_{0}'}{\left|\Lambda\right|\cdot m}\right)\\
 & \ge & \Omega\left(\frac{\rho_{1}\cdot\smooth'}{\left|\Lambda\right|\cdot m}\cdot\varepsilon'\right)\\
 & = & \Omega\left(\frac{\rho}{\left|\Lambda\right|^{2}\cdot m}\cdot\varepsilon'\right).
\end{eqnarray*}
The required rejection ratio follows.\qed

\section{\label{sec:dPCP-derandomized-parallel-repetition}Derandomized Parallel
Repetition of Decoding Graphs with Linear Structure}

In this section we prove Lemma~\ref{lem:decoding-derand-parallel-repetition}
(derandomized parallel repetition for dPCPs), restated below.
\begin{lemma*}
[\ref{lem:decoding-derand-parallel-repetition}, restated]There
exist a universal constant $h$ and a polynomial time procedure that
satisfy the following requirements: 
\begin{itemize}
\item \textbf{Input:}

\begin{itemize}
\item A finite field $\F$ of size $q$.
\item A decoding graph $G=\left(\F^{m},E\right)$ of size $n$ for input
circuit $\varphi:\Gamma^{t}\to\B$ with linear structure, alphabet
$\Sigma$, rejection ratio $\rho$, decision complexity $s$, and
smoothness $\smooth$.
\item The rejection ratio $\rho$ of $G$.
\item A parameter $d_{0}\in\N$ such that $d_{0}<m/h^{2}$ and $\rho\ge h\cdot d_{0}\cdot q^{-d_{0}/h}/\gamma$.
\end{itemize}
\item \textbf{Output:} A decoding graph $G'$ for $\varphi$ such that the
following holds:

\begin{itemize}
\item $G'$ has size $n^{O\left(d_{0}\right)}$. 
\item $G'$ has alphabet $\Sigma^{q^{O(d_{0})}}$. 
\item $G'$ is list-decoding with soundness error $\varepsilon\eqdef h\cdot d_{0}\cdot q^{-d_{0}/h}/\gamma$
and list size $L\eqdef q^{O(d_{0})}$.
\item $G'$ has the projection property.
\item $G'$ has decoding complexity $q^{O(d_{0})}\cdot\left(s+\poly\log\left|\Sigma\right|\right)$.
\end{itemize}
\end{itemize}
\end{lemma*}
The proof follows the proof of the corresponding lemma for constraint
graphs (Lemma~\ref{lem:derandomized-parallel-repetition}), with
the following modification: Recall that the proof of Lemma~\ref{lem:derandomized-parallel-repetition}
described the graph $G'$ by describing a \emph{verification} procedure
(the E-test, Figure~\ref{fig:E-test}). Moreover, recall that the
E-test works by choosing a random subspace $\EF$ of edges and verifying
that the edges in $F$ are satisfied by the assignment $\Pi\left(F\right)$.

In order to describe the graph $G'$ of Lemma~\ref{lem:decoding-derand-parallel-repetition},
we describe a \emph{decoding} procedure (the E-decoder, see Figure~\ref{fig:E-decoder}
below). The E-decoder is constructed by changing the E-test as follows.
Whenever the E-decoder is required to decode an index $k\in\left[t\right]$,
the E-decoder chooses a random edge $e$ that is associated with $k$,
and then chooses the subspace $F$ to be a random subspace that contains~$e$.
The E-decoder then checks, as before, that the edges in $F$ are satisfied
by the assignment $\Pi\left(F\right)$. If one of the edges in $F$
is unsatisfied, then the E-decoder rejects. If all the edges in $F$
are satisfied, then the E-decoder decodes the index $k$ by invoking
the circuit $\psi_{e}$ associated with $e$ on input $\Pi\left(F\right)_{|e}$.

The intuition that underlies the construction of the E-decoder is
as follows. Just as in the proof of Lemma~\ref{lem:derandomized-parallel-repetition},
we argue that the E-decoder contains an implicit S-test, and therefore
the assignment $\Pi$ needs to be roughly consistent with some assignment
$\pi$ to $G$ in order to be accepted. We now consider two cases:
\begin{enumerate}
\item If $G$ has high decoding error on $\pi$, then by the soundness of
$G$ it holds that many of the edges of $G$ reject $\pi$. By the
sampling property of $F$, there are many edges in $F$ that reject
$\pi$, and therefore the E-decoder must reject with high probability.
\item If $G$ has low decoding error on $\pi$, then due to the sampling
property of $F$, only few of the edges in $F$ err. In particular,
since $e$ is distributed like a random edge of $F$, it only errs
with low probability. Thus, in this case the E-decoder decodes correctly
with high probability.
\end{enumerate}
Thus, in both cases the soundness error of the E-decoder is small.

\subsection{The construction of $G'$ and its parameters}

\begin{figure}[h!t]
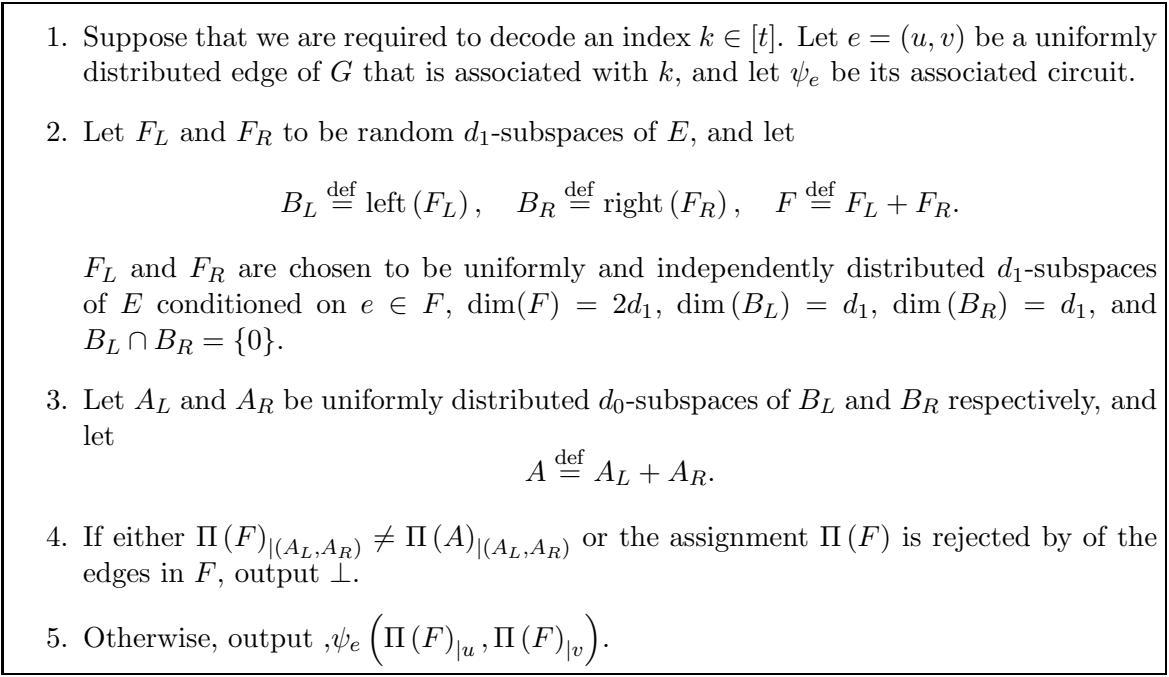

\centering \fbox{%
\begin{minipage}[c]{6in}%
\centering 
\begin{enumerate}
\item Suppose that we are required to decode an index $k\in\left[t\right]$.
Let $e=\left(u,v\right)$ be a uniformly distributed edge of $G$
that is associated with $k$, and let $\psi_{e}$ be its associated
circuit.
\item Let ${\EF}_{L}$ and ${\EF}_{R}$ to be random $d_{1}$-subspaces
of $\EE$, and let 
\[
B_{L}\eqdef\l\left({\EF}_{L}\right),\quad B_{R}\eqdef\r\left({\EF}_{R}\right),\quad{\EF}\eqdef{\EF}_{L}+{\EF}_{R}.
\]
${\EF}_{L}$ and ${\EF}_{R}$ are chosen to be uniformly and independently
distributed $d_{1}$-subspaces of $E$ conditioned on $e\in F$, $\dim({\EF})=2d_{1}$,
$\dim\left(B_{L}\right)=d_{1}$, $\dim\left(B_{R}\right)=d_{1}$,
and $B_{L}\nolinebreak\cap\nolinebreak B_{R}\nolinebreak=\nolinebreak\set 0$.
\item Let $A_{L}$ and $A_{R}$ be uniformly distributed $d_{0}$-subspaces
of $B_{L}$ and $B_{R}$ respectively, and let 
\[
A\eqdef A_{L}+A_{R}.
\]

\item If either $\Pi\left({\EF}\right)_{|\left(A_{L},A_{R}\right)}\ne\Pi\left(A\right)_{|\left(A_{L},A_{R}\right)}$
or the assignment $\Pi\left({\EF}\right)$ is rejected by of the edges
in~${\EF}$, output $\bot$.
\item Otherwise, output ,$\psi_{e}\left(\Pi\left(\EF\right)_{|u},\Pi\left(\EF\right)_{|v}\right)$.\end{enumerate}
\end{minipage}} \caption{\label{fig:E-decoder}The E-decoder}
\end{figure}

The decoding graph $G'$ is constructed as follows. Let $G=(\F^{m},\EE)$
and $d_{0}$ be as in Lemma~\ref{lem:decoding-derand-parallel-repetition}
(derandomized parallel repetition for dPCPs), and let $d_{1}=h\cdot d_{0}$
where $h$ is the universal constant from Lemma~\ref{lem:decoding-derand-parallel-repetition}
to be chosen later. As in the proof of the corresponding lemma for
constraint graphs (Lemma~\ref{lem:derandomized-parallel-repetition}),
the graph $G'$ is bipartite, the right vertices of $G'$ are the
$2d_{0}$-subspaces of $\F^{m}$ (the vertex-space of $G$), and the
left vertices of $G'$ are the $2d_{1}$-subspaces of the edge space
$E$ of $G$. An assignment $\Pi$ to $G'$ should label each $2d_{0}$-subspace
$A$ of~$\F^{m}$ with a function from $A$ to $\Sigma$, and each
$2d_{1}$-subspace $F$ of~$E$ with a function that maps the endpoints
of the edges in $F$ to $\Sigma$. The edges of $G'$ are constructed
such that they simulate the action of the ``E-decoder'' described
in Figure~\ref{fig:E-decoder}.

The completeness, size, and alphabet size of $G'$ is can be verified
in the same way as it was done in the proof of Lemma~\ref{lem:derandomized-parallel-repetition},
and so is the fact that $G'$ has the projection property. It remains
to analyze the soundness of $G'$, which is done in the following
section.

\subsection{The soundness of $G'$}

We turn to prove that $G'$ is list-decoding with $\varepsilon=h\cdot d_{0}\cdot q^{-d_{0}/h}/\gamma$
and list size $L=q^{O(d_{0})}$. Let $\Pi$ be an assignment to~$G'$.
That is, we prove that there exists a (possible empty) list of satisfying
assignments $x^{1},\ldots,x^{L}\in\Gamma^{t}$ to the input circuit~$\varphi$
such that when given as input a uniformly distributed index $k\in\left[t\right]$,
the probability that the output of the E-decoder is not in $\left\{ x_{k}^{1},\ldots,x_{k}^{L},\bot\right\} $
is at most $\varepsilon$.

Consider the distribution on the edges of $G'$ that results from
letting the edge $e$ of the E-decoder be chosen according to \emph{the
uniform distribution on the edges of $G$} instead of the decoding
distribution of $G$. We will refer to the above distribution as the
\emph{$G$-uniform distribution of $G'$}. It is straightforward to
show that the $G$-uniform distribution and decoding distribution
of $G'$ are $\smooth$-similar, by applying Claim~\ref{cla:similarity of derived variables}
with $X_{1}$ and $X_{2}$ being the choices of $e$ according the
the $G$-uniform distribution and the decoding distribution, and $Y_{1}$
and $Y_{2}$ being the $G$-uniform distribution and decoding distribution
of $G'$ respectively. In the following proof, all the probability
expressions are \emph{not} \emph{over the decoding distribution }of
$G'$, but rather over the \emph{$G$-uniform distribution of $G'$.}
We will later use the similarity between the distributions to argue
that $G'$ has small soundness error with respect to its decoding
distribution. 
\begin{notation}
We denote by $\mathcal{D}$ the random variable that equals to the
output of the E-decoder. As in the proof of Lemma~\ref{lem:derandomized-parallel-repetition}
(derandomized parallel repetition for constraint graphs), we denote
by~$\mathcal{T}$ the event in which the E-decoder accepts $\Pi$,
so $\mathcal{T}$ is the event $\mathcal{D}\ne\bot$. Moreover, as
in the proof of Lemma~\ref{lem:derandomized-parallel-repetition},
for an assignment $\pi:\F^{m}\to\Sigma$, we denote by $\Pi\left({\EF}\right)\apx{\alpha}\pi$
the claim that for at least $1-\alpha$~fraction of the edges $e$
of ${\EF}$ it holds that $\Pi\left({\EF}\right)$ is consistent with
$\pi$ on both the endpoints of $e$, and otherwise we denote $\Pi\left({\EF}\right)\napx{\alpha}\pi$.
\end{notation}
Our proof proceeds in two steps. We first show that there exists a
(possible empty) assignments $\pi^{1},\ldots,\pi^{L}:\F^{m}\to\Sigma$
such that whenever the E-decoder accepts $\Pi$, it almost always
does so while being roughly consistent with one of the assignments
$\pi^{1},\ldots,\pi^{L}$. We can then choose the assignments $x^{1},\ldots,x^{L}$
to be the assignments that minimize the decoding error of $\pi^{1},\ldots,\pi^{L}$
respectively. Next, we show that whenever $\Pi$ is roughly consistent
with $\pi^{i}$, the E-decoder either rejects~$\Pi$ with high probability
(if $\pi^{i}$ has high decoding error) or decodes $x^{i}$ successfully
with high probability (if $\pi^{i}$ has low decoding error). Thus,
the overall probability that the E-decoder fails is small.

The above strategy is made formal in the following three propositions.
Let $h'$ and $c$ be the universal constants defined in Theorem~\ref{thm:S-test-list-decoding}
below, and let $\alpha\eqdef h'\cdot d_{0}\cdot q^{-d_{0}/h'}$. Let
$\varepsilon_{0}\eqdef\varepsilon\cdot\smooth/3=h\cdot d_{0}\cdot q^{-d_{0}/h}/3$
and let $L=O\left(1/\varepsilon_{0}^{c}\right)$.
\begin{proposition}
\label{pro:E-decoder-list-decoding}There exists a (possibly empty)
list of assignments $\pi^{1},\ldots,\pi^{L}:\F^{m}\to\Sigma$ such
that 
\[
\Pr\left[\mathcal{T}\aand\not\exists i\in\left[L\right]\mbox{ s.t. }\Pi\left({\EF}\right)\apx{4\cdot\alpha}\pi^{i}\right]<\nolinebreak2\cdot\varepsilon_{0}.
\]

\end{proposition}

\begin{proposition}
\label{pro:inconsistency-with-errorneous-assignment}For every assignment
$\pi:\F^{m}\to\Sigma$ on which $G$ has decoding error at least $\varepsilon_{0}/2L$
it holds that $\Pr\left[\mathcal{T}\aand\Pi\left({\EF}\right)\apx{4\cdot\alpha}\pi\right]<\varepsilon_{0}/L$.
\end{proposition}

\begin{proposition}
\label{pro:decoding-on-good-assignment}For every assignment $\pi:\F^{m}\to\Sigma$
on which $G$ has decoding error less than $\varepsilon_{0}/2L$ with
respect to a satisfying assignment $x$ to the input circuit $\varphi$
it holds that 
\[
\Pr\left[\mathcal{D}\ne\ensuremath{x_{k}}\aand\Pi\left({\EF}\right)\apx{4\cdot\alpha}\pi\right]<\varepsilon_{0}/L,
\]
where $k$ is the index on which the E-decoder is invoked.
\end{proposition}
Propositions~\ref{pro:E-decoder-list-decoding} and~\ref{pro:decoding-on-good-assignment}
are proved in Sections~\ref{sub:E-decoder-list-decoding} and~\ref{sub:decoding-on-good-assignment}
respectively. Proposition~\ref{pro:inconsistency-with-errorneous-assignment}
can be proved in the same way as Proposition~\ref{pro:inconsistency-with-fixed-assignment},
by noting that due to the soundness of $G$, at least $\rho\cdot\varepsilon_{0}/2L$
of the edges of $G$ reject $\pi$.

We now prove that $G'$ is $\left(L,\varepsilon\right)$-list decoding
using Propositions~\ref{pro:E-decoder-list-decoding},~\ref{pro:inconsistency-with-errorneous-assignment},
and~\ref{pro:decoding-on-good-assignment}. Let $\pi^{1},\ldots,\pi^{L}$
be the assignments from Proposition~\ref{pro:E-decoder-list-decoding}.
For each $i\in\left[L\right]$, let $x^{i}$ be the assignment to
$\varphi$ that attains the decoding error of $\pi^{i}$. The decoding
error of $G'$ on $\Pi$ under the $G$-uniform distribution of $G'$
is as follows.
\begin{eqnarray}
\Pr\left[\mathcal{D}\notin\left\{ x_{k}^{1},\ldots,x_{k}^{L},\bot\right\} \right] & \le & \sum_{i=1}^{L}\Pr\left[\mathcal{D}\notin\left\{ x_{k}^{1},\ldots,x_{k}^{L},\bot\right\} \aand\Pi\left({\EF}\right)\apx{4\cdot\alpha}\pi^{i}\right]\nonumber \\
 &  & +\Pr\left[\mathcal{D}\notin\left\{ x_{k}^{1},\ldots,x_{k}^{L},\bot\right\} \aand\not\exists i\in\left[L\right]\mbox{ s.t. }\Pi\left({\EF}\right)\apx{4\cdot\alpha}\pi^{i}\right]\nonumber \\
 & \le & \sum_{i=1}^{L}\Pr\left[\ensuremath{\mathcal{D}\notin\left\{ x_{k}^{i},\bot\right\} }\aand\Pi\left({\EF}\right)\apx{4\cdot\alpha}\pi^{i}\right]\nonumber \\
 &  & +\Pr\left[\mathcal{T}\aand\not\exists i\in\left[L\right]\mbox{ s.t. }\Pi\left({\EF}\right)\apx{4\cdot\alpha}\pi^{i}\right]\nonumber \\
 & \le & \sum_{i=1}^{L}\varepsilon_{0}/L+2\cdot\varepsilon_{0}\label{eq:derand-par-rep-upper-bounds}\\
 & = & 3\cdot\varepsilon_{0},\nonumber 
\end{eqnarray}
where Inequality~\ref{eq:derand-par-rep-upper-bounds} follows from
Propositions~\ref{pro:E-decoder-list-decoding} and~\ref{pro:decoding-on-good-assignment}.
Finally, since the $G$-uniform distribution of $G'$ and the decoding
distribution of $G'$ are $\gamma$-similar, it follows that the decoding
error of $G'$ on $\Pi$ under the decoding distribution of $G'$
is at most $3\cdot\varepsilon_{0}/\gamma=\varepsilon$, as required.\qed

\subsubsection{\label{sub:E-decoder-list-decoding}Proof of Proposition~\ref{pro:E-decoder-list-decoding}}

Recall that in order to analyze the soundness of the E-test in Proposition~\ref{pro:test-passes-consistently},
we argued that the E-test contains an ``implicit S-test'', and then
relied on a theorem regarding the soundness of the S-test (Theorem~\ref{thm:S-test}).
The aforementioned theorem said that if the S-test accepts an assignment~$\Pi$
with some probability, then there exists an assignment $\pi$ such
that with some (smaller) probability, the S-test accepts $\Pi$ while
being consistent with the S-direct product of $\pi$. This can be
thought as a ``unique decoding'' theorem, that decodes $\pi$ from
$\Pi$.

In order to prove Proposition~\ref{pro:E-decoder-list-decoding}
for the E-decoder, we use a similar argument, but this time we use
a ``list decoding'' theorem for the S-test. The following theorem
says that there exists a short list of assignments $\pi_{1},\ldots,\pi_{L}$,
such that it is \emph{almost always} the case that if the S-test accepts
$\Pi$, it does so while being consistent with the S-direct product
of one of the assignments $\pi_{1},\ldots,\pi_{L}$.
\begin{theorem}
[\label{thm:S-test-list-decoding}List-decoding soundness of the
S-test]There exist universal constants $h',c\in\N$ such that for
every $d_{0}\in\N$, $d_{1}\ge h'\cdot d_{0}$, and $m\ge h'\cdot d_{1}$,
the following holds: Let $\varepsilon\ge h'\cdot d_{0}\cdot q^{-d_{0}/h'}$,
$\alpha\eqdef h'\cdot d_{0}\cdot q^{-d_{0}/h'}$. Let $\Pi$ be a
(possibly randomized) assignment to $2d_{0}$-subspaces of $\F^{m}$
and to pairs of $d_{1}$-subspaces of $\F^{m}$. Then, there exists
a (possibly empty) list of $L=O\left(1/\varepsilon^{c}\right)$ assignments
$\pi^{1},\ldots,\pi^{L}:\F^{m}\to\Sigma$ such that 

\[
\Pr\left[\Pi\left(B_{1},B_{2}\right)_{|\left(A_{1},A_{2}\right)}=\Pi\left(A\right)_{|\left(A_{1},A_{2}\right)}\aand\not\exists i\in\left[L\right]\mbox{ s.t. }\Pi\left(B_{1},B_{2}\right)\apx{\alpha}\pi_{|\left(B_{1},B_{2}\right)}^{i}\right]<\varepsilon.
\]

\end{theorem}
\noindent Theorem~\ref{thm:S-test-list-decoding} is proved in Section~\ref{sec:S-test-analysis}.

We turn to prove Proposition~\ref{pro:E-decoder-list-decoding} based
on Theorem~\ref{thm:S-test-list-decoding}. As in the proof of Proposition~\ref{pro:test-passes-consistently},
we begin by extending $\Pi$ to pairs of independent $d_{1}$-subspaces
of $\F^{m}$ in a randomized manner as follows: Given a pair of independent
$d_{1}$-subspaces $B_{1}$ and $B_{2}$, we choose ${\EF}_{1}$ and
${\EF}_{2}$ to be uniformly distributed and independent $d_{1}$-subspaces
of $\EE$ such that $\l\left({\EF}_{1}\right)=B_{1}$ and $\r\left({\EF}_{2}\right)=B_{2}$,
and set $\Pi\left(B_{1},B_{2}\right)=\Pi\left({\EF}_{1}+{\EF}_{2}\right)_{|\left(B_{1},B_{2}\right)}$.

Again as in the proof of Proposition~\ref{pro:test-passes-consistently},
we observe that the probability that the E-decoder accepts equals
to the probability that the S-test accepts the extended $\Pi$. The
reason is that the subspaces $B_{L}$, $B_{R}$, $A_{L}$, $A_{R}$
of the E-decoder are distributed like the subspaces $B_{1}$, $B_{2}$,
$A_{1}$, $A_{2}$ of the S-test. By choosing $h$ to be at least
the constant $h'$ we can invoke Theorem~\ref{thm:S-test-list-decoding}
(list-decoding soundness of the S-test), and conclude that there there
exists a list of $L=O\left(1/\varepsilon^{c}\right)$ assignments
$\pi^{1},\ldots,\pi^{L}:\F^{m}\to\Sigma$ such that for subspaces
$B_{1}$, $B_{2}$, $A_{1}$, $A_{2}$ as in the S-test it holds that
\[
\Pr\left[\Pi\left(B_{1},B_{2}\right)_{|\left(A_{1},A_{2}\right)}=\Pi\left(A\right)_{|\left(A_{1},A_{2}\right)}\aand\not\exists i\in\left[L\right]\mbox{ s.t. }\Pi\left(B_{1},B_{2}\right)\apx{\alpha}\pi_{|\left(B_{1},B_{2}\right)}^{i}\right]<\varepsilon_{0}.
\]
The latter inequality is equivalent to the following inequality:
\[
\Pr\left[\Pi\left({\EF}\right)_{|\left(B_{L},B_{R}\right)}=\Pi\left(A\right)_{|\left(A_{1},A_{2}\right)}\aand\not\exists i\in\left[L\right]\mbox{ s.t. }\Pi\left({\EF}\right)_{|\left(B_{L},B_{R}\right)}\apx{\alpha}\pi_{|\left(B_{L},B_{R}\right)}^{i}\right]<\varepsilon_{0},
\]
which in turn implies the inequality 
\begin{equation}
\Pr\left[\mathcal{T}\aand\not\exists i\in\left[L\right]\mbox{ s.t. }\Pi\left({\EF}\right)_{|\left(B_{L},B_{R}\right)}\apx{\alpha}\pi_{|\left(B_{L},B_{R}\right)}^{i}\right]<\varepsilon_{0}.\label{eq:derand-par-rep-S-test-list-decoding}
\end{equation}
In the rest of this section we show that this implies that
\begin{equation}
\Pr\left[\mathcal{T}\aand\not\exists i\in\left[L\right]\mbox{ s.t. }\Pi\left({\EF}\right)\apx{4\cdot\alpha}\pi^{i}\right]<\nolinebreak2\cdot\varepsilon_{0}\label{eq:E-decoder-list-decoding}
\end{equation}
To this end, we use Claim~\ref{cla:inconsistent-edges-spaces-inconsistent-on-S-test},
which was proved in Section~\ref{sub:test-passes-consistently} and
is restated below.
\begin{claim*}
[\ref{cla:inconsistent-edges-spaces-inconsistent-on-S-test}, restated]For
every fixed $2d_{0}$-subspace $F_{0}$ of $E$ such that $\Pi\left(F_{0}\right)\napx{4\alpha}\pi$,
it holds that
\[
\Pr\left[\left.\Pi\left({\EF}\right)_{|\left(B_{L},B_{R}\right)}\apx{\alpha}\pi_{|\left(B_{L},B_{R}\right)}\right|{F}=F_{0}\right]\le1/\left(q^{d_{1}-2}\cdot\alpha^{2}\right).
\]

\end{claim*}
\noindent Claim~\ref{cla:inconsistent-edges-spaces-inconsistent-on-S-test}
implies immediately the following corollary.
\begin{corollary}
\label{cor:inconsistent-lists-inconsistent-on-S-test}For every $i\in\left[L\right]$
it holds that
\[
\Pr\left[\left.\Pi\left({\EF}\right)_{|\left(B_{L},B_{R}\right)}\apx{\alpha}\pi_{i|\left(B_{L},B_{R}\right)}\right|\not\exists j\in\left[L\right]\mbox{ s.t. }\Pi\left({\EF}\right)\apx{4\cdot\alpha}\pi^{j}\right]<1/\left(q^{d_{1}-2}\cdot\alpha^{2}\right).
\]

\end{corollary}
\noindent In order to prove Inequality~\ref{eq:E-decoder-list-decoding},
we first show that
\begin{equation}
\Pr\left[\left.\mathcal{T}\aand\not\exists i\in\left[L\right]\mbox{ s.t. }\Pi\left({\EF}\right)_{|\left(B_{L},B_{R}\right)}\apx{\alpha}\pi_{|\left(B_{L},B_{R}\right)}^{i}\right|\not\exists i\in\left[L\right]\mbox{ s.t. }\Pi\left({\EF}\right)\apx{4\cdot\alpha}\pi^{i}\right]\ge\frac{1}{2}.\label{eq:inconsistent-S-test-conditioned-n-inconsistent-edge-space}
\end{equation}
To show it, we prove an upper bound on the complement event, that
is, we prove that 
\[
\Pr\left[\left.\mathcal{T}\aand\exists i\in\left[L\right]\mbox{ s.t. }\Pi\left({\EF}\right)_{|\left(B_{L},B_{R}\right)}\apx{\alpha}\pi_{|\left(B_{L},B_{R}\right)}^{i}\right|\not\exists i\in\left[L\right]\mbox{ s.t. }\Pi\left({\EF}\right)\apx{4\cdot\alpha}\pi^{i}\right]\le\frac{1}{2}.
\]
To see the latter inequality, observe that the right end side is upper
bounded by
\begin{eqnarray*}
\sum_{i\in\left[L\right]}\Pr\left[\left.\Pi\left({\EF}\right)_{|\left(B_{L},B_{R}\right)}\apx{\alpha}\pi_{|\left(B_{L},B_{R}\right)}^{i}\right|\not\exists j\in\left[L\right]\mbox{ s.t. }\Pi\left({\EF}\right)\apx{4\cdot\alpha}\pi^{j}\right] & \le & \sum_{i\in\left[L\right]}1/\left(q^{d_{1}-2}\cdot\alpha^{2}\right)\\
 & = & L\cdot/\left(q^{d_{1}-2}\cdot\alpha^{2}\right)\\
 & = & O\left(1/\varepsilon_{0}^{c}\cdot\left(q^{d_{1}-2}\cdot\alpha^{2}\right)\right)\\
 & \le & \frac{1}{2}.
\end{eqnarray*}
where the first inequality follows from Corollary~\ref{cor:inconsistent-lists-inconsistent-on-S-test},
and the second inequality follows for sufficiently large choice of
$h$. Now, it holds that
\begin{equation}
\Pr\left[\mathcal{T}\aand\not\exists i\in\left[L\right]\mbox{ s.t. }\Pi\left({\EF}\right)_{|\left(B_{L},B_{R}\right)}\apx{\alpha}\pi_{|\left(B_{L},B_{R}\right)}^{i}\aand\not\exists i\in\left[L\right]\mbox{ s.t. }\Pi\left({\EF}\right)\apx{4\cdot\alpha}\pi^{i}\right]\label{eq:both-S-test-and-edges-subspace-inconsistent}
\end{equation}
is upper bounded by
\[
\Pr\left[\mathcal{T}\aand\not\exists i\in\left[L\right]\mbox{ s.t. }\Pi\left({\EF}\right)_{|\left(B_{L},B_{R}\right)}\apx{\alpha}\pi_{|\left(B_{L},B_{R}\right)}^{i}\right]<\varepsilon_{0}.
\]
On the other hand, by writing the probability in~(\ref{eq:both-S-test-and-edges-subspace-inconsistent})
in conditional form and applying Inequality~\ref{eq:inconsistent-S-test-conditioned-n-inconsistent-edge-space},
we obtain that the probability in~(\ref{eq:both-S-test-and-edges-subspace-inconsistent})
is at least
\[
\frac{1}{2}\cdot\Pr\left[\mathcal{T}\aand\not\exists i\in\left[L\right]\mbox{ s.t. }\Pi\left({\EF}\right)\apx{4\cdot\alpha}\pi^{i}\right].
\]
By combining the two last bounds, we obtain that
\[
\Pr\left[\mathcal{T}\aand\not\exists i\in\left[L\right]\mbox{ s.t. }\Pi\left({\EF}\right)\apx{4\cdot\alpha}\pi^{i}\right]<2\cdot\varepsilon_{0},
\]
as required.\qed

\subsubsection{\label{sub:decoding-on-good-assignment}Proof of Proposition~\ref{pro:decoding-on-good-assignment}}

Fix an assignment $\pi:\F^{m}\to\Sigma$ on which $G$ has decoding
error less than $\varepsilon_{0}/2L$ with respect to a satisfying
assignment $x$ of the input circuit $\varphi$. We prove that $\Pr\left[\ensuremath{\mbox{\emph{D}}\ne x_{k}}\aand\Pi\left({\EF}\right)\apx{4\cdot\alpha}\pi\right]<\varepsilon_{0}/L$
Let us denote by $\mathcal{E}_{1}$ the event in which $\Pi\left({\EF}\right)\apx{4\cdot\alpha}\pi$
and by $\mathcal{E}_{2}$ the event in which $F$ contains less than
$\varepsilon_{0}/3L$ fraction of edges on which $G$ fails to decode
$x$ on~$\pi$. We will prove that 
\[
\Pr\left[\ensuremath{\mathcal{D}\ne x_{k}}\aand\mathcal{E}_{1}\right]=\Pr\left[\ensuremath{\mathcal{D}\ne x_{k}}\aand\Pi\left({\EF}\right)\apx{4\cdot\alpha}\pi\right]<\varepsilon_{0}/L.
\]
It holds that
\[
\Pr\left[\ensuremath{\mathcal{D}\ne x_{k}}\aand\mathcal{E}_{1}\right]=\Pr\left[\ensuremath{\mathcal{D}\ne x_{k}}\aand\mathcal{E}_{1}\aand\mathcal{E}_{2}\right]+\Pr\left[\ensuremath{\psi(a,b)\ne x_{k}}\aand\mathcal{E}_{1}\aand\neg\mathcal{E}_{2}\right].
\]
We upper bound both terms on the right hand side. The second term
is clearly upper bounded by $\Pr\left[\neg\mathcal{E}_{2}\right]$.
The latter probability can be shown to be at most~$O\left(L^{2}/q^{2\cdot d_{1}-2}\cdot\varepsilon_{0}^{2}+\cdot d_{1}/q^{m-2\cdot d_{1}}\right)$,
using the fact that~$F$ samples well the edges of $G$, and more
specifically using an argument similar to the one used in the proof
of Proposition~\ref{pro:inconsistency-with-fixed-assignment}. For
sufficiently large choice of $h$, the latter expression is upper
bounded by $\varepsilon/3L$.

We turn to upper bound the probability $\Pr\left[\ensuremath{\mathcal{D}\ne x_{j}}\aand\mathcal{E}_{1}\aand\mathcal{E}_{2}\right]$.
This probability is upper bounded by the probability $\Pr\left[\ensuremath{\mathcal{D}\ne x_{j}}|\mathcal{E}_{1}\aand\mathcal{E}_{2}\right]$.
Now, let $F_{0}$ be any $2d_{1}$-subspace of $E$ such that $\Pi\left(F_{0}\right)\apx{4\cdot\alpha}\pi_{i}$
and such that the fraction of edges of $F_{0}$ that fail to decode
$x$ on $\pi$ is at most $2\varepsilon_{0}/3L$. Let us consider
the probability $\Pr\left[\ensuremath{\mathcal{D}\ne x_{j}}|F=F_{0}\right]$.
Observe that conditioned on the choice~$F=F_{0}$, the edge $e$
chosen by the E-test is uniformly distributed among the edges of $F$.
Observe that $e$ fails to decode $x$ only if one of the endpoints
of $e$ is inconsistent with $\pi$ or if $e$ is one of the edges
in $F$ that fail to decode $x$ on $\pi$. The probability of the
first case is at most $4\cdot\alpha\le\varepsilon_{0}/3L$ (where
the latter inequality holds for sufficiently large choice of $h$),
and the probability of the second case is at most $\varepsilon_{0}/3L$.
It therefore holds that 
\[
\Pr\left[\ensuremath{\mathcal{D}\ne x_{k}}\aand\mathcal{E}_{1}\aand\mathcal{E}_{2}\right]\le\Pr\left[\ensuremath{\mathcal{D}\ne x_{j}}|F=F_{0}\right]\le\varepsilon_{0}/3L+\varepsilon_{0}/3L\le2\varepsilon_{0}/3L.
\]
All in all, it holds that $\Pr\left[\ensuremath{\mathcal{D}\ne x_{k}\aand}\mathcal{E}_{1}\right]$
is at most $2\varepsilon_{0}/3L+3\cdot\varepsilon_{0}/3L=\varepsilon_{0}/L$,
as required.\qed

\section{\label{sec:S-test-analysis}The Analysis of the Specialized Direct
Product Test}

In this section we provide the analysis of the S-test and prove Theorems~\ref{thm:S-test}
and~\ref{thm:S-test-list-decoding}, which are the theorems on the
soundness of the S-test that are used in Sections~\ref{sub:test-passes-consistently}
and~\ref{sub:E-decoder-list-decoding} respectively. The proof proceeds
in two steps. First, in Section~\ref{sub:P-squared-test}, we define
and analyze an intermediate direct product test, which we call the
$P^{2}$-test. Then, in Section \ref{sub:Proof-of-S-test}, we reduce
the analysis of the S-test to that of the $P^{2}$-test.

For the rest of this section, we let $\F$ be a finite field of size
$q$ and let $d_{0},d_{1}\in\N$.

\subsection{\label{sub:P-squared-test}The $P^{2}$-test}

In this section we define and analyze the $P^{2}$-test. Informally,
the $P^{2}$-test consists of two P-tests that are performed simultaneously.
Details follow.

Given two strings $\pi_{1},\pi_{2}:\F^{m}\to\Sigma$, we define their
\emph{$P^{2}$-direct product} $\Pi$ (with respect to $d_{0},d_{1}\in\N$)
as follows: $\Pi$ assigns each pair of $d_{0}$-subspaces $\left(A_{1},A_{2}\right)$
the pair of functions~$(\pi_{1|A_{1}},\pi_{2|A_{2}})$, and assigns
each pair of $d_{1}$-subspaces $\left(B_{1},B_{2}\right)$ to the
pair of functions $(\pi_{1|B_{1}},\pi_{2|B_{2}})$. We consider the
task of testing whether a given assignment $\Pi$ is the $P^{2}$-direct
product of some pair of strings $\pi_{1},\pi_{2}:\F^{m}\to\Sigma$.
That is, we are given an assignment $\Pi$ , and in order to check
whether $\Pi$ is a $P^{2}$-direct product, we invoke the $P^{2}$-test,
described in Figure~\ref{fig:P-squared-test}.

\begin{figure}
\centering\fbox{%
\begin{minipage}[c]{6in}%
\centering 
\begin{enumerate}
\item Choose two uniformly distributed $d_{1}$-subspaces $B_{1},B_{2}$
of $\F^{m}$.
\item Choose two uniformly distributed $d_{0}$-subspaces $A_{1}\subseteq B_{1}$,
$A_{2}\subseteq B_{2}$.
\item Accept if and only if $\Pi\left(B_{1},B_{2}\right)_{|\left(A_{1},A_{2}\right)}=\Pi\left(A_{1},A_{2}\right)$. \end{enumerate}
\end{minipage}} \caption{\label{fig:P-squared-test}The $P^{2}$-test}
\end{figure}

It is easy to see that if $\Pi$ is a $P^{2}$-direct product then
the $P^{2}$-test always accepts. Again, it can be shown that if $\Pi$
is ``far'' from being a $P^{2}$-direct product, then the $P^{2}$-test
rejects with high probability, and that this holds even if $\Pi$
is a randomized assignment. Formally, we have the following result.
\begin{theorem}
[\label{thm:P-squared-test}Soundness of the $P^{2}$-test]There
exist universal constants $\hps,\cps\in\N$ such that the following
holds: Let $\varepsilon\ge\hps\cdot d_{0}\cdot q^{-d_{0}/\hps}$,
$\aps\eqdef\hps\cdot d_{0}\cdot q^{-d_{0}/\hps}$. Assume that $d_{1}\ge\hps\cdot d_{0}$,
$m\ge\hps\cdot d_{1}$. Suppose that an assignment $\Pi$ passes the
$P^{2}$-test with probability at least $\varepsilon$. Then, there
exist two assignments $\pi_{1}$ and $\pi_{2}$ to $\F^{m}$ such
that for $B_{1}$, $B_{2}$, $A_{1}$, $A_{2}$, distributed as in
the $P^{2}$-test it holds that
\[
\Pr\left[\Pi\left(B_{1},B_{2}\right)_{|\left(A_{1},A_{2}\right)}=\Pi\left(A_{1},A_{2}\right)\aand\Pi\left(A_{1},A_{2}\right)\apx{\aps}\left(\pi_{1|A_{1}},\pi_{2|A_{2}}\right)\aand\Pi\left(B_{1},B_{2}\right)\apx{\aps}\left(\pi_{1|B_{1}},\pi_{2|B_{2}}\right)\right]
\]
is at least $\Omega\left(\varepsilon^{\cps}\right)$.
\end{theorem}
In the rest of this section we prove Theorem~\ref{thm:P-squared-test}.
We denote by $\mathcal{P}$ the event in which the $P^{2}$-test accepts,
that is, that $\Pi\left(B_{1},B_{2}\right)_{|\left(A_{1},A_{2}\right)}=\Pi\left(A_{1},A_{2}\right)$.
The core of the proof is the following lemma:
\begin{lemma}
\label{lem:one-side-consistency}There exist universal constants $h',c'\in\N$
such that the following holds: Let $\varepsilon\ge h'\cdot d_{0}\cdot q^{-d_{0}/h'}$,
$\alpha'\eqdef h'\cdot d_{0}\cdot q^{-d_{0}/h'}$. Assume that $d_{1}\ge h'\cdot d_{0}$,
$m\ge h'\cdot d_{1}$. If $\Pi$ passes the $P^{2}$-test with probability
at least $\varepsilon$ then there exists an assignment $\pi_{2}:\F^{m}\to\Sigma$
such that
\[
\Pr\left[\mathcal{P}\aand\Pi\left(A_{1},A_{2}\right)_{|A_{2}}\apx{\alpha'}\pi_{2|A_{2}}\aand\left(B_{1},B_{2}\right)_{|B_{2}}\apx{\alpha'}\pi_{2|B_{2}}\right]\ge\Omega(\eps^{c'}),
\]
 and symmetrically, there exists a function $\pi_{1}:\F^{m}\to\Sigma$
such that 
\[
\Pr\left[\mathcal{P}\aand\Pi\left(A_{1},A_{2}\right)_{|A_{1}}\apx{\alpha'}\pi_{1|A_{1}}\aand\left(B_{1},B_{2}\right)_{|B_{1}}\apx{\alpha'}\pi_{1|B_{1}}\right]\ge\Omega(\eps^{c'}).
\]

\end{lemma}
We prove Lemma~\ref{lem:one-side-consistency} in Section~\ref{sub:P-squared-test-one-side-consistency}.
We turn to derive Theorem~\ref{thm:P-squared-test} from Lemma~\ref{lem:one-side-consistency}.
\begin{myproof}
[Proof of Theorem~\ref{thm:P-squared-test}.]The following proof
is for the case where $\Pi$ is not randomized, but it can be easily
extended to the case where $\Pi$ is randomized (see Remark~\ref{rem:S-test-randomized-assignment}
for details). We will choose $\hps$ to be larger than the constant
$h'$ of Lemma~\ref{lem:one-side-consistency}, so we can apply this
lemma. Let $\pi_{2}:\F^{m}\to\Sigma$ be the assignment guaranteed
by Lemma~\ref{lem:one-side-consistency}, and let $\Pi'$ be an assignment
that is obtained from $\Pi$ as follows: 
\begin{enumerate}
\item For every pair $\left(A_{1},A_{2}\right)$ for which $\Pi\left(A_{1},A_{2}\right)_{|A_{2}}\apx{\alpha'}\pi_{2|A_{2}}$,
set $\Pi'\left(A_{1},A_{2}\right)=\Pi\left(A_{1},A_{2}\right)$. 
\item For every other pair $\left(A_{1},A_{2}\right)$, set $\Pi'\left(A_{1},A_{2}\right)=\bot$,
where $\bot$ is some special value on which the test never accepts.
\item Set the pairs $\left(B_{1},B_{2}\right)$ similarly. 
\end{enumerate}
The probability $\eps'$ that the assignment $\Pi'$ passes the $P^{2}$-test
is at least $\Omega(\varepsilon^{c'})$ by the definition of $\pi_{2}$.
By choosing $\hps$ to be sufficiently larger than the corresponding
constants of Lemma~\ref{lem:one-side-consistency}, we can make sure
that $\varepsilon'$ satisfies the requirements of Lemma~\ref{lem:one-side-consistency}.
Therefore, we can deduce by Lemma~\ref{lem:one-side-consistency}
that there exists an assignment $\pi_{1}:\F^{m}\to\Sigma$ such that
\[
\Pr\left[\mathcal{P}\aand\Pi'\left(A_{1},A_{2}\right)_{|A_{1}}\apx{\alpha'}\pi_{1|A_{1}}\aand\Pi'\left(B_{1},B_{2}\right)_{|B_{1}}\apx{\alpha'}\pi_{1|B_{1}}\right]\ge\Omega(\left(\varepsilon'\right)^{c'})=\Omega(\varepsilon^{\left(c'\right)^{2}}).
\]
We now choose $c=\left(c'\right)^{2}$. Since the test never accepts
when $\Pi'$ answers $\bot$, we deduce that 
\[
\Pr\left[\mathcal{P}\aand\Pi(A_{1},A_{2})\apx{\alpha'}\left(\pi_{1|A_{1}},\pi_{2|A_{2}}\right)\aand\Pi\left(B_{1},B_{2}\right)\apx{\alpha'}\left(\pi_{1|B_{1}},\pi_{2|B_{2}}\right)\right]\ge\Omega(\varepsilon^{c}).
\]
Choosing $\hps$ such that $\aps\ge\alpha'$ completes the proof. \end{myproof}
\begin{remark}
\label{rem:justifying_bot}Technically speaking, our use of the special
value~$\bot$ requires formal justification, since when defining
the $P^{2}$-test and stating Lemma~\ref{lem:one-side-consistency}
we did not allow the use of such a special symbol. To this end, we
observe that the use of $\bot$ can be implemented as follows: Let
$\Sigma'=\Sigma\cup\left\{ \bot_{A},\bot_{B}\right\} $, where $\bot_{A},\bot_{B}$
are symbols outside~$\Sigma$. We first observe that Lemma~\ref{lem:one-side-consistency}
works just as well if we replace the alphabet $\Sigma$ with the modified
alphabet $\Sigma'$, since Lemma~\ref{lem:one-side-consistency}
is oblivious to the choice of the alphabet. Now, whenever we wish
to set $\Pi'\left(A_{1},A_{2}\right)=\bot$ in the proof of Theorem~\ref{thm:P-squared-test},
we actually set $\Pi'\left(A_{1},A_{2}\right)$ to be the pair of
functions that map all the vectors of $A_{1}$ and $A_{2}$ respectively
to the symbol $\bot_{A}$. We deal with the case of $\Pi'\left(B_{1},B_{2}\right)=\bot$
similarly, this time using the symbol~$\bot_{B}$. It remains to
observe that when assigning $\Pi'\left(A_{1},A_{2}\right)$ this way,
the $P^{2}$-test will always reject $\Pi'\left(A_{1},A_{2}\right)$,
since the assignment $\Pi'$ never assigns pairs~$\left(B_{1},B_{2}\right)$
with the symbol $\bot_{A}$. The same holds for the case of $\Pi'\left(B_{1},B_{2}\right)=\bot$.
\end{remark}

\begin{remark}
\label{rem:S-test-randomized-assignment}If $\Pi$ is randomized,
then the definition of $\Pi'$ in the foregoing proof should be slightly
changed to consider the internal randomness of $\Pi$. That is, we
define $\Pi'$ to be a randomized assignment, and obtain it from $\Pi$
as follows. For every pair $\left(A_{1},A_{2}\right)$ and every internal
randomness $\omega$ of $\Pi$, let us denote by $\left(a_{1},a_{2}\right)$
the output of $\Pi$ on $\left(A_{1},A_{2}\right)$ and randomness
$\omega$. We define the output of $\Pi'$ on $\left(A_{1},A_{2}\right)$
and randomness $\omega$ to be $\left(a_{1},a_{2}\right)$ if~$a_{2}\apx{\alpha'}\pi_{2|A_{2}}$,
and define it to be $\bot$ otherwise. The definition for pairs $\left(B_{1},B_{2}\right)$
is again similar.
\end{remark}

\subsubsection{\label{sub:P-squared-test-one-side-consistency}The proof of Lemma~\ref{lem:one-side-consistency}}

We prove Lemma~\ref{lem:one-side-consistency} only for the assignment
$\pi_{2}$, and the conclusion $\pi_{1}$ can be proved analogously.
The proof proceeds in three steps. First, we rely on Theorem~\ref{thm:IKW}
(soundness of the P-test) to find for each pair of $A_{1},B_{1}$
a direct product function that agrees (on average) with a good fraction
of $\Pi(A_{1},\cdot)$ and $\Pi(B_{1},\cdot)$. Then, we show that
for each $A_{1}$ separately, the number of distinct such functions
is bounded. Next, we show that there is a single function $\pi$ such
that the probability that the test accepts and $\Pi\left(A_{1},A_{2}\right)_{|A_{2}}\approx\pi_{|A_{2}}$
is non-negligible (A priori there could have been a different $\pi$
for each $A_{1}$). Finally, we extend the latter result for $d_{1}$-subspaces
$B_{1}$, $B_{2}$. Let $h_{1}$ be the universal constant whose existence
is guaranteed in Theorem~\ref{thm:IKW}, and let $\alpha_{1}$ be
the corresponding value from Theorem~\ref{thm:IKW}.

\paragraph*{Step 1.}

Consider the bipartite graph corresponding to the $P$-test, that
is, the graph whose left vertices are $d_{0}$-subspaces and whose
right vertices are $d_{1}$-subspaces, and such that a $d_{0}$-subspace
$A_{1}$ is connected to a $d_{1}$-subspace $B_{1}$ by an edge if
and only if $A_{1}\subseteq B_{1}$. . We label an edge $(A_{1},B_{1})$
by $\pi:\F^{m}\to\Sigma$ if 
\[
\Pr_{A_{2},B_{2}}\left[\mathcal{P}\aand\Pi\left(B_{1},B_{2}\right)_{|B_{2}}\apx{\alpha_{1}}\pi_{|B_{2}}\aand\Pi\left(A_{1},A_{2}\right)_{|A_{2}}\apx{\alpha_{1}}\pi_{|A_{2}}\right]\ge\Omega\left(\varepsilon^{4}\right).
\]
If no such $\pi$ exists then do not label the edge.

Fix $A_{1},B_{1}$. We will choose the universal constant $h'$ to
be at least $2\cdot h_{1}$. If the probability of passing the $P^{2}$-test
conditioned on $A_{1},B_{1}$ is at least $\varepsilon/2$, then we
claim that the edge is labeled. Indeed, define an assignment $\Pi_{\left(A_{1},B_{1}\right)}$
by 
\[
\Pi_{\left(A_{1},B_{1}\right)}(A_{2})=\Pi\left(A_{1},A_{2}\right)_{|A_{2}}\aand\Pi_{\left(A_{1},B_{1}\right)}(B_{2})=\Pi\left(B_{1},B_{2}\right)_{|B_{2}}.
\]
If $\Pi_{\left(A_{1},B_{1}\right)}$ passes the $P$-test with probability
at least $\varepsilon/2$, then by Theorem~\ref{thm:IKW} (soundness
of the P-test) there is an assignment $\pi$ as needed (since $h'\ge2\cdot h_{1}$).

Furthermore, observe that by averaging at least $\varepsilon/2$ of
the edges $(A_{1},B_{1})$ have conditional success at least $\varepsilon/2$,
so $(A_{1},B_{1})$ is labeled.

\paragraph*{Step 2.}

Fix $B_{1}$ and let $L(B_{1})$ be the labels on edges touching $B_{1}$.
Consider the following ``pruning'' process: arbitrarily choose a
label $\pi\in L(B_{1})$ and remove all elements in $L(B_{1})$ that
are within relative Hamming distance $3\alpha_{1}$ of $\pi$. Repeat
until no more labels can be removed. Let $L'(B_{1})$ denote the remaining
set of labels. The set $L'(B_{1})$ has the following properties 
\begin{itemize}
\item Every pair of labels in $L'(B_{1})$ are at least $3\alpha_{1}$ apart,
and 
\item Every $f\in L(B_{1})$ is $3\alpha_{1}$-close to some label in $L'(B_{1})$. 
\end{itemize}
We prove that $\card{L'(B_{1})}\le O(1/\varepsilon^{4})$, using an
argument in the spirit of the Johnson bound: Suppose $L'(B_{1})=\{\pi_{1},\pi_{2},\ldots\}$
is non-empty. For every $\pi_{i}\ne\pi_{j}\in L'(B)$ let us denote
\begin{eqnarray*}
p_{i} & \eqdef & \Pr_{B_{2}}\left[\Pi\left(B_{1},B_{2}\right)_{|B_{2}}\apx{\alpha_{1}}\pi_{i|B_{2}}\right]\\
p_{i,j} & = & \Pr_{B_{2}}\left[\Pi\left(B_{1},B_{2}\right)_{|B_{2}}\apx{\alpha_{1}}\pi_{i|B_{2}}\aand\Pi\left(B_{1},B_{2}\right)_{|B_{2}}\apx{\alpha_{1}}\pi_{j|B_{2}}\right].
\end{eqnarray*}
By the definition of the labels $\pi_{i}$, we know that for some
universal constant $\eta$ it holds that $p_{i}\ge\eta\cdot\varepsilon^{4}$
for every $\pi_{i}$. We upper bound the fractions $p_{i,j}$: We
know that for every $\pi_{i}\ne\pi_{j}$ it holds that $\pi_{i}\napx{3\cdot\alpha_{1}}\pi_{j}$.
It follows that
\begin{eqnarray*}
p_{i,j} & \le & \Pr_{B_{2}}\left[\pi_{i|B_{2}}\apx{2\cdot\alpha_{1}}\pi_{j|B_{2}}\right]\\
 & \le & 1/\left(q^{d_{1}-2}\cdot\left(\alpha_{1}-q^{-d_{1}}\right)^{2}\right)\\
 & \le & \frac{1}{2}\cdot\eta^{2}\cdot\varepsilon^{8},
\end{eqnarray*}
where the second inequality follows by Lemma~\ref{lem:subspace-point-sampler}
(subspace-point sampler) and the third inequality holds for sufficiently
large choice of $h'$. Now, by the inclusion-exclusion principle that
\begin{eqnarray*}
\sum_{i}p_{i}-\sum_{i\ne j}p_{i,j} & \le & 1\\
\left|L'(B_{1})\right|\cdot\left(\eta\cdot\varepsilon^{4}\right)-\frac{1}{2}\left|L'(B_{1})\right|^{2}\cdot\left(\frac{1}{2}\cdot\eta^{2}\cdot\varepsilon^{8}\right) & \le & 1.
\end{eqnarray*}
 The last inequality immediately implies that $\left|L'(B_{1})\right|\le2/\left(\eta\cdot\varepsilon^{4}\right)=O(1/\varepsilon^{4})$.

We define $L(A_{1})$ similarly, and prune it to $L'(A_{1})$. Imagine
now choosing a random $\pi_{A_{1}}\in L'(A)$ for each $A_{1}$ and
a random $\pi_{B_{1}}\in L'(B_{1})$ for each $B_{1}$. An edge $(A_{1},B_{1})$
is called alive if it is labeled by a function $\pi$ that is $3\alpha'$-close
to both $\pi_{A_{1}}$ and $\pi_{B_{1}}$. We expect at least $1/\card{L'(A)}\card{L'(B)}=\Omega(\varepsilon^{8})$
fraction of edges to be alive. Fix a choice of $\pi_{A_{1}}$ and
$\pi_{B_{1}}$ for each $A_{1}$ and $B_{1}$ in a way that attains
this expectation.

\paragraph*{Step 3.}

Let $\D_{1}$ be the distribution of choosing a random $d_{1}$-subspace
$B_{1}$ and two neighbors $A_{1},A_{1}'$ of it in the graph. Let
$\D_{2}$ be the distribution of choosing two $d_{0}$-spaces $A_{1},A_{1}'$
independently and a random $B_{1}$ that is a common neighbor of them
in the graph. The statistical distance between $\D_{1}$ and $\D_{2}$
is small: 
\begin{claim}
\label{claim:D1D2}For every $\kappa\in\N$, if the constant $h'$
is sufficiently large then the distributions $\mathcal{D}_{1}$ and
$\mathcal{D}_{2}$ are $\delta$-close for $\delta<\varepsilon^{24}/\kappa$. 
\end{claim}
We defer the proof of this claim to Section~\ref{sub:P-squared-test-auxiliary-claims}.
Now choose a random triplet $A_{1},A_{1}',B_{1}$ according to $\D_{1}$.
We lower bound the probability that both edges $(A_{1},B_{1})$ and
$(A_{1}',B_{1})$ are alive. This certainly holds if (i) $\Omega(\varepsilon^{8})$
fraction of the edges adjacent to $B$ are alive, and (ii) both edges
$(A_{1},B_{1})$ and $(A_{1}',B_{1})$ are alive. Part (i) holds with
probability $\Omega(\varepsilon^{8})$ and conditioned on this, Part
(ii) holds with probability at least $\Omega(\varepsilon^{16})$.
Altogether 
\[
\Pr_{(B_{1},A_{1},A_{1}')\sim\D_{1}}\left[(A_{1},B_{1}),(A_{1}',B_{1})\mbox{ are both alive}\right]=\Omega(\varepsilon^{24}).
\]
Finally, if we let $\delta$ be the statistical distance of $\D_{1}$
and $\D_{2}$, and apply Claim~\ref{claim:D1D2} with sufficiently
large choices of $\kappa$ and $h'$, then we have that 
\[
\Pr_{(B_{1},A_{1},A_{1}')\sim\D_{2}}\left[(A_{1},B_{1}),(A_{1}',B_{1})\mbox{ are both alive}\right]\ge\Omega(\varepsilon^{24})-\delta=\Omega(\varepsilon^{24}).
\]
Now fix $A_{1}$ such that the above holds when conditioning on $A_{1}$.
This means that for at least $\Omega(\varepsilon^{24})$ fraction
of the $d_{0}$-subspaces $A_{1}'$ there exists a $d_{1}$-subspace
$B_{1}$ such that both the edges $\left(A_{1},B_{1}\right)$ and
$\left(A_{1}',B_{1}\right)$ are alive. For each such $A_{1}'$, it
holds that the label of $\left(A_{1}',B_{1}\right)$ is $3\alpha_{1}$-close
to $\pi_{B_{1}}$, which in turn is $3\alpha_{1}$-close to the label
of the edge $\left(A_{1},B_{1}\right)$, which is $3\alpha_{1}$-close
to $\pi_{A_{1}}$. Thus, the label of $\left(A_{1}',B_{1}\right)$
is is $9\alpha_{1}$-close to $\pi_{A_{1}}$. Let us denote by $\pi_{\left(A_{1}',B_{1}\right)}$
the label of the edge $\left(A_{1}',B_{1}\right)$. Recall that by
the definition of $\pi_{\left(A_{1}',B_{1}\right)}$ it holds that
\begin{equation}
\Pr_{A_{2},B_{2}}\left[\mathcal{P}\aand\Pi\left(A_{1}',A_{2}\right)_{|A_{2}}\apx{\alpha_{1}}\pi_{\left(A_{1}',B_{1}\right)|A_{2}}\right]\ge\Omega\left(\varepsilon^{4}\right).\label{eq:double-P-test-agreement-of-assignment-with-edge-label}
\end{equation}
Since $\pi_{\left(A_{1}',B_{1}\right)}\apx{9\cdot\alpha_{1}}\pi_{A}$
it holds by Lemma~\ref{lem:subspace-point-sampler} (subspace-point
sampler) that for a uniformly distributed $d_{0}$-subspace $A_{2}$:
\[
\Pr_{A_{2}}\left[\pi_{\left(A_{1}',B_{1}\right)|A_{2}}\napx{10\cdot\alpha_{1}}\pi_{A_{1}|A_{2}}\right]\le\frac{1}{q^{d_{0}-2}\cdot\left(\alpha_{1}-q^{-d_{0}}\right)^{2}}.
\]
The latter expression can be made smaller than any constant times
$\varepsilon^{4}$ by choosing $h'$ to be sufficiently large. By
subtracting that expression from Inequality~\ref{eq:double-P-test-agreement-of-assignment-with-edge-label},
we obtain that 
\[
\Pr_{A_{2},B_{2}}\left[\mathcal{P}\aand\Pi\left(A_{1}',A_{2}\right)_{|A_{2}}\apx{\alpha_{1}}\pi_{\left(A_{1}',B_{1}\right)|A_{2}}\aand\pi_{\left(A_{1}',B_{1}\right)|A_{2}}\apx{10\cdot\alpha_{1}}\pi_{A_{1}|A_{2}}\right]\ge\Omega\left(\varepsilon^{4}\right).
\]
By letting $\pi_{2}=\pi_{A_{1}}$ and choosing $c'=28$, we have by
the triangle inequality 
\begin{equation}
\Pr_{A_{1}',A_{2}}\left[\mathcal{P}\aand\Pi\left(A_{1}',A_{2}\right)_{|A_{2}}\apx{11\cdot\alpha_{1}}\pi_{2|A_{2}}\right]\ge\Omega(\varepsilon^{24})\cdot\Omega\left(\varepsilon^{4}\right)=\Omega(\varepsilon^{c'}).\label{eq:one-side-consistency-only-for-As}
\end{equation}

\paragraph*{Step 4.}

It remains to show that the assignment $\Pi$ agrees with $\pi_{2}$
on a non-negligible fraction of the $B$'s. To this end, we observe
that
\begin{equation}
\Pr\left[\left.\mathcal{P}\aand\Pi\left(A_{1},A_{2}\right)_{|A_{2}}\apx{11\cdot\alpha_{1}}\pi_{2|A_{2}}\right|\Pi\left(B_{1},B_{2}\right)_{|B_{2}}\napx{12\cdot\alpha_{1}}\pi_{2|B_{2}}\right]\le\frac{1}{q^{d_{0}-2}\cdot\left(\alpha_{1}/2\right)^{2}}.\label{eq:one-side-consistency-for-As-and-fails-for-Bs}
\end{equation}
 To see it, note that it suffices to prove that
\[
\Pr\left[\left.\Pi\left(B_{1},B_{2}\right)_{|A_{2}}\apx{11\cdot\alpha_{1}}\pi_{2|A_{2}}\right|\Pi\left(B_{1},B_{2}\right)_{|B_{2}}\napx{12\cdot\alpha_{1}}\pi_{2|B_{2}}\right]\le\frac{1}{q^{d_{0}-2}\cdot\left(\alpha_{1}-q^{-d_{0}}\right)^{2}}\le\frac{1}{q^{d_{0}-2}\cdot\left(\alpha_{1}/2\right)^{2}}.
\]
 The latter inequality is an immediate corollary of Lemma~\ref{lem:subspace-point-sampler}
(subspace-point sampler).

Now, by choosing $h'$ to be sufficiently large so that the upper
bound in Inequality~\ref{eq:one-side-consistency-for-As-and-fails-for-Bs}
is sufficiently smaller than $\varepsilon^{c'}$, and by combining
Inequality~\ref{eq:one-side-consistency-only-for-As} with Inequality~\ref{eq:one-side-consistency-for-As-and-fails-for-Bs},
we obtain that
\[
\Pr\left[\mathcal{P}\aand\Pi\left(A_{1},A_{2}\right)_{|A_{2}}\apx{11\cdot\alpha_{1}}\pi_{2|A_{2}}\aand\Pi\left(B_{1},B_{2}\right)_{|B_{2}}\apx{12\cdot\alpha_{1}}\pi_{2|B_{2}}\right]\ge\Omega(\varepsilon^{c'}).
\]
 By setting $h'$ such that $\alpha'\ge12\cdot\alpha_{1}$ this concludes
the proof of Lemma~\ref{lem:one-side-consistency}.\qed

\subsubsection{\label{sub:P-squared-test-auxiliary-claims}Proofs of Auxiliary Claim}


\begin{myproof}
[Proof of Claim~\ref{claim:D1D2}.]Fix $\kappa\in\N$. In order
to prove the claim, consider the event $J$ which holds if and only
if $A$ and $A'$ are independent. We argue that
\[
\D_{1}\quad\stackrel{\delta/2}{\approx}\quad\D_{1}|J\quad=\quad\D_{2}|J\quad\stackrel{\delta/2}{\approx}\quad D_{2}.
\]
The fact that $\D_{1}|J=\mathcal{D}_{2}|J$ is exactly Proposition~\ref{pro:triplet-distributions-equivalent}.
We show that $\D_{1}\stackrel{\delta/2}{\approx}\D_{1}|J$ and $\D_{2}\stackrel{\delta/2}{\approx}\D_{2}|J$.
The statistical distance between $\D_{1}$ and $\D_{1}|J$ (respectively,
$\D_{2}$ and $\D_{2}|J$) is exactly the probability that the event
$J$ does not occur under $\D_{1}$ (respectively $\D_{2}$). It follows
immediately from Proposition~\ref{pro:random-subspaces-disjoint}
that $\Pr_{\mathcal{D}_{1}}\left[\neg J\right]\le2\cdot d_{0}/q^{d_{1}-2\cdot d_{0}}$
and $\Pr_{\mathcal{D}_{2}}\left[\neg J\right]\le2\cdot d_{0}/q^{m-2\cdot d_{0}}$.
Both the latter expressions can indeed be made smaller than $\varepsilon^{24}/\kappa$
by choosing sufficiently large $h'$, as required.
\end{myproof}

\subsection{\label{sub:Proof-of-S-test}The proof of Theorems~\ref{thm:S-test}
and~\ref{thm:S-test-list-decoding}}

In the rest of this section we prove Theorems~\ref{thm:S-test} and~\ref{thm:S-test-list-decoding}.
\begin{theorem*}
[\ref{thm:S-test}, the soundness of the S-test, restated] There
exists a universal constants $\hs,\cs\in\N$ such that the following
holds: Let $\varepsilon\ge\hs\cdot d_{0}\cdot q^{-d_{0}/\hs}$, $\as\eqdef\hs\cdot d_{0}\cdot q^{-d_{0}/\hs}$.
Assume that $d_{1}\ge\hs\cdot d_{0}$, $m\ge\hs\cdot d_{1}$. Suppose
that a (possible randomized) assignment $\Pi$ passes the S-test with
probability at least~$\varepsilon$. There exists an assignment $\pi:\F^{m}\to\Sigma$
for which the following holds. Let $B_{1}$, $B_{2}$ be uniformly
distributed and independent $d_{1}$-subspaces of $\F^{m}$, let $A_{1}$
and $A_{2}$ be uniformly distributed $d_{0}$-subspaces of $B_{1}$
and $B_{2}$ respectively, and denote $A=A_{1}+A_{2}$. Then: 
\[
\Pr\left[\Pi\left(B_{1},B_{2}\right)_{|\left(A_{1},A_{2}\right)}=\Pi\left(A\right)_{|\left(A_{1},A_{2}\right)}\aand\Pi\left(B_{1},B_{2}\right)\apx{\as}\pi_{|\left(B_{1},B_{2}\right)}\right]=\Omega\left(\varepsilon^{\cs}\right).
\]
\end{theorem*}
\begin{remark}
Note that in the foregoing restatement of Theorem~\ref{thm:S-test}
we denote the first universal constant by $h$, while in its original
statement it was denoted by $h'$.
\end{remark}
The intuition that underlies the proof is the following. Consider
an adversary the chooses the proof $\Pi$. Since the S-test essentially
contains a $P^{2}$-test, the adversary must choose the assignment
$\Pi$ such that for random $d_{0}$-subspaces $A_{1}$ and $A_{2}$,
the assignment $\Pi\left(A_{1}+A_{2}\right)_{|\left(A_{1},A_{2}\right)}$
is consistent with two assignments $\pi_{1}$, $\pi_{2}$ on $A_{1}$,
$A_{2}$ respectively. On the other hand, given the sum $A_{1}+A_{2}$,
the adversary can not deduce the choices of $A_{1}$ and $A_{2}$,
and therefore he must label both of $A_{1}$ and $A_{2}$ with the
same assignment in order to make the S-test accept. We conclude that
$\pi_{1}$ and $\pi_{2}$ must be essentially the same. Details follow.

Let $h'$ be the universal constant whose existence guaranteed in
Theorem~\ref{thm:P-squared-test} (soundness of the $P^{2}$-test),
and let $\alpha'$ be the corresponding value from Theorem~\ref{thm:P-squared-test}.
We choose $\cs$ to be the same constant as in Theorem~\ref{thm:P-squared-test},
and will choose the universal constant~$\hs$ to be at least~$h'$.

Fix an assignment $\Pi$ that passes the S-test with probability at
least~$\varepsilon$. We define a new assignment $\Pi'$ that assigns
values to pairs of $d_{0}$-subspaces and to pairs of $d_{1}$-subspaces
of $\F^{m}$ (not necessarily independent) by choosing $\Pi'\left(B_{1},B_{2}\right)$
(respectively $\Pi'\left(A_{1},A_{2}\right)$) to be equal to $\Pi\left(B_{1},B_{2}\right)$
(respectively $\Pi\left(A_{1}+A_{2}\right)$) if $B_{1}$ and $B_{2}$
(respectively $A_{1}$ and $A_{2}$) are independent, and choosing
$\Pi'$ to be arbitrary otherwise. Observe that the assignment $\Pi'$
passes the $P^{2}$-test whenever $B_{1}$ and $B_{2}$ are independent
and $\Pi$ passes the S-test. Furthermore, the probability that two
uniformly distributed $d_{1}$-subspaces $B_{1}$ and $B_{2}$ of
$\F^{m}$ are not independent is at most $d_{1}/q^{m-2\cdot d_{1}}$
by Proposition~\ref{pro:random-subspaces-disjoint}, and therefore
$\Pi'$ passes the $P^{2}$-test with probability at least $\varepsilon-d_{1}/q^{m-2\cdot d_{1}}$.
For a sufficiently large choice of $\hs$, the latter probability
is at least $\Omega\left(\varepsilon\right)$, and also matches the
requirements of Theorem~\ref{thm:P-squared-test} (soundness of the
$P^{2}$-test), so we can apply this theorem. It follows that there
exist assignments $\pi_{1},\pi_{2}:\F^{m}\to\Sigma$ such that for
uniformly distributed (not necessarily independent) $B_{1}$, $B_{2}$,
$A_{1}\subseteq B_{1}$, $A_{2}\subseteq B_{2}$ it holds that
\begin{eqnarray}
 &  & \Pr[\Pi'\left(B_{1},B_{2}\right)_{|\left(A_{1},A_{2}\right)}=\Pi'\left(A_{1},A_{2}\right)\label{eq:S-test-applying-P-squared-test}\\
 &  & \aand\Pi'\left(A_{1},A_{2}\right)\apx{\alpha'}\left(\pi_{1|A_{1}},\pi_{2|A_{2}}\right)\nonumber \\
 &  & \aand\Pi'\left(B_{1},B_{2}\right)\apx{\alpha'}\left(\pi_{1|B_{1}},\pi_{2|B_{2}}\right)]\nonumber \\
 & = & \Omega\left(\varepsilon^{c}\right).\nonumber 
\end{eqnarray}
The probability that $B_{1}$ and $B_{2}$ are not independent is
at most $d_{1}/q^{m-2\cdot d_{1}}$, and the latter expression can
be made smaller than any constant factor times $\varepsilon^{c}$
by choosing $\hs$ to be sufficiently large. Thus, Inequality~\ref{eq:S-test-applying-P-squared-test}
also holds for uniformly distributed \emph{independent} $B_{1}$ and
$B_{2}$. We now argue that
\begin{claim}
\label{cla:S-test-agreement-of-assignments}For sufficiently large
choice of $\hs$, it holds that $\pi_{1}\apx{5\cdot\alpha'}\pi_{2}$.
\end{claim}
We defer the proof of Claim~\ref{cla:S-test-agreement-of-assignments}
to the end of this section. We turn to prove the theorem. By Inequality~\ref{eq:S-test-applying-P-squared-test}
it holds for uniformly distributed and independent $d_{1}$-subspaces
$B_{1}$ and $B_{2}$ of $\F^{m}$ that
\[
\Pr\left[\Pi'\left(B_{1},B_{2}\right)_{|\left(A_{1},A_{2}\right)}=\Pi'\left(A_{1},A_{2}\right)\aand\Pi\left(B_{1},B_{2}\right)\apx{\alpha'}\left(\pi_{1|B_{1}},\pi_{2|B_{2}}\right)\right]\ge\Omega\left(\varepsilon^{c}\right).
\]
By Claim~\ref{cla:S-test-agreement-of-assignments} it holds that
$\pi_{1}\apx{5\cdot\alpha'}\pi_{2}$. Since $B_{2}$ is a uniformly
distributed $d_{1}$-subspace of $\F^{m}$, this implies by Lemma~\ref{lem:subspace-point-sampler}
(subspace-point sampler) that
\[
\Pr\left[\pi_{1|B_{2}}\apx{6\cdot\alpha'}\pi_{2|B_{2}}\right]\ge1-\frac{1}{q^{d_{1}-2}\cdot\left(\alpha'-q^{-d_{1}}\right)^{2}}\ge1-\frac{1}{q^{d_{1}-2}\cdot\left(\alpha'/2\right)^{2}}.
\]
We conclude that
\begin{eqnarray*}
 &  & \Pr\left[\Pi'\left(B_{1},B_{2}\right)_{|\left(A_{1},A_{2}\right)}=\Pi'\left(A_{1},A_{2}\right)\aand\Pi\left(B_{1},B_{2}\right)\apx{7\cdot\alpha'}\left(\pi_{1|B_{1}},\pi_{1|B_{2}}\right)\right]\\
 & \ge & \Pr\left[\Pi'\left(B_{1},B_{2}\right)_{|\left(A_{1},A_{2}\right)}=\Pi'\left(A_{1},A_{2}\right)\aand\Pi\left(B_{1},B_{2}\right)\apx{\alpha'}\left(\pi_{1|B_{1}},\pi_{2|B_{2}}\right)\aand\pi_{1|B_{2}}\apx{6\cdot\alpha'}\pi_{2|B_{2}}\right]\\
 & = & \Omega\left(\varepsilon^{c}\right)-\frac{1}{q^{d_{1}-2}\cdot\left(\alpha'/2\right)^{2}}\\
 & = & \Omega\left(\varepsilon^{c}\right),
\end{eqnarray*}
where the last equality holds for sufficiently large choice of $\hs$.
the theorem now follows by defining $\pi=\pi_{1}$ and setting $\hs$
to be sufficiently large such that $\alpha=7\cdot\alpha'.$\qed
\begin{myproof}
[Proof of Claim~\ref{cla:S-test-agreement-of-assignments}.]For
the sake of contradiction, assume that $\pi_{1}\napx{5\cdot\alpha'}\pi_{2}$.
Let $A$ be a uniformly distributed $2\cdot d_{0}$-subspace $A$
of $\F^{m}$ and let $A_{1}$ and $A_{2}$ be uniformly distributed
and independent $d_{0}$-subspaces of $A$. By Lemma~\ref{lem:subspace-point-sampler},
it holds that
\[
\Pr\left[\pi_{1|A}\napx{4\cdot\alpha'}\pi_{2|A}\right]\ge1-\frac{1}{q^{2\cdot d_{0}-2}\cdot\left(\alpha'-q^{-2d_{0}}\right)^{2}}\ge1-\frac{1}{q^{2\cdot d_{0}-2}\cdot\left(\alpha'/2\right)^{2}}.
\]
If $\pi_{1|A}\napx{4\cdot\alpha'}\pi_{2|A}$ then by the triangle
inequality it either holds that $\Pi\left(A\right)\napx{2\cdot\alpha'}\pi_{1|A}$
or that $\Pi\left(A\right)\napx{2\cdot\alpha'}\pi_{2|A}$. Since $A_{1}$
is a uniformly distributed $d_{0}$-subspace of $A$, it holds by
Lemma~\ref{lem:subspace-point-sampler} (subspace-point sampler)
that
\[
\Pr\left[\left.\Pi\left(A\right)_{|A_{1}}\napx{\alpha'}\pi_{1|A_{1}}\right|\Pi\left(A\right)\napx{2\cdot\alpha'}\pi_{1|A}\right]\ge1-\frac{1}{q^{2\cdot d_{0}-2}\cdot\left(\alpha'/2\right)^{2}}.
\]
A similar claim can be made for $\pi_{2}$ and $A_{2}$. Now, if either
$\Pi\left(A\right)_{|A_{1}}\napx{\alpha'}\pi_{1|A_{1}}$ or $\Pi\left(A\right)_{|A_{2}}\napx{\alpha'}\pi_{2|A_{2}}$
then by definition it holds that $\Pi\left(A\right)_{|\left(A_{1},A_{2}\right)}\napx{\alpha'}\left(\pi_{1|A_{1}},\pi_{2|A_{2}}\right)$.
We conclude that
\[
\Pr\left[\left.\Pi\left(A\right)_{|\left(A_{1},A_{2}\right)}\napx{\alpha'}\left(\pi_{1|A_{1}},\pi_{2|A_{2}}\right)\right|\pi_{1|A}\napx{4\cdot\alpha'}\pi_{2|A}\right]\ge1-\frac{1}{q^{2\cdot d_{0}-2}\cdot\left(\alpha'/2\right)^{2}},
\]
and therefore by lifting the conditioning and substituting $A=A_{1}+A_{2}$
we obtain that for a uniformly distributed and independent $d_{0}$-subspaces
$A_{1}$ and $A_{2}$ of $\F^{m}$ it holds that 
\[
\Pr\left[\Pi\left(A_{1}+A_{2}\right)_{|\left(A_{1},A_{2}\right)}\apx{\alpha'}\left(\pi_{1|A_{1}},\pi_{2|A_{2}}\right)\right]\le\frac{2}{q^{2\cdot d_{0}-2}\cdot\left(\alpha'/2\right)^{2}}.
\]

On the other hand, by the definition of $\Pi'$, Inequality~\ref{eq:S-test-applying-P-squared-test}
implies that for uniformly distributed and independent $d_{0}$-subspaces
$A_{1}$ and $A_{2}$ of $\F^{m}$ it holds that 
\[
\Pr\left[\Pi\left(A_{1}+A_{2}\right)_{|\left(A_{1},A_{2}\right)}\apx{\alpha'}\left(\pi_{1|A_{1}},\pi_{2|A_{2}}\right)\right]\ge\Omega\left(\varepsilon^{c}\right).
\]
By choosing $\hs$ to be sufficiently large, the latter lower bound
can be made larger than $2/\left(q^{2\cdot d_{0}-2}\cdot\left(\alpha'\right)^{2}\right)$,
and this is a contradiction. \end{myproof}
\begin{theorem}
[\ref{thm:S-test-list-decoding}, list-decoding soundness of the
S-test, restated]There exist universal constants $h,c\in\N$ such
that for every $d_{0}\in\N$, $d_{1}\ge h\cdot d_{0}$, and $m\ge h\cdot d_{1}$,
the following holds: Let $\varepsilon\ge h\cdot d_{0}\cdot q^{-d_{0}/h}$,
$\alpha\eqdef h\cdot d_{0}\cdot q^{-d_{0}/h}$. Let $\Pi$ be a (possibly
randomized) assignment to $2d_{0}$-subspaces of $\F^{m}$ and to
pairs of $d_{1}$-subspaces of $\F^{m}$. Then, there exists a (possibly
empty) list of $L=O\left(1/\varepsilon^{c}\right)$ assignments $\pi^{1},\ldots,\pi^{L}:\F^{m}\to\Sigma$
such that 

\[
\Pr\left[\Pi\left(B_{1},B_{2}\right)_{|\left(A_{1},A_{2}\right)}=\Pi\left(A\right)_{|\left(A_{1},A_{2}\right)}\aand\not\exists i\in\left[L\right]\mbox{ s.t. }\Pi\left(B_{1},B_{2}\right)\apx{\alpha}\pi_{|\left(B_{1},B_{2}\right)}^{i}\right]<\varepsilon
\]
\end{theorem}
\begin{remark}
Note that in the foregoing restatement of Theorem~\ref{thm:S-test-list-decoding}
we denote the first universal constant by $h$, while in its original
statement it was denoted by $h'$.
\end{remark}
The basic idea of the proof is as follows. We apply Theorem~\ref{thm:S-test}
to $\Pi$, thus ``decoding'' from it an assignment $\pi^{1}$. We
then remove from $\Pi$ the places at which it roughly agrees with
$\pi^{1}$, resulting in an assignment $\Pi^{2}$. If the assignment~$\Pi^{2}$
is accepted by the S-test with probability less than $\varepsilon$,
then we are finished - the required list of assignments in this case
consists only of $\pi^{1}$. Otherwise, the assignment $\Pi^{2}$
is accepted by the S-test with probability at least~$\varepsilon$,
and we can therefore ``decode'' a second assignment $\pi^{2}$ from
$\Pi^{2}$. Next, we remove from $\Pi^{2}$ the places at which it
roughly agrees with $\pi^{2}$, resulting in an assignment~$\Pi^{3}$.
We proceed in this manner, each time obtaining new assignments $\Pi^{i}$
and $\pi^{i}$, until the conclusion of Theorem~\ref{thm:S-test-list-decoding}
holds.

We prove Theorem~\ref{thm:S-test-list-decoding} only for non-randomized
assignments $\Pi$, but the proof can easily be extended to randomized
assignments, see Remark~\ref{rem:randomized-S-test-list-decoding}
for details. We choose the constants $h$ and $c$ to be the same
as in Theorem~\ref{thm:S-test}. If the S-test accepts $\Pi$ with
probability less than $\varepsilon$ then the theorem holds vacuously.
We thus assume that the S-test accepts $\Pi$ with probability at
least $\varepsilon$. We show that for $L=O\left(1/\varepsilon^{c}\right)$
there exist assignments $\pi^{1},\ldots,\pi^{L}:\F^{m}\to\Sigma$
such that
\begin{eqnarray}
 &  & \Pr\left[\Pi\left(B_{1},B_{2}\right)_{|\left(A_{1},A_{2}\right)}=\Pi\left(A\right)_{|\left(A_{1},A_{2}\right)}\right]\label{eq:list-decoding}\\
 &  & -\Pr\left[\Pi\left(B_{1},B_{2}\right)_{|\left(A_{1},A_{2}\right)}=\Pi\left(A\right)_{|\left(A_{1},A_{2}\right)}\aand\exists i\in\left[L\right]:\,\Pi\left(B_{1},B_{2}\right)\apx{\alpha}\pi_{|\left(B_{1},B_{2}\right)}^{i}\right]\nonumber \\
 & \le & \varepsilon.\nonumber 
\end{eqnarray}
We construct the assignments $\pi^{1},\ldots,\pi^{L}$ as follows.
We begin by applying Theorem~\ref{thm:S-test} to $\Pi$, obtaining
the assignment $\pi^{1}$, and set $\Pi^{1}\eqdef\Pi$. Then, for
each $i\ge1$ we define an assignment $\Pi^{i+1}$ as follows.
\begin{enumerate}
\item For every pair of $d_{1}$-subspaces $B_{1},B_{2}$ such that $\Pi^{i}\left(B_{1},B_{2}\right)\apx{\alpha}\pi_{|\left(B_{1},B_{2}\right)}^{i}$,
we set $\Pi^{i+1}\left(B_{1},B_{2}\right)=\bot$, where $\bot$ is
a special symbol that the test always rejects. This is our formal
way of ``removing'' $\Pi^{i}\left(B_{1},B_{2}\right)$.
\item For every pair of $d_{1}$-subspaces $B_{1},B_{2}$ such that $\Pi^{i}\left(B_{1},B_{2}\right)\napx{\alpha}\pi_{|\left(B_{1},B_{2}\right)}^{i}$,
we set $\Pi^{i+1}\left(B_{1},B_{2}\right)=\Pi^{i}\left(B_{1},B_{2}\right)$.
\item For every $2d_{0}$-subspace $A$, we set $\Pi^{i+1}\left(A\right)=\Pi^{i}\left(A\right)$.
\end{enumerate}
Now, observe that 
\begin{eqnarray}
 &  & \Pr\left[\Pi^{i+1}\left(B_{1},B_{2}\right)_{|\left(A_{1},A_{2}\right)}=\Pi^{i+1}\left(A\right)_{|\left(A_{1},A_{2}\right)}\right]\label{eq:list-decoding-iteration}\\
 & = & \Pr\left[\Pi^{i}\left(B_{1},B_{2}\right)_{|\left(A_{1},A_{2}\right)}=\Pi^{i}\left(A\right)_{|\left(A_{1},A_{2}\right)}\right]\nonumber \\
 &  & -\Pr\left[\Pi^{i}\left(B_{1},B_{2}\right)_{|\left(A_{1},A_{2}\right)}=\Pi^{i}\left(A\right)_{|\left(A_{1},A_{2}\right)}\wedge\Pi^{i}\left(B_{1},B_{2}\right)\apx{\alpha}\pi_{|\left(B_{1},B_{2}\right)}^{i}\right],\nonumber 
\end{eqnarray}
since we must have $\Pi^{i+1}\left(B_{1},B_{2}\right)_{|\left(A_{1},A_{2}\right)}\ne\Pi^{i+1}\left(A\right)_{|\left(A_{1},A_{2}\right)}$
whenever $\Pi^{i+1}\left(B_{1},B_{2}\right)_{|\left(A_{1},A_{2}\right)}=\bot$,
and the latter occurs whenever $\Pi^{i}\left(B_{1},B_{2}\right)\apx{\alpha}\pi_{|\left(B_{1},B_{2}\right)}^{i}$.
If $\Pr\left[\Pi^{i+1}\left(B_{1},B_{2}\right)_{|\left(A_{1},A_{2}\right)}=\Pi^{i+1}\left(A\right)_{|\left(A_{1},A_{2}\right)}\right]<\nolinebreak\varepsilon$
then we set $L=i$ and finish the construction. Otherwise, we construct
$\pi^{i+1}$ by applying Theorem~\ref{thm:S-test} to the assignment
$\Pi^{i+1}$ and setting $\pi^{i+1}$ to be the resulting assignment.

It is easy to prove by induction that for every $i\in\left[L\right]$
it holds that
\begin{eqnarray}
 &  & \Pr\left[\Pi^{i+1}\left(B_{1},B_{2}\right)_{|\left(A_{1},A_{2}\right)}=\Pi^{i+1}\left(A\right)_{|\left(A_{1},A_{2}\right)}\right]\label{eq:list-decoding-induction}\\
 & = & \Pr_{A\subseteq B}\left[\Pi\left(B_{1},B_{2}\right)_{|\left(A_{1},A_{2}\right)}=\Pi\left(A\right)_{|\left(A_{1},A_{2}\right)}\right]\nonumber \\
 &  & -\Pr_{A\subseteq B}\left[\Pi\left(B_{1},B_{2}\right)_{|\left(A_{1},A_{2}\right)}=\Pi\left(A\right)_{|\left(A_{1},A_{2}\right)}\aand\exists i\in\left[L\right]:\,\Pi_{i}\left(B_{1},B_{2}\right)\apx{\alpha}\pi_{|\left(B_{1},B_{2}\right)}^{i}\right].\nonumber 
\end{eqnarray}
The proof of the Equality~\ref{eq:list-decoding-induction} goes
essentially by summing over the probabilities of events of the form
\[
\Pi^{i}\left(B_{1},B_{2}\right)_{|\left(A_{1},A_{2}\right)}=\Pi^{i}\left(A\right)_{|\left(A_{1},A_{2}\right)}\aand\Pi^{i}\left(B_{1},B_{2}\right)\apx{\alpha}\pi_{|\left(B_{1},B_{2}\right)}^{j}\aand\not\exists j<i\mbox{ s.t. }\Pi^{j}\left(B_{1},B_{2}\right)\apx{\alpha}\pi_{|\left(B_{1},B_{2}\right)}^{j},
\]
for different values of $i$.

Finally, by combining Equality~\ref{eq:list-decoding-induction}
with the fact that 
\[
\Pr\left[\Pi^{L+1}\left(B_{1},B_{2}\right)_{|\left(A_{1},A_{2}\right)}=\Pi^{L+1}\left(A\right)_{|\left(A_{1},A_{2}\right)}\right]<\varepsilon,
\]
it follows that the assignments $\pi^{1},\ldots,\pi^{L}$ satisfy
Inequality~\ref{eq:list-decoding}. To see that $L=O\left(1/\varepsilon^{c}\right)$,
observe that for each $i$ we have that
\[
\Pr\left[\Pi^{i}\left(B_{1},B_{2}\right)_{|\left(A_{1},A_{2}\right)}=\Pi^{i}\left(A\right)_{|\left(A_{1},A_{2}\right)}\aand\Pi^{i}\left(B_{1},B_{2}\right)\apx{\alpha}\pi_{|\left(B_{1},B_{2}\right)}^{i}\right]=\Omega\left(\varepsilon^{c}\right).
\]
By Equality~\ref{eq:list-decoding-iteration}, this implies that
the acceptance probability of~$\Pi^{i+1}$ is smaller than the acceptance
probability of $\Pi^{i}$ by at least $\varepsilon^{c}$, and therefore
that the number of iterations can be at most $O\left(1/\varepsilon^{c}\right)$,
as required.
\begin{remark}
As in the proof of Theorem~\ref{thm:P-squared-test} (soundness of
the $P^{2}$-test), the use of the special symbol~$\bot$ requires
formal justification. This can be done as explained in Remark~\ref{rem:justifying_bot}.
\end{remark}

\begin{remark}
\label{rem:randomized-S-test-list-decoding}As in the proof of Theorem~\ref{thm:P-squared-test}
(soundness of the $P^{2}$-test), if $\Pi$ is randomized, then for
each $i$ the definition of $\Pi^{i+1}$ should be slightly changed
to consider the internal randomness of $\Pi^{i}$. That is, we define
$\Pi^{i+1}$ to be a randomized assignment, and obtain it from $\Pi$
as follows. For every pair $\left(B_{1},B_{2}\right)$ and every internal
randomness $\omega$ of $\Pi^{i}$, let us denote by $\left(b_{1},b_{2}\right)$
the output of $\Pi_{i}$ on $\left(B_{1},B_{2}\right)$ and randomness~$\omega$.
We define the output of $\Pi^{i+1}$ on $\left(B_{1},B_{2}\right)$
and randomness $\omega$ to be $\bot$ if~$\left(b_{1},b_{2}\right)\apx{\alpha'}\pi_{|\left(B_{1},B_{2}\right)}^{i}$,
and define it to be $\left(b_{1},b_{2}\right)$ otherwise. The definition
for $2d_{0}$-spaces $A$ can be changed similarly to include the
internal randomness of $\Pi^{i}$.\qed\end{remark}
\begin{acknowledgement*}
We would like to thank Eli Ben Sasson for a useful discussion, and
to anonymous referees for comments that improved the presentation
of this work.
\end{acknowledgement*}

\newcommand{\etalchar}[1]{$^{#1}$}

\appendix

\section{\label{sec:Proof-of-IKW}Proof of Theorem~\ref{thm:IKW}, soundness
of the P-test}

In this section we prove Theorem~\ref{thm:IKW}, restated below,
by adapting the analysis of~\cite{IKW09} (in particular, Sections~3.4
and~4) to the setting of the $P$-test, while relying on a lemma
of~\cite{IKW09}. Let $\F$ be a finite field of size $q$, let $m,d_{0},d_{1}\in\N$,
and consider a (possible randomized) assignment $\Pi$ that assigns
values to $d_{0}$- and $d_{1}$-subspaces of $\F^{m}$.
\begin{theorem}
[\ref{thm:IKW}, soundness of the P-test, restated]There exists
a universal constant $\hp\in\N$ such that the following holds: Let
$\varepsilon\ge\hp\cdot d_{0}\cdot q^{-d_{0}/\hp}$, $\ap\eqdef\hp\cdot d_{0}\cdot q^{-d_{0}/\hp}$.
Assume that $d_{1}\ge\hp\cdot d_{0}$, $m\ge\hp\cdot d_{1}$. Suppose
that an assignment $\Pi$ passes the P-test with probability at least
$\varepsilon$. Then, there exists an assignment $\pi$ such that
\[
\Pr\left[\Pi\left(B\right)_{|A}=\Pi\left(A\right)\aand\Pi\left(B\right)\apx{\ap}\pi_{|B}\aand\Pi\left(A\right)\apx{\ap}\pi_{|A}\right]=\Omega(\varepsilon^{4}),
\]
where the probability is over $A,B$ chosen as in the $P$-test.
\end{theorem}
We begin by recalling the required preliminaries from~\cite{IKW09},
and then turn to prove Theorem~\ref{thm:IKW}.
\begin{definition}
[\label{def:goodness}Good]Let $A$ be a $d_{0}$-subspace of $\F^{m}$
and let $\varepsilon\in\left(0,1\right)$. We say that $A$ is \emph{$\eps$-good}
(with respect to an assignment $\Pi$) if for a uniformly distributed
$d_{1}$-dimensional subspace $B$ that contains $A$ it holds that
\[
\Pr\left[\Pi\left(B\right)_{|A}=\Pi\left(A\right)\right]\ge\varepsilon,
\]
where the randomness is over the choice of $B$ and over the randomness
of $\Pi$.
\end{definition}

\begin{definition}
[\label{def:plurality}Plurality function]Let $A$ be a $d_{0}$-subspace
of $\F^{m}$. We denote by $\pi_{A}:\F^{m}\to\Sigma$ the \emph{plurality
function} of~$A$ (with respect to $\Pi$). In other words, for every
$x\in\F^{m}$ we define $\pi_{A}(x)$ to be the value $v\in\Sigma$
that maximizes 
\[
\Pr_{B\supseteq A}\left[\Pi\left(B\right)_{|x}=v\left|\Pi\left(B\right)_{|A}=\Pi\left(A\right)\right.\right],
\]
where $B$ is a uniformly distributed $d_{1}$-dimensional subspace
that contains $A$.
\end{definition}

\begin{definition}
[\label{def:DP-consistent}DP-consistent]Let $A$ be a $d_{0}$-subspace
of $\F^{m}$ and let $\alpha,\gamma\in\left(0,1\right)$. We say that
$A$ is $\left(\varepsilon,\alpha,\gamma\right)$\emph{-direct product
consistent }(abbreviated $\left(\varepsilon,\alpha,\gamma\right)$\emph{
-DP-consistent}) if $A$ is $\varepsilon$-good and it holds that
\[
\Pr_{B\supseteq A}\left[\Pi\left(B\right)\apx{\alpha}\pi_{A|B}\left|\Pi\left(B\right)_{|A}=\Pi\left(A\right)\right.\right]\ge1-\gamma.
\]

\end{definition}
The following lemma is a direct corollary of the proofs of \cite[Lemma 4.2]{IKW09}
and \cite[Lemma 4.4]{IKW09}.
\begin{lemma}
\label{lem:DP-consistent}There exists a universal constant $h_{0}\in\N$
such that the following holds: Let $\varepsilon\ge h_{0}\cdot q^{-\left(d_{1}/h_{0}-d_{0}\right)}$
and $\alpha,\gamma\in\left(0,1\right)$. The probability that a uniformly
distributed $A$ is $\eps$-good but not $\left(\varepsilon,\alpha,\gamma\right)$-DP-consistent
is at most O$\left(1/\left(\alpha\cdot\gamma\cdot\varepsilon^{2}\cdot q^{d_{0}-2}\right)\right)$.
\end{lemma}

\subsection*{Proof of Theorem~\ref{thm:IKW}}

We will choose the universal constant $\hp$ to be larger than $h_{0}$
(where $h_{0}$ is the constant from Lemma~\ref{lem:DP-consistent}).
Assume that the P-test accepts with probability at least $\eps$ as
in the statement of the theorem. Let $\eps_{1}=\frac{1}{3}\cdot\varepsilon$
and $\gamma_{1}=\varepsilon_{1}^{3}/\hp$ . Choose $\alpha_{1}=O\left(1/\eps_{1}^{3}\cdot\gamma_{1}\cdot q^{d_{0}-2}\right)$
such that the probability in Lemma~\ref{lem:DP-consistent} that
$A$ is $\eps_{1}$-good but not $\left(\eps_{1},\alpha_{1},\gamma_{1}\right)$-DP-consistent
is at most~$\eps_{1}$, which is indeed possible for sufficiently
large choice of $\hp$. We will later choose $\ap=O\left(\alpha_{1}\right)$,
by choosing again $h$ to be sufficiently large.

We consider the following sequence of events. Let $A_{1},A_{2}$ denote
random $d_{0}$-subspaces, and let $B$ denote a random $d_{1}$-subspace,
and define events $\mathcal{S}_{1},\mathcal{S}_{2},\mathcal{S}_{3}$
as follows: 
\begin{enumerate}
\item $\mathcal{S}_{1}(A_{1},A_{2},B):$ $A_{1}$ and $A_{2}$ are $\left(\eps_{1},\alpha_{1},\gamma_{1}\right)$-DP-consistent
and $\Pi\left(B\right)_{|A_{1}}=\Pi\left(A_{1}\right)$, $\Pi\left(B\right)_{|A_{2}}=\Pi\left(A_{2}\right)$. 
\item $\mathcal{S}_{2}(A_{1},A_{2},B):$ The event $\mathcal{S}_{1}\left(A_{1},A_{2},B\right)$
occurs and $\pi_{A_{1}|B}\apx{2\alpha_{1}}\pi_{A_{2}|B}$ (recall
that $\pi_{A_{1}}$ and $\pi_{A_{2}}$ are the plurality assignments
of $A_{1}$ and $A_{2}$ respectively). 
\item $\mathcal{S}_{3}(A_{1},A_{2})$: $A_{1}$ and $A_{2}$ are $\left(\eps_{1},\alpha_{1},\gamma_{1}\right)$-DP-consistent
and $\pi_{A_{1}}\apx{3\alpha_{1}}\pi_{A_{2}}$. 
\end{enumerate}
In the next three claims we choose $A_{1}$, $A_{2}$ and $B$ according
to the following distribution: choose $A_{1}$ and $A_{2}$ to be
uniformly distributed and independent $d_{0}$-spaces $A_{1},A_{2}$,
and a choose $B$ to be a uniformly distributed $d_{1}$-subspace
that contains them. We show that the probability of events $\mathcal{S}_{1},\mathcal{S}_{2},\mathcal{S}_{3}$
under this distribution is non-negligible.
\begin{claim}
\label{cla:double-DP-consistency} $Pr[\mathcal{S}_{1}]\ge\Omega\left(\eps_{1}^{3}\right)$.\end{claim}
\begin{myproof}
Let $B'$ be a uniformly distributed $d_{1}$-subspace of $\F^{m}$
and let $A'$ be a $d_{0}$-uniformly distributed subspace of $B'$.
We begin by lower bounding the probability
\begin{equation}
\Pr\left[\Pi\left(B'\right)_{|A'}=\Pi\left(A'\right)\aand A'\mbox{ is }\left(\eps_{1},\alpha_{1},\gamma_{1}\right)\mbox{-DP-consistent}\right].\label{eq:P-test-passes-and-DP-consistent}
\end{equation}
To this end, let us denote by $\mathcal{P}$ the event that $\Pi\left(B'\right)_{|A'}=\Pi\left(A'\right)$,
by $\mathcal{D}$ the event that $A'$ is $\left(\eps_{1},\alpha_{1},\gamma_{1}\right)$-DP-consistent,
and by $\mathcal{G}$ the event that $A'$ is $\eps_{1}$-good. Observe
that$\Pr\left[\mathcal{\mathcal{P}}\aand\neg\mathcal{G}\right]\le\Pr\left[\mathcal{P}|\neg\mathcal{G}\right]\le\eps_{1}$.
Furthermore, $A'$ is a uniformly distributed $d_{0}$-subspace of
$\F^{m}$ and thus by Lemma~\ref{lem:DP-consistent} and our choice
of $\alpha_{1}$, it holds that $\Pr\left[\mathcal{G}\aand\neg\mathcal{D}\right]\le\eps_{1}$.
Finally, it holds that the probability in~(\ref{eq:P-test-passes-and-DP-consistent})
is
\begin{eqnarray*}
\Pr\left[\mathcal{P}\aand\mathcal{D}\right] & \ge & \Pr\left[\mathcal{P}\aand\mathcal{G}\aand\mathcal{D}\right]\\
 & = & \Pr\left[\mathcal{P}\aand\mathcal{G}\right]-\Pr\left[\mathcal{P}\aand\mathcal{G}\aand\neg\mathcal{D}\right]\\
 & = & \Pr\left[\mathcal{P}\right]-\Pr\left[\mathcal{P}\aand\neg\mathcal{G}\right]-\Pr\left[\mathcal{P}\aand\mathcal{G}\aand\neg\mathcal{D}\right]\\
 & \ge & \Pr\left[\mathcal{P}\right]-\Pr\left[\mathcal{P}\aand\neg\mathcal{G}\right]-\Pr\left[\mathcal{G}\aand\neg\mathcal{D}\right]\\
 & \ge & \varepsilon-\eps_{1}-\eps_{1}\\
 & \ge & \eps_{1}.
\end{eqnarray*}
So the probability in (\ref{eq:P-test-passes-and-DP-consistent})
is at least $\eps_{1}$. By averaging, this implies that for $\Omega\left(\eps_{1}\right)$~fraction
of the $d_{1}$-subspaces $B'$ it holds that at least $\Omega\left(\eps_{1}\right)$~fraction
of the $d_{0}$-subspaces $A'$ of $B'$ are $\left(\eps_{1},\alpha_{1},\gamma_{1}\right)$-DP-consistent
and satisfy $\Pi\left(B'\right)_{|A'}=\Pi\left(A'\right)$.

Now, observe that by Proposition~\ref{pro:triplet-distributions-equivalent},
the distribution over $A_{1}$, $A_{2}$, $B$ is equivalent to choosing
$B$ to be a uniformly distributed $d_{1}$-subspace of $\F^{m}$
and then choosing $A_{1}$ and $A_{2}$ to be independent uniformly
distributed $d_{0}$-subspaces of $B$. With probability at least
$\Omega\left(\eps_{1}\right)$ it holds for $B$ that at least $\Omega\left(\eps_{1}\right)$~fraction
of the $d_{0}$-subspaces $A$ of $B$ are $\left(\eps_{1},\alpha_{1},\gamma_{1}\right)$-DP-consistent
and satisfy $\Pi\left(B\right)_{|A}=\Pi\left(A\right)$. We condition
on the latter event, and claim that under this conditioning the event
$\mathcal{S}_{1}(A_{1},A_{2},B)$ occurs with probability at least
$\Omega\left(\eps_{1}^{2}\right)$. To see it, consider two uniformly
distributed (\emph{not necessarily independent}) $d_{0}$-subspaces
$A_{1}'$ and $A_{2}'$ of~$B$. Then, by our conditioning, it holds
that $\mathcal{S}_{1}(A_{1}',A_{2}',B)$ occurs with probability at
least $\Omega\left(\eps_{1}^{2}\right)$. Furthermore, by Proposition~\ref{pro:random-subspaces-disjoint}
it holds with probability at least $1-2\cdot d_{0}/q^{d_{1}-2\cdot d_{0}}$
that $A_{1}'$ and $A_{2}'$ are independent. It therefore follows
under the foregoing conditioning on $B$ that
\begin{eqnarray*}
\Pr\left[\mathcal{S}_{1}(A_{1},A_{2},B)\right] & = & \Pr\left[\left.\mathcal{S}_{1}(A_{1}',A_{2}',B)\right|A_{1}',A_{2}'\mbox{ are disjoint}\right]\\
 & \ge & \Pr\left[\mathcal{S}_{1}(A_{1}',A_{2}',B)\and A_{1}',A_{2}'\mbox{ are disjoint}\right]\\
 & \ge & \Pr\left[\mathcal{S}_{1}(A_{1}',A_{2}',B)\right]-\Pr\left[A_{1}',A_{2}'\mbox{ are disjoint}\right]\\
 & \ge & \Omega\left(\varepsilon_{1}^{2}\right)-2\cdot d_{0}/q^{d_{1}-2\cdot d_{0}}\\
 & \ge & \Omega\left(\eps_{1}^{2}\right),
\end{eqnarray*}
where the last inequality holds for sufficiently large $\hp$. Lifting
the conditioning on $B$, we get that for a uniformly distributed
$d_{1}$-subspace $B$ of $\F^{m}$ and two independent uniformly
distributed $d_{0}$-subspaces $A_{1}$ and $A_{2}$ of $B$, it holds
with probability at least $\Omega\left(\eps_{1}^{3}\right)$ that
both $A_{1}$ and $A_{2}$ are $\left(\eps_{1},\alpha_{1},\gamma_{1}\right)$-DP-consistent
and that $\Pi\left(B\right)_{|A_{1}}=\Pi\left(A_{1}\right)$, $\Pi\left(B\right)_{|A_{2}}=\Pi\left(A_{2}\right)$,
as required.\end{myproof}
\begin{claim}
\label{cla:agreement-on-B}$\Pr[\mathcal{S}_{2}]\ge\Omega\left(\eps_{1}^{3}\right)$.\end{claim}
\begin{myproof}
Let $\mathcal{E}_{1}$ be the event in which $A_{1}$ is $\left(\eps_{1},\alpha_{1},\gamma_{1}\right)$-DP-consistent,
$\Pi\left(B\right)_{|A_{1}}=\Pi\left(A_{1}\right)$ and $\Pi\left(B\right)\napx{\alpha_{1}}\pi_{A_{1}|B}$,
and let $\mathcal{E}_{2}$ be the corresponding event for $A_{2}$.
We begin by noting that the probabilities of both $\mathcal{E}_{1}$
and $\mathcal{E}_{2}$ are upper bounded by~$\gamma_{1}$. To see
it for $\mathcal{E}_{1}$, note that conditioned on $A_{1}$ being
$\left(\eps_{1},\alpha_{1},\gamma_{1}\right)$-DP-consistent and on
$\Pi\left(B\right)_{|A_{1}}=\Pi\left(A_{1}\right)$ it holds that
$B$ is a uniformly distributed $d_{1}$-subspace satisfying $\Pi\left(B\right)_{|A_{1}}=\Pi\left(A_{1}\right)$,
and therefore it holds that $\Pi\left(B\right)\napx{\alpha_{1}}\pi_{A_{1}|B}$
with probability at most $\gamma_{1}$ (by the DP-consistency of $A_{1}$).
The probability of $\mathcal{E}_{2}$ can be upper bounded similarly.

It now follows by Claim~\ref{cla:double-DP-consistency} that
\begin{eqnarray*}
\Pr\left[\mathcal{S}_{2}\right] & = & \Pr\left[\mathcal{S}_{1}\aand\pi_{A_{1}|B}\apx{2\alpha_{1}}\pi_{A_{2}|B}\right]\\
 & \ge & \Pr\left[\mathcal{S}_{1}\aand\neg\mathcal{E}_{1}\aand\neg\mathcal{E}_{2}\right]\\
 & \ge & \Pr\left[\mathcal{S}_{1}\right]-\Pr\left[\mathcal{E}_{1}\right]-\Pr\left[\mathcal{E}_{2}\right]\\
 & \ge & \Omega\left(\varepsilon_{1}^{3}\right)-2\cdot\gamma_{1}\\
 & \ge & \Omega\left(\varepsilon_{1}^{3}\right),
\end{eqnarray*}
where the last inequality holds for sufficiently large choice of $\hp$.
The required result follows.\end{myproof}
\begin{claim}
\label{cla:agreement-of-pluralities} $\Pr[\mathcal{S}_{3}]\ge\Omega\left(\eps_{1}^{3}\right)$.\end{claim}
\begin{myproof}
Let us say that $A_{1}$ and $A_{2}$ are ``agree on a random $B$''
if both $A_{1}$ and $A_{2}$ are $\left(\eps_{1},\alpha_{1},\gamma_{1}\right)$-DP-consistent
and $\Pr_{B\supset A_{1},A_{2}}\left[\pi_{A_{1}|B}\apx{2\cdot\alpha_{1}}\pi_{A_{2}|B}\right]\ge\Omega\left(\eps_{1}^{3}\right)$.
By Claim~\ref{cla:agreement-on-B} and by averaging, we know that
with probability at least $\Omega\left(\eps_{1}^{3}\right)$ it holds
that $A_{1}$ and $A_{2}$ agree on a random $B$. We show that for
every $A_{1}$ and $A_{2}$ that are $\left(\eps_{1},\alpha_{1},\gamma_{1}\right)$-DP-consistent
such that $\pi_{A_{1}}\napx{3\cdot\alpha_{1}}\pi_{A_{2}}$ it holds
that $A_{1}$ and $A_{2}$ do not agree on a random $B$. This will
imply that if $A_{1}$ and $A_{2}$ agree on a random $B$ then it
must hold that $\pi_{A_{1}}\apx{3\cdot\alpha_{1}}\pi_{A_{2}}$. Since
we know that the probability of $A_{1}$ and $A_{2}$ to agree on
a random~$B$ is at least $\Omega\left(\eps_{1}^{3}\right)$ the
required result will follow.

Fix $A_{1}$ and $A_{2}$ to be any $\left(\eps_{1},\alpha_{1},\gamma_{1}\right)$-DP-consistent
independent $d_{0}$-subspaces such that $\pi_{A_{1}}\napx{3\cdot\alpha_{1}}\pi_{A_{2}}$.
Now, by Lemma~\ref{lem:subspace-point-sampler} (subspace-point sampler)
and by sufficiently large choice of $\hp$, the probability that a
uniformly distributed $d_{1}$-subspace $B$ that contains $A_{1}$
and $A_{2}$ contains at most $2\cdot\alpha_{1}\le3\cdot\alpha_{1}-1/q^{d_{0}-2}-1/q^{d_{1}-2\cdot d_{0}}$~fraction
of coordinates on which $\pi_{A_{1}}$ and $\pi_{A_{2}}$ disagree
is at most $1/\left(q^{d_{1}-4\cdot d_{0}-6}\right)$, and the latter
expression can be made smaller than any constant factor times~$\varepsilon_{1}^{3}$.
Thus, it holds that $\Pr_{B\supset A_{1},A_{2}}\left[\pi_{A_{1}|B}\apx{2\cdot\alpha_{1}}\pi_{A_{2}|B}\right]$
can be made sufficiently small such that $A_{1}$ and $A_{2}$ do
not agree on a random $B$, as required.
\end{myproof}
We now find a global assignment $\pi$ and show that it agrees with
$\Pi$ on many $B$'s, and then on many $A$'s.
\begin{claim}
There exists an assignment $\pi:\F^{m}\to\Sigma$ such that $\Pr_{B}[\Pi(B)\apx{5\cdot\alpha_{1}}\pi_{|B}\aand\Pi\left(B\right)_{|A}=\Pi\left(A\right)]\ge\Omega\left(\eps_{1}^{4}\right)$.\end{claim}
\begin{myproof}
By Claim~\ref{cla:agreement-of-pluralities} and by averaging, we
get that for at least $\Omega\left(\eps_{1}^{3}\right)$~fraction
of the $d_{0}$-subspaces $A_{1}$ it holds that $A_{1}$ is $\left(\eps_{1},\alpha_{1},\gamma_{1}\right)$-DP-consistent
and 
\[
\Pr_{A_{2}:A_{2}\mbox{ is disjoint from }A_{1}}\left[A_{2}\mbox{ is }\left(\eps_{1},\alpha_{1},\gamma_{1}\right)\mbox{-DP-consistent and }\pi_{A_{1}}\apx{3\cdot\alpha_{1}}\pi_{A_{2}}\right]\ge\Omega\left(\eps_{1}^{3}\right)
\]
 Fix such $d_{0}$-subspace $A_{1}$, and set $\pi=\pi_{A_{1}}$.
Consider choosing a uniformly distributed $d_{0}$-space $A_{2}$
and a uniformly distributed $d_{1}$-space $B\supset A_{2}$. We show
that $\Pi(B)\apx{5\cdot\alpha_{1}}\pi_{|B}$ with probability at least~$\Omega\left(\eps_{1}^{4}\right)$.

Let us denote by $\mathcal{D}$ the event in which $A_{2}$ is independent
from $A_{1}$, by $\mathcal{P}$ the event in which $\Pi\left(B\right)_{|A_{2}}=\Pi\left(A_{2}\right)$,
and by $\mathcal{C}$ the event in which $A_{2}$ is $\left(\eps_{1},\alpha_{1},\gamma_{1}\right)$-DP-consistent
and $\pi_{A_{1}}\apx{3\cdot\alpha_{1}}\pi_{A_{2}}$.

By Proposition \ref{pro:random-subspaces-disjoint}, it holds that
$\Pr\left[\mathcal{D}\right]\ge1-2\cdot d_{0}/q^{m-2\cdot d_{0}}\ge\frac{1}{2}$
(where the second inequality holds for sufficiently large~$\hp$).
Furthermore, conditioned on $\mathcal{D}$, the subspace $A_{2}$
is a uniformly distributed $d_{0}$-subspace of $\F^{m}$ \emph{that
is independent from $A_{1}$}, and thus by the choice of $A_{1}$
it holds that $\Pr\left[\mathcal{C}|\mathcal{D}\right]\ge\Omega\left(\eps_{1}^{3}\right)$.
Lifting the conditioning, it follows that $\Pr\left[\mathcal{C}\right]\ge\Omega\left(\eps_{1}^{3}\right)$.
Next, observe that $B$ is distributed uniformly over the $d_{1}$-subspaces
that contain $A_{2}$, and thus (since in particular $A_{2}$ is $\eps_{1}$-good)
$\Pr\left[\mathcal{P}|\mathcal{C}\right]\ge\eps_{1}$. It therefore
holds that $\Pr\left[\mathcal{C}\aand\mathcal{P}\right]\ge\Omega\left(\varepsilon_{1}^{4}\right)$

Now, let us condition on the events $\mathcal{C}$ and $\mathcal{P}$.
By Lemma~\ref{lem:subspace-point-sampler} (subspace-point sampler)
and for sufficiently large $\hp$, it holds with probability at least
$1-1/\left(q^{d_{1}-3\cdot d_{0}-6}\right)\ge\frac{3}{4}$ that $B$
contains at most $4\cdot\alpha_{1}\ge3\alpha_{1}+1/q^{d_{0}-2}+1/q^{d_{1}-2\cdot d_{0}}$~fraction
of coordinates on which $\pi_{A_{1}}$ and $\pi_{A_{2}}$ disagree.
Furthermore, by the DP-consistency of $A_{2}$ and for sufficiently
large choice of $\hp$, it holds with probability at least~$1-\gamma_{1}\ge\frac{3}{4}$
that $\Pi\left(B\right)\apx{\alpha_{1}}\pi_{A_{2}|B}$. By the union
bound and the triangle inequality, it follows that with probability
at least~$\frac{1}{2}$ it holds that $\Pi\left(B\right)$ disagrees
with $\pi_{A_{1}|B}$ on at most $5\cdot\alpha_{1}$~fraction of
the coordinates. Lifting the conditioning on $\mathcal{C}$ and $\mathcal{P}$,
we obtain that with probability at least $\Omega\left(\eps_{1}^{4}\right)$
it holds that $\Pi\left(B\right)\apx{5\cdot\alpha_{1}}\pi_{A_{1}|B}$,
and $\Pi\left(B\right)=\Pi\left(A\right)$ as required.
\end{myproof}
Finally, we turn to prove the theorem. Let $\pi$ be the assignment
whose existence is guaranteed by the previous claim. Let us denote
by $\mathcal{P}$ the event in which $\Pi\left(B\right)_{|A}=\Pi\left(A\right)$
(i.e., the P-test accepts $A$ and $B$), by $\mathcal{E}_{1}$ the
event in which $\Pi(B)\apx{5\cdot\alpha_{1}}\pi_{|B}$, by $\mathcal{E}_{2}$
the event in which $\Pi(A)\apx{6\cdot\alpha_{1}}\pi_{|A}$, and by
$\mathcal{E}_{3}$ the event in which $\Pi\left(B\right)_{|A}\apx{6\cdot\alpha_{1}}\pi_{|A}$.
Using this notation, it suffices to prove that
\[
\Pr\left[\mathcal{P}\aand\mathcal{E}_{1}\aand\mathcal{E}_{2}\right]=\Omega\left(\eps_{1}^{4}\right).
\]
 By the definition of $\pi$, it holds that 
\[
\Pr\left[\mathcal{P}\aand\mathcal{E}_{1}\right]=\Omega\left(\eps_{1}^{4}\right).
\]
 The subspace $A$ is a uniformly distributed $d_{0}$-subspace of
$B$, and therefore it holds by Lemma~\ref{lem:subspace-point-sampler}
(subspace-point sampler) that
\[
\Pr\left[\neg\mathcal{E}_{3}\left|\mathcal{E}_{1}\right.\right]=O\left(1/q^{d_{0}/2-2}\right).
\]
 This implies that
\begin{eqnarray*}
\Pr\left[\mathcal{P}\aand\mathcal{E}_{1}\aand\mathcal{E}_{3}\right] & = & \Pr\left[\mathcal{P}\aand\mathcal{E}_{1}\right]-\Pr\left[\mathcal{P}\aand\mathcal{E}_{1}\aand\neg\mathcal{E}_{3}\right]\\
 & \ge & \Pr\left[\mathcal{P}\aand\mathcal{E}_{1}\right]-\Pr\left[\neg\mathcal{E}_{3}|\mathcal{E}_{1}\right]\\
 & = & \Omega\left(\eps_{1}^{4}\right)-O\left(1/q^{d_{0}/2-2}\right)\\
 & = & \Omega\left(\eps_{1}^{4}\right),
\end{eqnarray*}
where the last inequality holds for sufficiently large $\hp$. Now,
observe that whenever both the events $\mathcal{P}$ and $\mathcal{E}_{3}$
occur, the event $\mathcal{E}_{2}$ also occurs. It follows that 
\[
\Pr\left[\mathcal{P}\aand\mathcal{E}_{1}\aand\mathcal{E}_{2}\right]\ge\Pr\left[\mathcal{P}\aand\mathcal{E}_{1}\aand\mathcal{E}_{3}\right]=\Omega\left(\eps_{1}^{4}\right),
\]
 as required.\qed

\section{\label{sec:Routing-on-deBruijn}Routing on de~Bruijn graphs}

In this section we prove the routing property of de~Bruijn graph
given in Fact~\ref{fac:routing}. Recall the following.
\begin{definition*}
[\ref{def:deBruijn}, restated]Let $\Lambda$ be a finite alphabet
and let $m\in\N$. The \emph{de~Bruijn graph} $\DBs$ is the directed
graph whose vertices set is $\Lambda^{m}$ such that each vertex $\left(\alpha_{1},\ldots,\alpha_{t}\right)\in\Lambda^{m}$
has outgoing edges to all the vertices of the form $\left(\alpha_{2},\ldots,\alpha_{t},\beta\right)$
for $\beta\in\Lambda$.\end{definition*}
\begin{fact*}
[\ref{fac:routing}, restated]Let $\DBs$ be a de-Bruijn graph.
Then, given a permutation $\mu$ on the vertices of $\DBs$ one can
find a set of undirected paths of length $l=2m$ which connect each
vertex $v$ to $\mu(v)$ and which have the following property: For
every $j\in\left[l\right]$, each vertex $v$ is the $j$-th vertex
of exactly one path. Furthermore, finding the paths can be done in
time that is polynomial in the size of $\DBs$.
\end{fact*}
We actually prove the following slightly stronger result, which says
that if the permutation $\mu$ acts only on the $i$ last coordinates
of its input then the routing can be done in only $2i$ steps.
\begin{claim}
\label{cla:generalized-routing}Let $\DBs$ be a de-Bruijn graph and
let $i\in\left[m\right]$. Then, given a permutation $\mu$ on $\Lambda^{i}$
one can find a set of undirected paths of length $2\cdot i$ that
connect each vertex $\left(\alpha_{1},\ldots,\alpha_{m}\right)$ of
$\DBs$ to the vertex $\left(\alpha_{1},\ldots,\alpha_{m-i},\mu\left(\alpha_{m-i+1},\ldots,\alpha_{m}\right)\right)$
and that have the following two property: For every $j\in\left[l\right]$,
each vertex $v$ is the $j$-th vertex of exactly one path. Furthermore,
finding the paths can be done in time that is polynomial in the size
of $\DBs$.
\end{claim}
The proof works by induction on $i$. For $i=0$ the claim is obvious.
Assume that the claim holds for some $0\le i<m$. We prove that the
claim holds for~$i+1$. Let $\DB=\DBs$, and let $\mu$ be a permutation
on $\Lambda^{i+1}$. For convenience, let us define the action of
$\mu$ on each $\left(\alpha_{1},\ldots,\alpha_{m}\right)\in\F^{m}$
as $\mu\left(\alpha_{1},\ldots,\alpha_{m}\right)=\left(\alpha_{1},\ldots,\alpha_{m-i-1},\mu\left(\alpha_{m-i},\ldots,\alpha_{m}\right)\right)$.

Let $G$ be the directed graph whose vertices are the set $\Lambda^{m}$
and whose edges are all the pairs of the form $\left(v,\mu(v)\right)$.
Let $G'$ be the graph that is obtained from $G$ by contracting each
$\left|\Lambda\right|$ vertices of $G$ that agree on their last
coordinate to one vertex. Clearly, every vertex in $G'$ has in-degree
and out-degree exactly $\left|\Lambda\right|$, and each edge of $G'$
corresponds to an edge of $G$. Furthermore, observe that the vertices
of $G'$ can be identified with the vertices of $\Lambda^{m-1}$.

The $\left|\Lambda\right|$-regularity of $G$ implies that the edges
of $G'$ can be partitioned to $\left|\Lambda\right|$ perfect matchings
$\left\{ G_{\sigma}'\right\} _{\sigma\in\Lambda}$ in polynomial time
(see, e.g.,~\cite[Proposition 18.1.2]{C98}). Fix a matching $G_{\sigma}'$,
and consider an edge $e'$ in $G_{\sigma}'$. Observe that if $e$
is coming out of a vertex $\left(\alpha_{1},\ldots,\alpha_{m-1}\right)$
of $G'$, then it must enter a vertex of the form $\left(\alpha_{1},\ldots,\alpha_{m-i},\alpha_{m-i+1}',\ldots,\alpha_{m-1}'\right)$.
Thus, we can define a permutation $\nu_{\sigma}$ on $\Lambda^{i}$
that maps $\left(\alpha_{m-i},\ldots,\alpha_{m-1}\right)$ to $\left(\alpha_{m-i}',\ldots,\alpha_{m-1}'\right)$
for each such edge $e'$(since $G_{\sigma}'$ is a perfect matching,
this is well defined). We now invoke the induction hypothesis on the
graph $\DB=\DBs$ to find a set of paths $\mathcal{P}_{\sigma}$ of
length $2i$ for each permutation $\nu_{\sigma}$.

We construct the required paths for $\mu$ as follows. Let $v=\left(\alpha_{1},\ldots,\alpha_{m}\right)\in\Lambda^{m}$,
and suppose that $\mu\left(\alpha_{m-i},\ldots,\alpha_{m}\right)=\left(\alpha_{m-i}',\ldots,\alpha_{m}'\right)$.
We wish to construct a path $p$ in $\DB$ that connects $v$ to $\mu\left(v\right)$.
The edge $\left(v,\mu\left(v\right)\right)$ corresponds to some edge
$e'$ in $G'$, so let $G_{\beta}'$ be the matching to which $e'$
belongs. We turn to construct the path $p$: The first edge in the
path $p$ connects $v=\left(\alpha_{1},\ldots,\alpha_{m}\right)$
to the vertex $\left(\beta,\alpha_{1},\ldots,\alpha_{m-1}\right)$.
The next $2i$ edges of $p$ will be the edges of the path in $\mathcal{P}_{\beta}$
that connects $\left(\beta,\alpha_{1},\ldots,\alpha_{m-1}\right)$
to $\left(\beta,\alpha_{1},\ldots,\alpha_{m-i-1},\alpha_{m-i}',\ldots,\alpha_{m-1}'\right)$.
Finally, the last edge of $p$ will go from the vertex $\left(\beta,\alpha_{1},\ldots,\alpha_{m-i-1},\alpha_{m-i}',\ldots,\alpha_{m-1}'\right)$
to the vertex $\left(\alpha_{1},\ldots,\alpha_{m-i-1},\alpha_{m-i}',\ldots,\alpha_{m}'\right)=\mu\left(v\right)$.
Observe that $p$ indeed connects $v$ to $\mu\left(v\right)$ and
is of length $2\cdot\left(i+1\right)$

It remains to show that for each $j\in\left[2i+2\right]$ it holds
that every vertex $v$ is the $j$-th vertex of exactly one path.
The cases of $j=1$ and $j=2\cdot i+2$ are trivial. We analyze the
case of $j=2$, and the rest of the cases will follow from the induction
hypothesis. Let $u=\left(\beta,\alpha_{1},\ldots,\alpha_{m-1}\right)\in\Lambda^{m}$.
We show that $u$ is the second vertex of a unique path $p$ by constructing
$p$. Let $e'$ be the unique edge of $G'$ that comes out of the
vertex $\left(\alpha_{1},\ldots,\alpha_{m-1}\right)$ and that belongs
to the matching $G_{\beta}'$. The edge $e'$ of $G'$ corresponds
to some unique edge $\left(v,\mu\left(v\right)\right)$ of $G$. Now,
by construction, the only path $p$ such that $u$ is the second vertex
of $p$ is the path that connects $v$ to $\mu\left(v\right)$. The
required result follows.\qed

\section{\label{sec:proof-of-projection-to-sides}Proof of Claim~\ref{cla:projection-to-sides}}

In this section, we prove Claim~\ref{cla:projection-to-sides}, restated
below. Recall that $G=\left(\F^{m},E\right)$ is a graph with linear
structure and in particular $E$ is a linear subspace of edges.
\begin{claim*}
[\ref{cla:projection-to-sides}, restated]Let $d\in\N$ and let
$E_{a}$ be a uniformly distributed $d$-subspace of $\EE$. Then,
$\Pr\left[\dim\left(\l\left(E_{a}\right)\right)=d\right]\ge1-d/q^{m-d}$,
and conditioned on $\dim\left(\l\left(E_{a}\right)\right)=d$, it
holds that $\l\left(E_{a}\right)$ is a uniformly distributed $d$-subspace
of $\F^{m}$. The same holds for $\r\left(E_{a}\right)$.

More generally, let $E_{b}$ be a fixed subspace of $\EE$ such that
$\dim\left(E_{b}\right)>d$ and $\dim\left(\l\left(E_{b}\right)\right)>d$.
Let $E_{a}$ be a uniformly distributed $d$-subspace of $E_{b}$.
Then, $\Pr\left[\dim\left(\l\left(E_{a}\right)\right)=d\right]\ge1-d/q^{\dim\left(\l\left(E_{b}\right)\right)-d}$,
and conditioned on $\dim\left(\l\left(E_{a}\right)\right)=d$, it
holds that $\l\left(E_{a}\right)$ is a uniformly distributed $d$-subspace
of $\l\left(E_{b}\right)$. Again, the same holds for $\r\left(E_{a}\right)$.\end{claim*}
\begin{myproof}
We prove the proposition only for special case in which $E_{b}=E$
and only for $\l\left(E_{a}\right)$. The proof of the general case
and of the case of for $\r\left(E_{a}\right)$ is analogous. Let $e_{1},\ldots,e_{d}$
be independent and uniformly distributed vectors of $\EE$, and let
$E_{a}'=\span\left\{ e_{1},\ldots,e_{d}\right\} $. We prove Proposition~\ref{cla:projection-to-sides}
by showing that $E_{a}$ is distributed similarly to $E_{a}'$, and
analyzing the distribution of $E_{a}'$.

Observe that by Proposition~\ref{pro:random-vectors-have-full-dim},
it holds that conditioned on $\dim\left(E_{a}'\right)=d$, the subspace
$E_{a}'$ is a uniformly distributed $d$-subspace of $\EE$. It therefore
holds that
\begin{eqnarray*}
\Pr\left[\dim\left(\l\left(E_{a}\right)\right)=d\right] & = & \Pr\left[\dim\left(\l\left(E_{a}'\right)\right)=d|\dim\left(E_{a}'\right)=d\right]\\
 & \ge & \Pr\left[\dim\left(\l\left(E_{a}'\right)\right)=d\aand\dim\left(E_{a}'\right)=d\right]\\
 & = & \Pr\left[\dim\left(\l\left(E_{a}'\right)\right)=d\right],
\end{eqnarray*}
where the last equality holds since clearly $\dim\left(\l\left(E_{a}'\right)\right)=d$
implies $\dim\left(E_{a}'\right)=d$. Now, since $\l\left(\cdot\right)$
is a linear function, it holds that $\l\left(e_{1}\right),\ldots\l\left(e_{d}\right)$
are independent and uniformly distributed vectors of $\l\left(\EE\right)=\F^{m}$,
and therefore by Proposition~\ref{pro:random-vectors-have-full-dim}
it holds that $\Pr\left[\dim\left(\l\left(E_{a}'\right)\right)=d\right]\ge1-d/q^{m-d}$.
It thus follows that $\Pr\left[\dim\left(\l\left(E_{a}\right)\right)=d\right]\ge1-d/q^{m-d}$,
as required.

It remains to show that conditioned on $\Pr\left[\dim\left(\l\left(E_{a}\right)\right)=d\right]$
it holds that $\l\left(E_{a}\right)$ is a uniformly distributed $d$-subspace
of $\F^{m}$. To see it, observe that for every fixed $d$-subspace
$D$ of $\F^{m}$, it holds that
\begin{eqnarray*}
\Pr\left[\l\left(E_{a}\right)=D|\dim\left(\l\left(E_{a}\right)\right)=d\right] & = & \Pr\left[\l\left(E_{a}'\right)=D|\dim\left(E_{a}'\right)=d\aand\dim\left(\l\left(E_{a}'\right)\right)=d\right]\\
 & = & \Pr\left[\l\left(E_{a}'\right)=D|\dim\left(\l\left(E_{a}'\right)\right)=d\right],
\end{eqnarray*}
where the first equality again holds since conditioned on $\dim\left(E_{a}'\right)=d$
it holds that $E_{a}'$ is a uniformly distributed $d$-subspace,
and the second equality again holds since $\dim\left(\l\left(E_{a}'\right)\right)=d$
implies $\dim\left(E_{a}'\right)=d$. Now, it holds that $\l\left(E_{a}'\right)$
is the span of $d$ uniformly distributed vectors of $\F^{m}$, and
therefore by Proposition~\ref{pro:random-vectors-have-full-dim}
it holds that conditioned on $\dim\left(\l\left(E_{a}'\right)\right)=d$
the subspace $\l\left(E_{a}'\right)$ is a uniformly distributed $d$-subspace
of $\l\left(E_{b}\right)$. This implies that the probability 
\[
\Pr\left[\l\left(E_{a}'\right)=D|\dim\left(\l\left(E_{a}'\right)\right)=d\right]
\]
 is the same for all possible choices of $D$, and therefore the probability
\[
\Pr\left[\l\left(E_{a}\right)=D|\dim\left(\l\left(E_{a}\right)\right)=d\right]
\]
 is the same for all possible choices of $D$, as required.
\end{myproof}

\section{\label{sec:any-PCP-decoder-implies-decoding-graphs}Proof of Proposition~\ref{pro:any-PCP-decoder-implies-decoding-graphs}}

In this section we prove Proposition~\ref{pro:any-PCP-decoder-implies-decoding-graphs},
restated below.
\begin{proposition*}
[\ref{pro:any-PCP-decoder-implies-decoding-graphs}, restated]Let
$\iab$, $\oab$, $r(n)$, $q(n)$, $\ell(n)$, $s(n)$, and $\rho(n)$
be as in Definition~\ref{def:Unique-Decodable-PCPs}, and let $h_{0}$
and $d_{0}$ be the constants from Fact~\ref{fac:expanders-exist}.
If there exists a udPCP $D$ for $\csat_{\iab}$ with the foregoing
parameters, then there exists a polynomial time procedure that acts
as follows. When given a circuit $\varphi:\Gamma^{t}\to\B$ of size
$n$, the procedure outputs a corresponding decoding graph $G=\left(V,E\right)$
$q(n)\cdot d_{0}\cdot t\cdot2^{r(n)}$ with randomness complexity
$r(n)+\log\left(d_{0}\cdot q(n)\right)$, alphabet $\oab^{q(n)}$,
decoding complexity $s(n)+\poly\log\left|\Sigma(n)\right|$, and rejection
ratio $\Omega\left(\rho(n)/\left(q(n)\right)^{2}\right)$. Furthermore,
$G$ is~$\left(q(n)\cdot d_{0}\right)$-regular, and has $t\cdot2^{r(n)}$
vertices and smoothness~$1$.
\end{proposition*}
Fix $n\in\N$ and let $r=r(n)$, $q=q(n)$, $\ell=\ell\left(n\right)$,
$\Sigma=\Sigma(n)$, and $s=s(n)$. We describe the output of the
procedure on fixed circuit $\varphi:\Gamma^{t}\to\B$ of size $n$.
The procedure outputs a decoding graph $G$ defined as follows: 
\begin{itemize}
\item The vertices set of $G$ is the set $\left[t\right]\cdot\B^{r}$,
whose elements are identified with all the pairs $\left(k,\omega\right)$
where $k\in\left[t\right]$ is an index to be decoded and $\omega$
is a sequence of coin tosses of $D$ on input $\left(\varphi,k\right)$.
We denote by $I_{\left(k,\omega\right)}$ and $\psi_{\left(k,\omega\right)}$
are the queries tuple and circuit that are output by $D$ on input
$\left(\varphi,k\right)$ and coin tosses $\omega$. 
\item The alphabet of $G$ is $\Sigma^{q}$. 
\item The edges of $G$ are constructed as follows. For every $i\in\left[\ell\right]$,
we let $C_{i}$ be the set of pairs $\left(k,\omega\right)$ such
that on $I_{\left(k,\omega\right)}$ contains $i$. For each $i\in\left[\ell\right]$,
we consider the expander $G_{\left|C_{i}\right|}$ over $\left|C_{i}\right|$
vertices from Fact~\ref{fac:expanders-exist}, and identify its vertices
with the elements of $C_{i}$. Now, for each undirected edge of $G_{\left|C_{i}\right|}$,
we put two directed edges between the corresponding vertices in $C_{i}$,
one edge per direction. 
\item If an edge is coming out from a vertex $\left(k,\omega\right)$, then
it is associated with the index $k$. 
\item The circuits $\psi_{e}$ associated with the edges are constructed
as follows. Let $e$ be an edge going from $\left(k_{1},\omega_{1}\right)$
to $\left(k_{2},\omega_{2}\right)$, let $\psi_{e}$ be the associated
circuit. Suppose that $\left(k_{1},\omega_{1}\right)$ and $\left(k_{2},\omega_{2}\right)$
belong to $C_{i}$, so there exist $j_{1},j_{2}\in\left[q\right]$
such that ${(I_{(k_{1},\omega_{1})})}_{j_{1}}={(I_{(k_{2},\omega_{2})})}_{j_{2}}=i$.
Now, the circuit $\psi_{e}$ is given as input two tuples $a,b\in\Sigma^{q}$,
outputs $\bot$ if $a_{j_{1}}\ne b_{j_{2}}$, and otherwise outputs
$\psi_{(k_{1},\omega_{1})}(a)$. 
\end{itemize}
Let $\ell'$ and $n'$ denote the numbers of vertices and edges of
$G$. It is easy to see that the decoding graph $G$ has the correct
size, randomness complexity, alphabet, decoding complexity, and number
of vertices, and also that it is $q\cdot d_{0}$-regular. To see that
it has smoothness~$1$, consider an edge $\left(u,v\right)$ that
is chosen under the decoding distribution and observe that 
\begin{itemize}
\item $u$ is uniformly distributed among the vertices of $G$. 
\item Conditioned on the choice of $u$, the edge $\left(u,v\right)$ is
uniformly distributed among the edges of $u$. 
\end{itemize}
Combining the two above observations with the regularity of $G$ implies
that the decoding distribution of $G$ is the uniform distribution
over the edges.

We turn to show the completeness of $G$. Let $x$ be a satisfying
assignment for $\varphi$, and let $\pi=\pi_{x}$ be the corresponding
proof string for $D$. We define an assignment $\Pi$ to the vertices
of $G$ by defining $\Pi_{\left(k,\omega\right)}$ to be $\pi_{|I_{\left(k,\omega\right)}}$.
It should be clear that this choice of $\Pi$ satisfies the requirements.

It remains to analyze the rejection ratio of $G$. Let $\Pi$ be an
assignment to $G$. For each vertex $\left(k,\omega\right)$, if for
some $j\in\left[q\right]$ it holds that ${(I_{\left(j,\omega\right)})}_{j}=i$,
then we refer to $\left(\Pi_{\left(k,\omega\right)}\right)_{j}$ as
the opinion of $\left(k,\omega\right)$ on $i$, and also as the $j$-th
opinion of $\left(k,\omega\right)$. Let $\pi$ be the proof string
for $D$ defined by setting $\pi_{i}$ to be the most popular opinion
of a vertex of $G$ on $i$. Suppose that $D$ has decoding error~$\varepsilon$
on $\pi$ and let $x$ be the satisfying assignment to $\varphi$
that achieves this decoding error. Let $\varepsilon'$ be the decoding
error of $G$ on $\Pi$ with respect to $x$. We show that at least
$\frac{\rho}{q}\cdot\varepsilon'$~fraction of the edges of $G$
reject $\Pi$, and this will establish the rejection ratio of $G$.

Let $\eta$ be the fraction of vertices of $G$ that have an opinion
that is inconsistent with $\pi$. Clearly, $\varepsilon'\le\varepsilon+\eta$:
To see it, note that for at least $1-\varepsilon-\eta$ of the vertices
$\left(k,\omega\right)$ of $G$ it holds that all the opinions of
of $\left(k,\omega\right)$ are consistent with $\pi$ and that $D$
does not err on proof string $\pi$ and on $\left(k,\omega\right)$
(i.e. $\psi_{\left(k,\omega\right)}\left(\pi_{|I_{\left(j,\omega\right)}}\right)\in\left\{ \bot,x_{k}\right\} $).
Then, observe that all the outgoing edges of such a vertex $\left(k,\omega\right)$
do not err.

Let $k$ be uniformly distributed over $\left[t\right]$. We consider
two possible cases. First, consider the case in which $\eta\le\rho\cdot\varepsilon/2$.
By the soundness of $D$, it holds that $D$ rejects $\pi$ with probability
at least~$\rho\cdot\varepsilon$. Thus, at least $\rho\cdot\varepsilon$~fraction
of the vertices $\left(k,\omega\right)$ of $G$, it holds that $D$
rejects $\pi$ on $\left(k,\omega\right)$. This implies that at least
$\left(\rho\cdot\varepsilon-\eta\right)$~fraction of the vertices
$\left(k,\omega\right)$ of $G$, it holds that both $D$ rejects
$\pi$ on $\left(k,\omega\right)$ \emph{and} all the opinions of
$\left(k,\omega\right)$ are consistent with $\pi$, in which case
all the outgoing edges of $\left(k,\omega\right)$ reject $\Pi$.
It follows that the fraction of edges of $G$ that reject $\Pi$ is
at least
\[
\rho\cdot\varepsilon-\eta\ge\rho\cdot\varepsilon/2\ge\frac{1}{2}\cdot\eta+\frac{\rho}{4}\cdot\varepsilon\ge\frac{\rho}{4}\left(\eta+\varepsilon\right)\ge\frac{\rho}{4}\cdot\varepsilon',
\]
 as required.

We turn to consider the case in which $\eta\ge\rho\cdot\varepsilon/2$.
By averaging, there exists some $j\in\left[q\right]$ such that for
at least $\eta/q$~fraction of the vertices $\left(k,\omega\right)$
of $G$ it holds that the $j$-th opinion of $\left(k,\omega\right)$
is inconsistent with $\pi$. For every $i\in\left[\ell\right]$, denote
by $S_{i}$ the set of vertices of $C_{i}$ whose $j$-th opinion
is an opinion on~$i$ that is inconsistent with $\pi_{i}$, and observe
that
\[
\frac{1}{\ell'}\cdot\sum_{i=1}^{\ell}\left|S_{i}\right|\ge\frac{\eta}{q}.
\]
Fix $i\in\left[\ell\right]$ and denote $\overline{S}_{i}=C_{i}\backslash S_{i}$,
and note that since $\pi_{i}$ is the plurality vote it holds that
$\left|S_{i}\right|\le\left|C_{i}\right|/2$. Now, observe that every
edge that goes from $S_{i}$ to $\overline{S}_{i}$ or vice versa
must reject $\Pi$. By the edge expansion of $G_{\left|C_{i}\right|}$,
the number of such edges is at least $h_{0}\cdot d_{0}\cdot\left|S\right|$.
Since this holds for every $i\in\left[\ell\right]$, it follows that
the fraction of edges of $G$ that reject $\Pi$ is at least 
\begin{eqnarray*}
\frac{1}{n'}\cdot\sum_{i=1}^{\ell}h_{0}\cdot d_{0}\cdot\left|S_{i}\right| & = & \frac{1}{q\cdot d_{0}\cdot\ell'}\cdot\sum_{i=1}^{\ell}h_{0}\cdot d_{0}\cdot\left|S_{i}\right|\\
 & = & \frac{h_{0}}{q\cdot\ell'}\cdot\sum_{i=1}^{\ell}\left|S_{i}\right|\\
 & \ge & \frac{h_{0}}{q}\cdot\frac{\eta}{q}\\
 & \ge & \frac{h_{0}}{2\cdot q^{2}}\cdot\rho\cdot\varepsilon,
\end{eqnarray*}
where the first equality follows since $G$ is $\left(q\cdot d_{0}\right)$-regular.
The required result follows.\qed

\section{\label{sec:decoding-increasing-vertices-number}Proof of Proposition~\ref{pro:decoding-increasing-vertices-number}}

In this section we prove Proposition~\ref{pro:decoding-increasing-vertices-number},
restated below.
\begin{proposition*}
[\ref{pro:decoding-increasing-vertices-number}, restated]There
exists a polynomial time procedure that acts as follows:
\begin{itemize}
\item \textbf{Input:}

\begin{itemize}
\item A vertex-decoding graph $G$ of size $n$ for input circuit $\varphi:\Gamma^{t}\to\B$
with $\ell$ vertices, alphabet $\Sigma$, rejection ratio~$\rho$,
decoding complexity $s$, degree bound~$d$, and smoothness~$\smooth$.
\item A number $\ell'\in\N$ such that $\ell'\ge\ell$ (given in unary).
\end{itemize}
\item \textbf{Output: }Let $c\eqdef\left\lfloor \frac{\ell'}{\ell}\right\rfloor $
and let $d_{0}$ and $h_{0}$ be the constants from Fact~\ref{fac:expanders-exist}.
The procedure outputs a vertex-decoding graph $G'$ of size at most
$2\cdot(c+1)\cdot d_{0}\cdot n$ for input circuit~$\varphi$ that
has exactly $\ell'$ vertices and also has alphabet $\Sigma$, output
size $s+\poly\log\left|\Sigma\right|$, rejection ratio~$\Omega\left(\smooth^{2}\cdot\rho/d^{2}\right)$,
degree bound $2\cdot d_{0}\cdot d$, and smoothness $\frac{1}{2}\cdot\smooth$.
\end{itemize}
Furthermore, if $G$ is $d$-regular then $G'$ is $\left(2\cdot d_{0}\cdot d\right)$-regular
and has rejection ratio $\Omega\left(\smooth^{2}\cdot\rho\right)$.
\end{proposition*}
Let $G=\left(V,E\right)$, $\varphi$, $\ell$, and $\ell'$ be as
in the proposition and let $z=\ell'\mbox{ mod }\ell$. We construct
$G'$ as follows. Choose an arbitrary set $T\subseteq V$ of size
$z$. The vertices of $G'$ consist of a set $C_{v}$ of vertices
for each $v\in V$, where $\left|C_{v}\right|=c+1$ if $v\in T$ and
$\left|C_{v}\right|=c$ otherwise. Observe that $G'$ indeed has $\ell'$
vertices. For each $v\in V$ let us denote $C_{v}=\left\{ v_{1},\ldots,v_{\left|C_{v}\right|}\right\} $.
The edges of $G'$ are defined as follows:
\begin{enumerate}
\item For each edge $\left(u,v\right)$ of $G$ and for each $l\in\left[c\right]$,
the graph $G'$ has $d_{0}$ edges $\left(u_{l},v_{l}\right)$ that
are associated with the same index $k_{\left(u,v\right)}$ and circuit
$\psi_{\left(u,v\right)}$ as the edge $\left(u,v\right)$ of $G$.
We call such edges ``$G$-edges''.
\item \label{enu:increasing-vertices-num-trivial-edges}For each edge $\left(u,v\right)$
for which $u\in T$, the graph $G'$ contains the following ``trivial''
edges: Let $jk=k_{\left(u,v\right)}$ and $\psi=\psi_{\left(u,v\right)}$
be the index and circuit associated with $\left(u,v\right)$. Recall
that since $G$ is vertex-decoding, there exists a function $f:\Sigma\to\Gamma$
such that for every $a,b\in\Sigma$ on which $\psi\left(a,b\right)\ne\bot$,
it holds that $\psi\left(a,b\right)=f(a)$. Let $\psi':\Sigma^{2}\to\Gamma\cup\left\{ \bot\right\} $
be the circuit that for every input $\left(a,b\right)\in\Sigma^{2}$
outputs $f(a)$. The graph $G'$ contains $d_{0}$ edges $\left(u_{c+1},u_{c+1}\right)$
that are associated with the index $k$ and with the circuit $\psi'$.
\item For each edge $\left(u,v\right)$ of $G$ the graph $G'$ contains
the following edges, which correspond to ``equality constraints'':
Let $k=k_{\left(u,v\right)}$ and $\psi=\psi_{\left(u,v\right)}$
be the index and circuit associated with $\left(u,v\right)$, and
let $f:\Sigma\to\Gamma$ as in Item~\ref{enu:increasing-vertices-num-trivial-edges}.
Let $\psi'$ be the circuit that on input $\left(a,b\right)\in\Sigma^{2}$
outputs $\bot$ if $a\ne b$ and outputs $f(a)$ otherwise. We now
identify the vertices of $C_{u}$ with the vertices of the expander
$G_{\left|C_{u}\right|}$ from Fact~\ref{fac:expanders-exist}, and
for every (undirected) edge of $G_{\left|C_{u}\right|}$ we put two
directed edges between the corresponding vertices of $C_{u}$, where
the directed edges are associated with the index $k$ and with the
circuit $\psi'$. We call such edges ``consistency edges'' of $u$.
\end{enumerate}
Let $n'$ be the size of $G'$. It is easy to see that $G'$ has the
correct size, alphabet, decoding complexity, and degree bound,  and
also that $G'$ satisfies the completeness requirement. It can also
be verified that $G'$ has smoothness $\left(1-\frac{1}{c+1}\right)\cdot\smooth\ge\frac{1}{2}\cdot\gamma$
using the smoothness criterion (Proposition~\ref{pro:smoothness-criterion})
and a straightforward calculation.

It remains to analyze the rejection ratio of $G'$. Let $\pi'$ be
an assignment to the vertices of $G'$, and let $\pi$ be the corresponding
plurality assignment to $G$. That is, $\pi$ is the assignment that
assigns each vertex $v$ of $G$ the most popular value among the
values that $\pi'$ assigns to vertices in $C_{v}$. Suppose that
$G$ has decoding error $\varepsilon$ on $\pi$ and let $x\in\Gamma^{t}$
be an assignment that attains this decoding error. Let $\varepsilon'$
be the decoding error of $G'$ on $\pi'$ with respect to $x$. We
will show that $G'$ rejects $\pi'$ with probability at least $\frac{h_{0}\cdot\smooth^{2}}{64}\cdot\rho\cdot\varepsilon'$
under the decoding distribution, and this clearly suffices since $\varepsilon'$
is an upper bound on the decoding error of $G'$. To this end, we
will analyze the decoding error and rejection probability of $G'$
under the uniform distribution on the edges, and then use the smoothness
of $G'$ to derive conclusions on the decoding distribution.

By the smoothness of $G'$, the probability that a uniformly distributed
edge of $G'$ fails to decode $x$ on $\pi'$ is at least $\varepsilon_{1}'\eqdef\frac{1}{2}\cdot\smooth\cdot\varepsilon'$.
Furthermore, a uniformly distributed edge of $G$ fails to decode
$x$ on $\pi$ with probability at least $\varepsilon_{1}\eqdef\smooth\cdot\varepsilon$
and rejects with probability at least $\rho\cdot\varepsilon_{1}=\smooth\cdot\rho\cdot\varepsilon$.
Let $\eta$ be the fraction of vertices of $G'$ on which $\pi'$
is inconsistent with $\pi$. We begin the analysis by expressing $\varepsilon_{1}'$
in terms of $\varepsilon_{1}$ and $\eta$.

Let $F$ be the set of edges of $G$ that fail to decode $x$ on $\pi$,
let $F'$ be the set of edges of $G'$ that fail to decode $x$ on
$\pi'$, and let $S'$ be the set of vertices of $G'$ on which $\pi'$
is inconsistent with plurality assignment $\pi$, so $\eta\eqdef\left|S'\right|/\ell'$.
An edge $e'=\left(u,v\right)$ of $G'$ is in $F'$ if and only if
$e'$ corresponds to some $e\in F$ or if $u$ is in $S'$ (note that
since $G'$ is vertex-decoding, we need not consider the case where
$v$ is in $S'$). Now, every edge in $F$ has $d_{0}\cdot c$ corresponding
$G$-edges in $G'$, and every vertex in $S'$ has at most $2\cdot d_{0}\cdot d$
outgoing edges. Thus, it holds that
\[
\left|F'\right|\le d_{0}\cdot c\cdot\left|F\right|+2\cdot d_{0}\cdot d\cdot\left|S'\right|
\]
Observe that since every vertex of $G$ has at least one outgoing
edge (since $G$ is vertex-decoding), it holds that every vertex in
$G'$ has at least $2\cdot d_{0}$ outgoing edges, and therefore $n'\ge2\cdot d_{0}\cdot\ell'$
. It follows that 
\begin{eqnarray}
\varepsilon_{1}' & = & \frac{\left|F'\right|}{n'}\label{eq:increasing-vertices-num-expressing-error-of-G'}\\
 & \le & \frac{d_{0}\cdot c\cdot\left|F\right|+2\cdot d_{0}\cdot d\cdot\left|S'\right|}{n'}\nonumber \\
 & \le & \frac{d_{0}\cdot c\cdot\left|F\right|}{2\cdot d_{0}\cdot c\cdot n}+\frac{2\cdot d_{0}\cdot d\cdot\left|S'\right|}{2\cdot d_{0}\cdot\ell'}\nonumber \\
 & \le & \varepsilon_{1}+d\cdot\eta.\nonumber 
\end{eqnarray}
Observe that the last inequality implies that if $\eta$ is small
compared to $\varepsilon_{1}'$ then $\varepsilon_{1}$ must be large,
and vice versa. We turn to consider each of the cases separately.

\paragraph*{The case where $\eta$ is small.}

First, consider the case where $\eta\le\rho\cdot\varepsilon_{1}'/16\cdot d$.
In this case, we argue that $\pi'$ is roughly consistent with $\pi$,
and therefore the action of $G'$ on $\pi'$ is similar to the action
of $G$ on $\pi$. In particular, we argue that the fraction of edges
of $G'$ that reject $\pi'$ must be related to the fraction of edges
of $G$ that reject $\pi$, which is at least $\rho\cdot\varepsilon_{1}$.
However, since by Inequality~\ref{eq:increasing-vertices-num-expressing-error-of-G'}
it holds that $\varepsilon_{1}$ is large compared to $\varepsilon_{1}'$,
it will follow that the fraction of edges of $G'$ that reject $\pi'$
is roughly $\rho\cdot\varepsilon_{1}'$, as required.

More formally, it holds that the fraction of edges touching $S'$
(both incoming and outgoing) is at most 
\begin{eqnarray*}
\frac{2\cdot d_{0}\cdot d\cdot\left|S'\right|}{n'} & = & \frac{2\cdot d_{0}\cdot d\cdot\eta\cdot\ell'}{n'}\\
\mbox{(Since \ensuremath{n'\ge2\cdot d_{0}\cdot\ell'})} & \le & \frac{2\cdot d_{0}\cdot d\cdot\eta}{2\cdot d_{0}}\\
\mbox{(By assumption on \ensuremath{\eta})} & \le & \frac{d_{0}\cdot d\cdot\rho\cdot\varepsilon_{1}'}{d_{0}\cdot16d}\\
 & = & \frac{\rho\cdot\varepsilon_{1}'}{16}
\end{eqnarray*}
On the other hand, it holds that the size of $F$ (the set of edges
of $G$ that reject $\pi$) is at least $\rho\cdot\varepsilon_{1}\cdot n$.
Each such edge has at least $d_{0}\cdot c$ corresponding $G$-edges
in $G'$, and since $n'\le2\cdot d_{0}\cdot(c+1)\cdot n$, it follows
that the fraction of edges of $G'$ that correspond to edges in $F$
is at least $\left(\frac{d_{0}\cdot c\cdot\left|F\right|}{2\cdot d_{0}\cdot\left(c+1\right)\cdot n}\right)\ge\rho\cdot\varepsilon_{1}/4$.
Furthermore, it holds that 
\[
\varepsilon_{1}\ge\varepsilon_{1}'-d\cdot\eta\ge\varepsilon_{1}'-\rho\cdot\varepsilon_{1}'/16\ge\varepsilon_{1}'/2.
\]
So in fact the fraction of edges in $G'$ that correspond to edges
in $F$ is at least $\rho\cdot\varepsilon_{1}/4\ge\rho\cdot\varepsilon_{1}'/8$.
This implies that the fraction of edges of $G'$ that both correspond
to edges in $F$ and whose endpoints are consistent with $\pi$ is
at least $\rho\cdot\varepsilon_{1}'/8-\rho\cdot\varepsilon_{1}'/16\ge\rho\cdot\varepsilon_{1}'/16$.
Since all of these edges reject $\pi'$, it follows that the fraction
of edges of $G'$ that reject $\pi'$ is at least $\rho\cdot\varepsilon_{1}'/16\ge\rho\cdot\frac{1}{2}\cdot\smooth\cdot\varepsilon'/16\ge\smooth\cdot\rho\cdot\varepsilon'/32$.
This implies that the rejection probability of $\pi'$ under the decoding
distribution of $G'$ is at least $\Omega\left(\smooth^{2}\cdot\rho\cdot\varepsilon'\right)$.
as required.

\paragraph*{The case where $\eta$ is large.}

We turn to consider the case where $\eta\ge\rho\cdot\varepsilon_{1}'/16\cdot d$.
In this case, the assignment $\pi'$ is quite inconsistent with $\pi$,
and we argue that a significant fraction of the consistency edges
reject $\pi'$. More formally, using similar considerations as in
the proof of Proposition~\ref{pro:any-PCP-decoder-implies-decoding-graphs},
every set $C_{v}$ contributes at least $h_{0}\cdot d_{0}\cdot\left|S'\cap C_{v}\right|$
rejecting consistency edges. Thus, there are at least $h_{0}\cdot d_{0}\cdot\left|S'\right|$
rejecting edges. This implies that the fraction of rejecting edges
is at least
\begin{eqnarray*}
\frac{h_{0}\cdot d_{0}\cdot\left|S'\right|}{n'} & \ge & \frac{h_{0}\cdot d_{0}\cdot\left|S'\right|}{2\cdot d_{0}\cdot d\cdot\ell'}\\
 & = & \frac{h_{0}}{2\cdot d}\cdot\eta\\
 & \ge & \frac{h_{0}}{32\cdot d^{2}}\cdot\rho\cdot\varepsilon_{1}'\\
 & \ge & \frac{h_{0}}{32\cdot d^{2}}\cdot\rho\cdot\frac{1}{2}\cdot\smooth\cdot\varepsilon'\\
 & \ge & \frac{h_{0}\cdot\smooth}{64\cdot d^{2}}\cdot\rho\cdot\varepsilon',
\end{eqnarray*}
which implies that the rejection probability under the decoding distribution
is at least $\Omega\left(\smooth^{2}\cdot\rho\cdot\varepsilon'/d^{2}\right)$,
as required.\medskip{}

\paragraph*{The ``furthermore'' part.}

For the ``furthermore'' part of the lemma, first observe that it
is easy to see from the definition of $G'$ that if $G$ is $d$-regular
then $G'$ is $\left(2\cdot d_{0}\cdot d\right)$-regular. For the
rejection ratio part, note that in the foregoing analysis we lose
a $1/d$ factor in two places:
\begin{enumerate}
\item We lose a factor of $1/d$ in the proof of Inequality~\ref{eq:increasing-vertices-num-expressing-error-of-G'},
where our upper bound on the number of edges that go out of $S$ is
$2\cdot d_{0}\cdot d\cdot\left|S\right|$ while our lower bound on
$n'$ is only $2\cdot d_{0}\cdot\ell'$. However, if $G$ is $d$-regular,
then $G'$ is $\left(2\cdot d_{0}\cdot d\right)$-regular, and thus
the lower bound on $n'$ can be improved to $2\cdot d_{0}\cdot d\cdot\ell'$.
This implies that Inequality~\ref{eq:increasing-vertices-num-expressing-error-of-G'}
becomes $\varepsilon_{1}'\le\varepsilon_{1}+\eta$.\\
As a result, the case of ``small $\eta$'' can be extended to all
the cases where $\eta\le\rho\cdot\varepsilon_{1}'/16$, and in the
case of ``large $\eta$'' we can assume that $\eta\ge\rho\cdot\varepsilon_{1}'/16$.
This saves a factor of $1/d$ in the case of ``large $\eta$''.
\item We lose a factor of $1/d$ in the case of ``large $\eta$'', since
the lower bound on the number of rejecting consistency edges for a
set $C_{v}$ is only $h_{0}\cdot d_{0}\cdot\left|S\cap C_{v}\right|$,
while the upper bound on the number of consistency edges in the graph
is $d_{0}\cdot d\cdot n$. However, if $G$ is $d$-regular then the
foregoing lower bound can be improved to $h_{0}\cdot d_{0}\cdot d\cdot\left|S\cap C_{v}\right|$,
regaining the factor of $1/d$.\qed\end{enumerate}

\end{document}